\documentclass[%
 reprint,
 amsmath,amssymb,
 aps,
floatfix,
]{revtex4-2}
\usepackage[utf8]{inputenc}
\usepackage[T1]{fontenc}

\usepackage{graphicx}
\usepackage{dcolumn}
\usepackage{bm}
\usepackage[printonlyused,nohyperlinks,smaller,nolist]{acronym}
\usepackage{subcaption}
\usepackage{alphalph}

\usepackage{amssymb}
\usepackage{amsmath}
\usepackage{amsfonts}
\newcommand{\diff}{\mathop{}\!\mathrm{d}}
\usepackage[frozencache,%
            cachedir=.]{minted}
\newcommand{\Lim}[1]{\raisebox{0.5ex}{\scalebox{0.8}{$\displaystyle \lim_{#1}\;$}}}
\makeatletter
\def\tagform@#1{\maketag@@@{\ignorespaces#1\unskip\@@italiccorr}}
\let\orgtheequation\theequation
\def\theequation{(\orgtheequation)}
\makeatother

\usepackage{hyperref}
\hypersetup{colorlinks,allcolors=black}

\usepackage{bookmark}

\begin{document}

\title{Pore-network models and effective medium theory:\\A convergence analysis}

\author{Jack Edwards}
 \email{jdedward@ualberta.ca}
\affiliation{%
 Department of Mechanical Engineering,\\ University of Alberta
}%

\author{Peter Berg}
 \email{pberg@ualberta.ca}
\affiliation{%
 Theoretical Physics Institute,\\ University of Alberta
}%

\date{\today}

\begin{abstract}
    The convergence between effective medium theory and pore-network modelling is examined. Electrical conductance on two and three-dimensional cubic resistor networks is used as an example of transport through composite materials or porous media. Effective conductance values are calculated for the networks using effective medium theory and pore-network models. The convergence between these values is analyzed as a function of network size. Effective medium theory results are calculated analytically and numerically. Pore-network results are calculated numerically using Monte Carlo sampling. The reduced standard deviations of the Monte Carlo sampled pore-network results are examined as a function of network size. Finally, a ``quasi-two-dimensional'' network is investigated to demonstrate the limitations of effective medium theory when applied to thin porous media. Power law fits are made to these data to develop simple models governing convergence. These can be used as a guide for future research that uses both effective medium theory and pore-network models.
\end{abstract}

\maketitle

\acrodef{EMT}{effective medium theory}
\acrodefplural{EMT}{effective medium theories}
\acrodef{PDF}{probability density function}
\acrodef{PNM}{pore-network model}
\acrodef{RD}{relative difference}
\acrodef{RSD}{reduced standard deviation}

\acresetall

\section{Introduction}
\label{sec:intro}

\Ac{EMT} is a useful tool for calculating the macroscopic properties of a composite material based upon the properties of its constituent parts. As defined by \citet{choy-2015-emt}, at its most basic, \ac{EMT}  ``functions by being able to define averages, which one hopes will be
representative of the system and be connected with experimental measurements.'' While the ability to directly calculate macro- or meso-scale behaviour based upon microscopic properties has improved with increasing computational power, it remains unfeasible to do so for all applications. Replacing a composite material with a uniform effective medium that replicates its larger-scale properties can greatly simplify calculations involving that material.

\Ac{PNM} techniques are another tool for analyzing the macroscopic behaviour of materials based upon the physics at the microscopic level. They are used to model porous media in a variety of fields, most notably Earth sciences, though it has seen more widespread use in recent years~\cite{xiong-2016-pnm}. The underlying logic of \ac{PNM} is to model porous media as a system of interconnected pores, on which transport equations can be calculated exactly. The porous medium can then be treated similarly to a network of resistors, and reduced to a linear system of equations analogous to Kirchhoff's Circuit Laws or a more complex, nonlinear system of equations. Ideally, the aggregate behaviour of the medium can be calculated through this method~\cite{fatt-1956-network}. 

The goal of this report is to examine the convergence behaviour between \ac{EMT} and \ac{PNM} models. To help contextualize these techniques, a brief history of \ac{EMT} is recounted in \autoref{sec:emt}. The development of \ac{PNM} is similarly detailed in \autoref{sec:rnm-pnm}. In \autoref{sec:pnm-emt}, the relationship between the two is explored and their continued development and use is presented. The central questions about this relationship that we hope to answer are detailed in \autoref{sec:problem}. 

The methodology used in this analysis is detailed in \autoref{sec:methods}. To narrow the scope, electrical conductance on resistor networks is examined using a particular \ac{EMT} derived by \citet{kirkpatrick-1973-percolation} and \ac{PNM} techniques. The open-source \ac{PNM} Python package Open\acs{PNM} is used for the pore-network calculations~\cite{gostick-2016-openpnm}. Effective conductances are calculated for both two-dimensional and three-dimensional networks with a variety of resistor conductance distributions. Through examining networks with different properties, the effect of differing network properties on the convergence between \ac{EMT} and \ac{PNM} techniques can be determined. Additionally, a ``quasi-two-dimensional'' network is examined, which is very large in two dimensions but small in the third. This is to simulate how a thin porous medium can affect the convergence between these techniques. The results are presented in \autoref{sec:results} and a discussion into their implications is presented in \autoref{sec:discussion}. 

The ultimate aim of this report is to guide the development of research into composite and porous materials, especially in regards to transport problems. Through the examination of the convergence behaviours between the macroscopic models of \ac{EMT} and the discretized, microscopic approaches used in \ac{PNM}, underlying trends governing their relationship are revealed. This can be used as a framework for future efforts in these fields, in determining what problems are likely suitable for \ac{EMT} treatment and what domain sizes can be considered sufficient for \ac{PNM} to adequately model macroscopic properties. 

\section{Literature Review}
\label{sec:lit-review}

\subsection{Effective medium theory}
\label{sec:emt}

Attempts to describe the macroscopic properties of composite materials can be traced back to the nineteenth century. Early work in the field can be attributed to the physicists \citet{lorenz-1880-emt} and \citet{lorentz-1880-emt}. This research related the refractive indices of a particular medium to its density, modelling the transport of light through the medium as it pertains to its average properties. They associated the transport in the dilute limit, where the obstacles to transport can be considered small in comparison to the distance between them. Lord Rayleigh built upon this foundation, expanding it to transport obstacles whose sizes are not negligible in regards to their spacing~\cite{rayleigh-1892-emt}. In this treatment, Rayleigh looked at the case of spherical or cylindrical objects with uniform conductance arranged in a ``rectangular or square'' (cubic) order, embedded within a conducting medium. Rayleigh's solution for the effective transport coefficient of this two-phase medium is 
\begin{equation}
    \label{eq:rayleigh-emt}
    \frac{D_{\mathrm{eff}} - D_{2}}{D_{\mathrm{eff}} + m D_{2}}
    =
    \frac{\nu_{1}}{\nu_{2}}
    \frac{D_{1} - D_{2}}{D_{1} + m D_{2}}
\end{equation}
when $\nu_{1} \ll \nu_{2}$, where $D_{i}$ and $\nu_{i}$ refers to the transport coefficient and volume fraction of the two phases, respectively (adapted from \citet{tjaden-2016-origin}). The variable $m$ is set to a value of 1 for cylinders and 2 for spheres.

These initial approaches, though often framed from the standpoint of optics and light refraction, had more general applications to heat or electrical conduction. These developments were important, but limited in their applicability. In the cases of Lorenz and Lorentz, their models were valid only in the dilute limit. For Rayleigh, this model was only valid for objects arranged in cubic lattices in which the flux of the system is parallel to one of the lattice constants.

The first traditional \ac{EMT} was the Maxwell Garnett formula, itself an extension of the earlier Clausius--Mossotti formula~\cite{choy-2015-emt,garnett-1904-colours}. It provides a method for calculating an effective (or bulk) dielectric constant, $D_{\mathrm{eff}}$, of a composite medium. If considering spherical inclusions of material 1 embedded into material 0, the equation takes the form
\begin{equation}
    \label{eq:maxwell-garnett-emt}
    \frac{D_{\mathrm{eff}} - D_{0}}
    {D_{\mathrm{eff}} + 2 D_{0}}
    =
    \nu_{1}
    \frac{D_{1} - D_{0}}
    {D_{1} + 2 D_{0}}.
\end{equation}
This is similar to \autoref{eq:rayleigh-emt}. However, the coefficient in front of the right-hand-side fraction in \autoref{eq:maxwell-garnett-emt} is only the volume fraction of the inclusions, as opposed to the ratio of the volume fractions of both materials in \autoref{eq:rayleigh-emt}. This allows the equation to be generalized to $n$ materials,
\begin{equation}
    \label{eq:maxwell-garnett-sum}
    \frac{D_{\mathrm{eff}} - D_{0}}
    {D_{\mathrm{eff}} + 2D_{0}}
    =
    \sum_{i=1}^{n}
    \nu_{i}
    \frac{D_{i} - D_{0}}
    {D_{i} + 2 D_{0}}.
\end{equation}
However, this \ac{EMT} was only accurate at small values of $\nu_{i}$. Taken to the extreme, if $\nu_{1}=1$ in \autoref{eq:maxwell-garnett-emt} (i.e. only material 1 is present), the calculated $D_{\mathrm{eff}}$ is still dependent on the value of $D_{0}$ despite material 0 being completely absent from the composite. It also does not model critical threshold behaviour.

The next major development of \ac{EMT}s came from \citet{bruggeman-1935-berechnung}. Bruggeman expanded Rayleigh's approach from strictly cubic lattices to random structures, and did so in a way that improved upon the shortcomings present in the Maxwell Garnett formula~\cite{tjaden-2016-origin,choy-2015-emt}. Bruggeman examined a sample with two distinct phases, 1 and 2, embedded within some homogeneous medium. Bruggeman's model was based upon three underlying assumptions:
\begin{enumerate}
    \item The two phases are homogeneous and isotropic.
    \item The particle sizes of the phases are small in comparison to the overall sample size.
    \item The phases have been randomly distributed within the sample.
\end{enumerate}
Through an iterative embedding procedure, the entire sample volume is filled with increasingly greater amounts of phases 1 and 2, completely replacing the embedding medium. Eventually, as the entire volume is filled with inclusions of these phases, the following result is found:
\begin{equation}
    \label{eq:bruggeman-emt}
    \nu_{1}
    \frac{D_{1} - D_{\mathrm{eff}}}
    {D_{1} + m D_{\mathrm{eff}}}
    +
    \nu_{2}
    \frac{D_{2} - D_{\mathrm{eff}}}
    {D_{2} + m D_{\mathrm{eff}}}
    =
    0.
\end{equation}
Much like the Maxwell Garnett formula, the terms in \autoref{eq:bruggeman-emt} are dependent only on the volume fraction $\nu_{i}$ of a single phase. This allows it to be generalized to an arbitrary number of phases, similar to \autoref{eq:maxwell-garnett-sum}:
\begin{equation}
    \label{eq:bruggeman-sum}
    \sum_{i=1}^{n}
    \nu_{i}
    \frac{D_{i} - D_{\mathrm{eff}}}
    {D_{i} + m D_{\mathrm{eff}}}
    =
    0.
\end{equation}
This is the main result from Bruggeman's analysis, which has become a widely used \ac{EMT} across many disciplines. 

Considering how important this result has been for the development of \ac{EMT}s, the barriers to accessing the original publication have hindered discussions of the result. It was published in German and, to our knowledge, has not been translated into other languages. However, excellent explanations of the theory are given by both \citet{tjaden-2016-origin} and \citet{choy-2015-emt}.

\subsection{Pore-network modelling}
\label{sec:rnm-pnm}

The techniques of \ac{EMT} can also be applied to transport through porous media, analogously to their applications in optics and electrical conductance. However, transport through porous media has been studied for longer than \ac{EMT}s have been available, and there exists a history of distinct techniques in this field. The earliest empirical studies can be traced to Henry Darcy and the development of Darcy's Law~\cite{darcy-1856-fontaines}. This research, based upon the flow of water through soil and sand, yielded a simple transport equation for a fluid with a known viscosity flowing through a medium with a known permeability under the influence of a known pressure gradient. One limitation of Darcy's Law is that it only applies to laminar flows, reducing its applicability in certain scenarios (for instance, turbulent flows through gravel). Nonetheless, Darcy's Law continues to be widely used. 

Attempts to describe the transport through porous media in more detail followed. Models involving systems of packed spheres were used, notably the Kozeny--Carman equation~\cite{fatt-1956-network}. Through this method, the permeability of the medium was related to the porosity and geometry of the sphere packs, with some proportionality constant. However, the complexity of the pore-geometry within this sphere-packed model prevented accurate mathematical derivations. This method, though a step toward more complex models, proved to be of limited use outside of specific fields (e.g. estimating the surface area of powders)~\cite{fatt-1956-network}.

The next attempts involved modelling porous media as bundles of tubes through which fluid flows~\cite{fatt-1956-network}. This geometry allowed for comparatively simpler and rigorous mathematical solutions. These gains made in mathematical rigour, however, were offset by the model's inability to accurately model several key features of porous media. For example, porous media tend to be isotropic, whereas the bundle of tubes model is anisotropic. Additionally, these approaches were incapable of modelling multiphase flow phenomena. 

In order to address some of the shortcomings present in these models, \citet{fatt-1956-network} took elements from both the packed-spheres models and bundle of tubes models to develop what he referred to as the network model, or network of tubes. This is now most-commonly referred to as a \acf{PNM}. \citet{fatt-1956-network} took the interconnectedness of the packed-spheres models, which better represents the isotropic properties of most porous media; and combined it with the cylindrical tubes of the bundle of tubes models, which are much simpler mathematically and allow for exact flow calculations to be performed. In this treatment, porous media can essentially be modelled as networks of resistors. \citet{fatt-1956-network} discusses how manipulating the distribution of cylinder radii within the network, lattice structure of the network, and other geometric properties can be used to model the properties of a variety of different materials. Indeed, it is this aspect of \ac{PNM}s that makes them useful to this day. 

\ac{PNM}s were initially mostly used within fields of Earth sciences, modelling the flow of fluids through porous rock or soil. Since then, their uses and applications have spread into many other fields, from nuclear waste management to electrochemical energy systems~\cite{xiong-2016-pnm,eikerling-1997-pnm,berg-2021-pnm}. 

\subsection{Pore-network modelling and effective medium theory}
\label{sec:pnm-emt}

When \citet{fatt-1956-network} first introduced \ac{PNM}s, it was an important achievement in the modelling of porous media. The ability to introduce random physical characteristics to the network, as well as the notions of ``coordination number'' to describe the connectivity of individual nodes within the network, was a pioneering achievement. However, Fatt's treatments were limited by the computing power available at the time (the initial calculations were performed by hand)~\cite{larson-1981-percolation,fatt-1956-network}. The networks considered were necessarily small and results were dominated by boundary effects. Additionally, randomized characteristics had to be described through discrete distributions, as opposed to more physically-accurate continuous distributions. Much like in Bruggeman's time, \ac{EMT}s remained a necessity in describing macroscopic properties of porous or composite media. Direct calculations through \ac{PNM} techniques were not feasible; only effective medium approximations could be practically used.

As computing technology advanced, \ac{PNM}s were able to grow larger and more complex. Extensions to Fatt's models were developed, adding more detail and increasing the size of the domains examined~\cite{payatakes-1973-porous,dullien-1975-porous}. Comparisons between the predictions of \ac{PNM}s and \ac{EMT} were made, examining the applicability of each in various situations~\cite{kirkpatrick-1973-percolation,eikerling-1997-pnm,bernasconi-1976-emt,burganos-1987-diffusion}. Today, computational power is such that many problems can be ``brute-forced'' in \ac{PNM} and similar models. 

Nevertheless, \ac{EMT}s are still extensively used. The Bruggeman model continues to be wide-spread, recently seeing greater applications in fields such as fuel cell and battery research~\cite{tjaden-2016-origin,vadakkepatt_bruggemans_2015,das_effective_2010,baschuk_modelling_2000}. When applied to porous media, transport through one of the phases is often negligible (e.g. water will flow in the spaces between grains of sand, but not through the sand itself). This allows \autoref{eq:bruggeman-emt} to be simplified to
\begin{equation}
    \label{eq:bruggeman-simple}
    D_{\mathrm{eff}}
    =
    D \frac{\nu}{\tau},
\end{equation}
where $D$ and $\nu$ are the transport coefficient and volume fraction of the transporting phase. $\tau$ is the tortuosity factor, given by
\begin{equation}
    \label{eq:tau}
    \tau = 
    \begin{cases}
    \nu^{-1/2}, & \text{for spherical inclusions} \\
    \nu^{-1}, & \text{for cylindrical inclusions}
    \end{cases}.
\end{equation}
\autoref{eq:bruggeman-simple} is the \ac{EMT} that is most often applied to porous media transport~\cite{tjaden-2016-origin}. 

\section{Problem Statement}
\label{sec:problem}

In this report, we will examine the applicability of \ac{EMT} to a given transport problem within a composite material. Specifically, we will calculate the effective electrical conductance within a composite material, though the results will be applicable to other problems which can be modelled with similar \ac{PNM} or resistor techniques. This is to answer several related questions: When calculating effective transport properties in a composite medium, how large does the system have to be for \ac{EMT} to model it accurately? How large of a network is needed for \ac{PNM} techniques to agree with the macroscale limit that \ac{EMT} describes? What is the rate of convergence between \ac{PNM} and \ac{EMT} results as the system size grows? What effect does the specific random conductance distribution have on this behaviour?

\section{Methods}
\label{sec:methods}

\subsection{Effective medium theory}
\label{sec:methods-emt}

\begin{figure}
    \centering
    \includegraphics[width=0.6\linewidth]{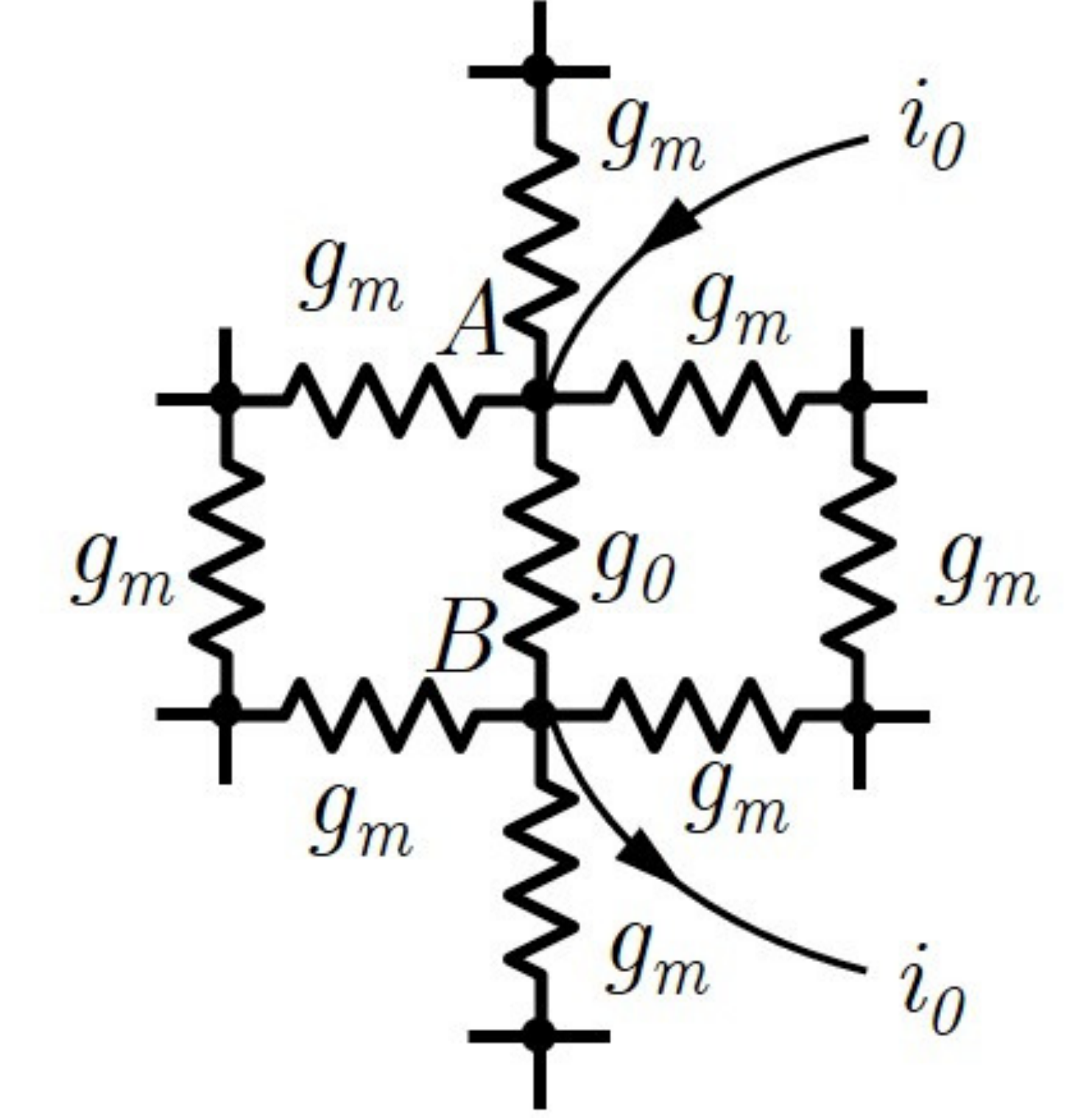}
    \caption{Example resistor network. The random network has been replaced with an effective medium of uniform conductance.}
    \label{fig:kirkpatrick-network}
\end{figure}

To address the problem in \autoref{sec:problem}, it is necessary to compare \ac{EMT} transport solutions to those directly calculated by \ac{PNM} techniques. To narrow the focus of this work, we will specifically look at an \ac{EMT} developed by \citet{kirkpatrick-1973-percolation}, a formulation based upon Bruggeman's work. \citeauthor{kirkpatrick-1973-percolation}'s formulation considers the case of a network of random resistors, arranged in a square (2D) or cubic (3D) lattice. An example network is shown in \autoref{fig:kirkpatrick-network}. 

Within such a network, the electric potential differences across the various resistors are generated by an external electric field. The potentials can then be treated as consisting of two contributions: the portion supplied by the external field, which increases the voltage at a given node by a uniform amount per row; and the portion from a fluctuating local field, which averages to zero over a sufficiently large area. The average effects of the random resistors are represented through an effective medium, where the total field inside it is equal to the external field. For this to hold, the medium must be homogeneous, represented as every resistor having an equal conductance $g_m$, as seen in \autoref{fig:kirkpatrick-network}. This formulation, used by Kirkpatrick, only considers identical resistors in nearest-neighbour arrangement on a cubic lattice \footnote{``For simplicity we shall consider [the network] to be made up of a set of equal conductances, $g_m$, connecting nearest neighbors on the cubic mesh,'' \citet{kirkpatrick-1973-percolation}.}. As Kirchhoff's Laws stipulate that the total current entering and exiting a node in the network must be zero, this allows the network to be described as a linear system of equations, 
\begin{equation}
    \label{eq:kirchhoff}
    \sum_{i} g_{ij} \left( V_i - V_j \right) = 0,
\end{equation}
where $g_{ij}$ represents conductance, $V_k$ is electric potential, and the subscripts $i$ refer to the nodes adjacent to node $j$ on the cubic network.

Now, the issue is to determine the value of $g_m$ that replaces the random conductances of the network such that the predicted effective properties are accurate for the given network. To do this, it must be ensured that the local fields in the effective medium average to zero. Kirkpatrick achieves this by considering a single resistor in the network, oriented along the external field, with a conductance of $g_{AB}=g_0$. This is seen in \autoref{fig:kirkpatrick-network}. The uniform field solution of this network has the voltages increasing by a constant amount per row, $V_m$, caused by the external field. The effects of a fictitious current $i_0$ are added to this system, introduced through node $A$ and extracted from node $B$. This causes an imbalance in the current conservation described in \autoref{eq:kirchhoff}. The new current equation caused by this imbalance is
\begin{equation}
    \label{eq:i0}
    V_m \left( g_m - g_0 \right) = i_0.
\end{equation}

\begin{figure}
    \centering
    \includegraphics[width=0.6\linewidth]{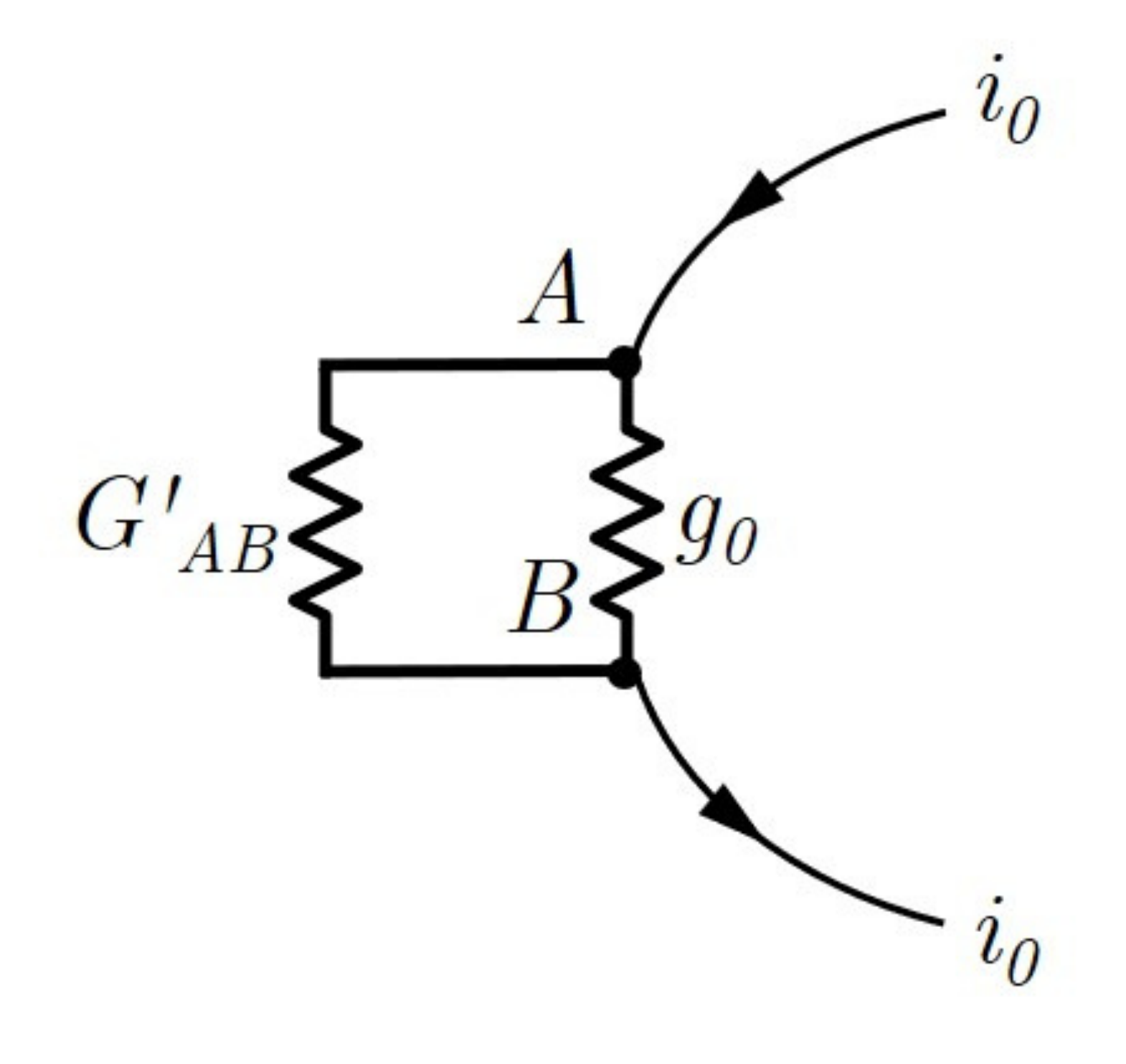}
    \caption{Example resistor network. The random network has been replaced with an equivalent resistor with conductance $G'_{AB}$.}
    \label{fig:kirkpatrick-network-equivalent}
\end{figure}

There is an extra voltage induced between $A$ and $B$, introduced by $i_0$. It can be calculated using the conductance $G'_{AB}$, the value for the equivalent resistor replacing the network with the resistor $g_0$ removed. This equivalent network is shown in \autoref{fig:kirkpatrick-network-equivalent}. The extra voltage is given by
\begin{equation}
    \label{eq:extra-voltage}
    V_0 = \frac{i_0}{g_m + G'_{AB}}.
\end{equation}
The conductance of the total effective medium, with the missing resistor reintroduced with $g_m$, is given by $G_{AB}=G'_{AB}+g_m$. 

Examining the network as shown in \autoref{fig:kirkpatrick-network}, the conductance $g_0$ can be replaced with the effective medium $g_m$. The current distribution in this network can be described as the sum of two components: a current $i_0$ introduced at $A$ and extracted at infinity, and an equal current introduced at infinity and extracted at $B$. In both cases, the current flowing through the $z$ equivalent bonds at the point the current enters the $AB$ resistor is $i_0/z$. This gives a total current of $2i_0/z$ flowing through the $AB$ bond. From this, we know that $G_{AB}=(z/2)g_m$, and subsequently that $G'_{AB}=(z/2 - 1)g_m$. Knowing now the current across this resistor, and knowing its conductance to be $g_m$, the voltage across this resistor can be determined. Substituting \autoref{eq:i0} into \autoref{eq:extra-voltage}, and substituting for $G'_{AB}$, we obtain the expression
\begin{equation}
    \label{eq:v0}
    V_0 
    =
    \frac{V_m \left( g_m - g_0 \right)}{g_0 + \left( \frac{z}{2} - 1 \right)g_m}.
\end{equation}

As the conductances in the resistor networks being examined are random, we can distribute the conductances with a \ac{PDF} $f(g)$, which may be continuous or discrete. We can then define a function $F(g_m)$,
\begin{equation}
    \label{eq:kirkpatrick-integral}
    F(g_m)
    =
    \int
     f(g) \frac{V_0 (g)}{V_m}
    \diff g
    =
    \int 
    \frac{ f(g) (g_m - g)}
    {g + \left( \frac{z}{2} - 1 \right) g_m}
    \diff g.
\end{equation} 

As stated previously, it is required that $V_0$ averages to zero. This requirement leads to the expression
\begin{equation}
    \label{eq:kirkpatrick-vanish}
    F(g_m)
    =
    \int 
    \frac{ f(g) (g_m - g)}
    {g + \left( \frac{z}{2} - 1 \right) g_m}
    \diff g
    =
    0,
\end{equation}
which can be used to determine the value of $g_m$. In essence, the solution to \autoref{eq:kirkpatrick-vanish} determines the effective conductance $g_m$ that can replace the random network with a uniform effective medium. Given any idealized square or cubic infinite lattice, with a known distribution of conductances, transport calculations can be performed exactly on it using \ac{EMT}. 

For certain discrete distributions, \autoref{eq:kirkpatrick-vanish} has analytic solutions. For example, take a discrete binary distribution
\begin{equation}
    \label{eq:binary}
    f(g) = p \delta(g - \alpha) + (1-p) \delta(g - \beta), 
\end{equation}
where $\delta(x)$ is the Dirac delta function, $p$ is the probability that a given bond has conductance $\alpha$, and $(1-p)$ is the probability that a given bond has conductance $\beta$. Substituting \autoref{eq:binary} into \autoref{eq:kirkpatrick-vanish}, we obtain the quadratic equation
\begin{multline}
    \label{eq:kirkpatrick-quadratic}
    \left(\frac{z}{2} - 1 \right)g_m^2
    +
    \left\{
    \beta \left[ \frac{z}{2} \left( p - 1 \right) + 1 \right]
    +
    \alpha \left[ 1 - \frac{z}{2} p \right]
    \right\} g_m \\
    -
    \alpha \beta
    =
    0.
\end{multline}
There are two roots in this equation, positive and negative. The appropriate value of $g_m$ is the positive root. For any discrete distribution of $N$ positive delta functions (e.g. a ternary distribution), \autoref{eq:kirkpatrick-vanish} will result in a polynomial equation of order $N$.

For a continuous \ac{PDF}, numerical techniques can be used to determine the solution. To show that \autoref{eq:kirkpatrick-integral} has exactly one positive root, the derivative with respect to $g_m$ is needed. This is given by
\begin{equation}
    \label{eq:kirkpatrick-derivative}
    F'(g_m) = \frac{\diff F(g_m)}{\diff g_m}
    =
    \int \frac{f(g) \frac{z}{2}g}
    {\left(g + \left(\frac{z}{2} - 1 \right)g_m \right)^2}
    \diff g.
\end{equation}
The numerator of \autoref{eq:kirkpatrick-derivative} is the product of a \ac{PDF}, $f(g)$, which is always non-negative; a positive integer, $z/2$; and a conductance, $g$, which is always non-negative. The denominator is the square of a positive real number, which is always positive. As a result, the value of \autoref{eq:kirkpatrick-derivative} is never negative, and is, in fact, always positive for any physically meaningful \ac{PDF}, $f(g)$. Moreover, we know that $F(0)=-1$ and that $\Lim{g_m \rightarrow \infty} F(g_m) > 0$. So, $F(g_m)$ in \autoref{eq:kirkpatrick-integral} always has a positive slope and always has exactly one intercept with the positive $g_m$ axis. This guarantees that the \ac{EMT} solution to \autoref{eq:kirkpatrick-vanish} is unique. Therefore, root finding methods always converge (e.g. Newton's method). 

To solve for the root, \autoref{eq:kirkpatrick-integral} and \autoref{eq:kirkpatrick-derivative} were written as Python functions. These functions accept a \ac{PDF} corresponding to $f(g)$ as an argument and numerically calculate the integral function utilizing Python's native numerical integration capabilities (namely, the \texttt{scipy.\hspace{0pt}integrate} package). The Python codes implementing $F(g)$, $F'(g)$, and the root finding methods are available in Appendices~\hyperref[app:code-emt]{\ref*{app:code-emt}} and \hyperref[app:code-root]{\ref*{app:code-root}}. 

It should be noted that Kirkpatrick's effective-medium approach breaks down near percolation thresholds but, to the best of our knowledge, this breakdown has not been investigated in detail in the literature to-date. 

\subsection{Pore-network modelling}
\label{sec:methods-pnm}

To assess the accuracy of \ac{EMT} solutions, we need to compare them to some ``empirical'' value. To that end, \ac{PNM} techniques are used to simulate resistor networks and directly calculate the effective transport property of interest, namely the effective conductance of the network. Open\ac{PNM}, an open-source \ac{PNM} package for Python, was used to perform these direct simulations~\cite{gostick-2016-openpnm}. Using Open\ac{PNM}, square or cubic lattice resistor networks can be simulated, with random conductances distributed according to specified \ac{PDF}s. This leads to the linear system of equations from \autoref{eq:kirchhoff}, which can be solved directly. Technically, Open\ac{PNM} solves for an effective conductivity for the network. This is easily converted into a conductance, however, by multiplying by the length of the bonds. 

\begin{figure}[htb]
    \centering
    \begin{subfigure}[b]{.45\linewidth}
        \includegraphics[width=\linewidth]{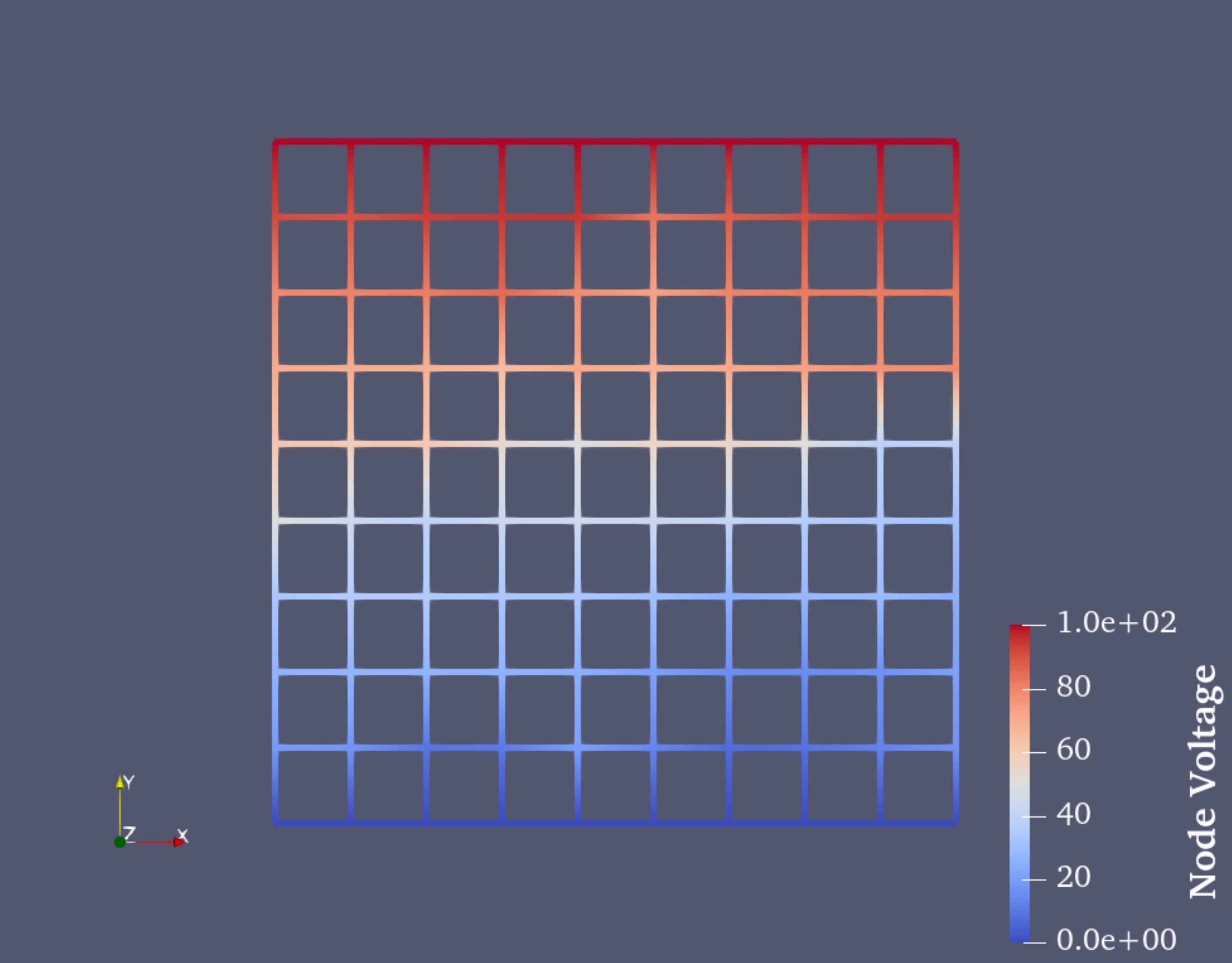}
        \caption{2D network.}
        \label{fig:network-voltage-2d}
    \end{subfigure}
    \begin{subfigure}[b]{.45\linewidth}
        \includegraphics[width=\linewidth]{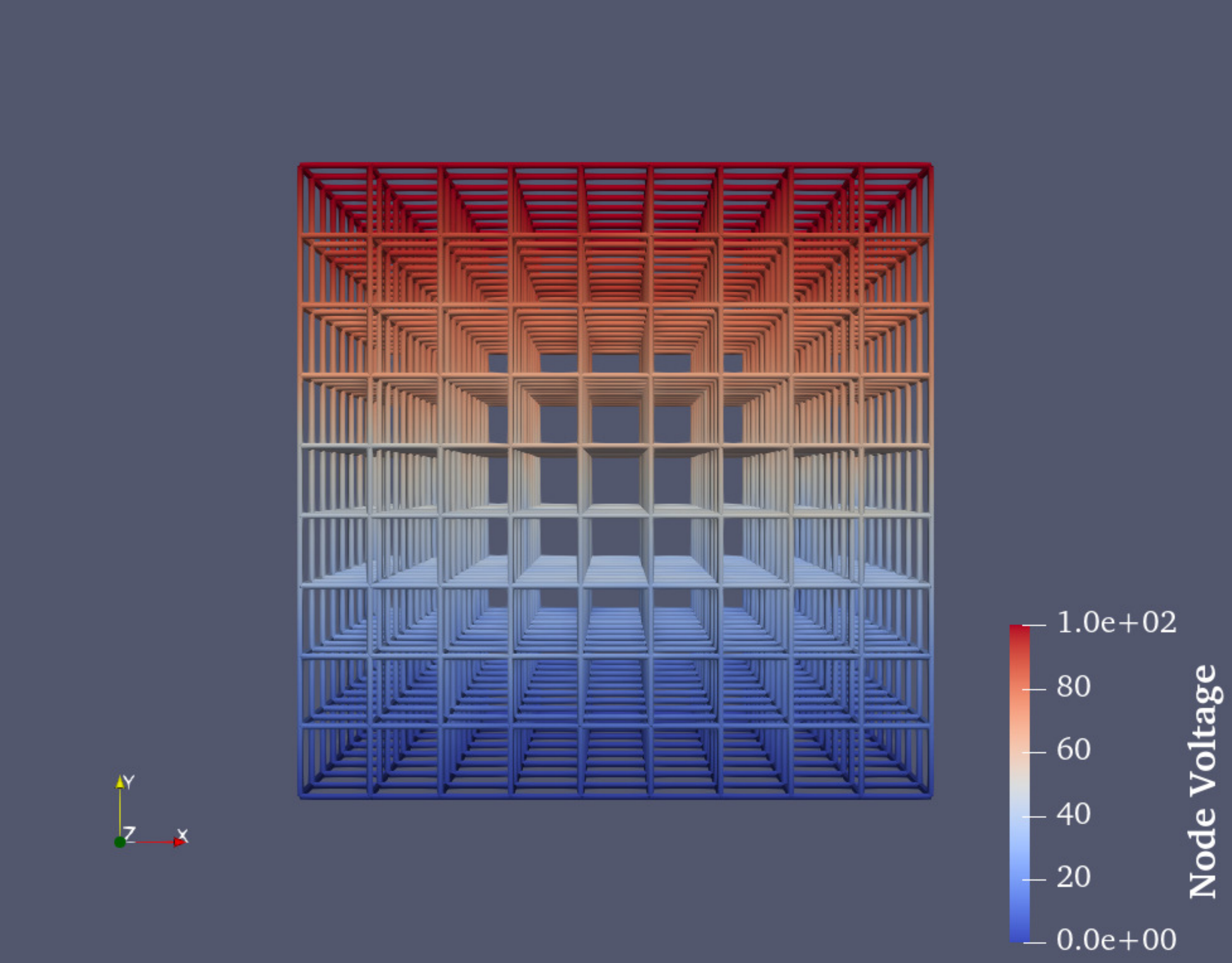}
        \caption{3D network.}
        \label{fig:network-voltage-3d}
    \end{subfigure}
    \begin{subfigure}[b]{.45\linewidth}
        \includegraphics[width=\linewidth]{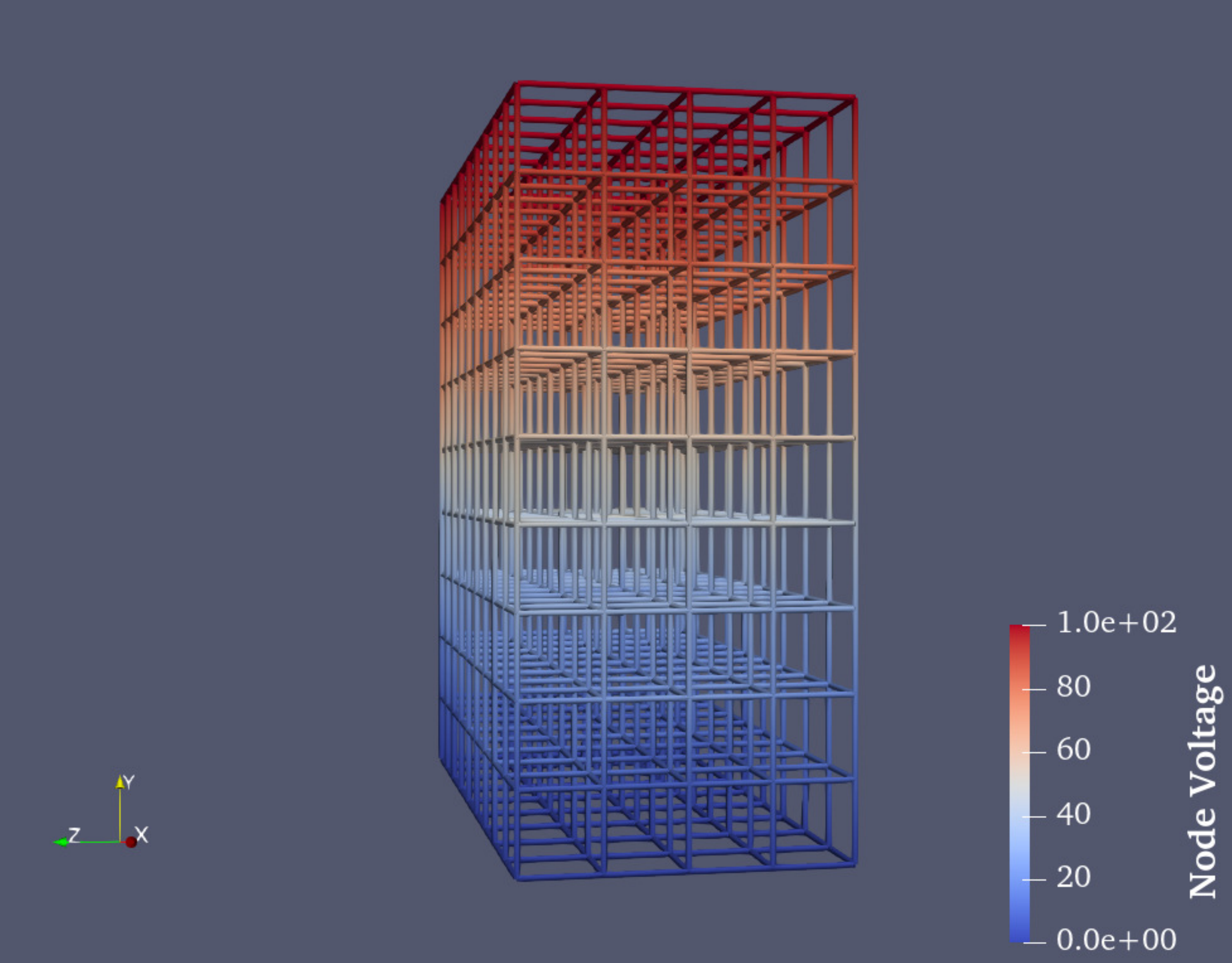}
        \caption{Quasi-2D network.}
        \label{fig:network-voltage-quasi-2d}
    \end{subfigure}
    \caption{Examples of \acs{PNM} networks with network voltages, $n=10$.}
    \label{fig:network-voltages}
\end{figure}

For convenience, all simulated networks have a lattice constant of one, so the effective conductivity and the effective conductance of the network have the same numerical value. Networks examined were two and three-dimensional, with an equal number of nodes $n$ in each direction. A third ``quasi-two-dimensional'' case was also looked at, in which the network was very large in two dimensions and small in the third. Specifically, the network had dimensions of $n \times n \times 5$. This case is to represent thin porous media, such as ionomer membranes in fuel cells. These membranes can be millimetres or centimetres wide, but only nanometres or micrometres thick, with channel lengths on the order of a few nanometres~\cite{kusoglu-2017-pfsa,allen-2015-nafion}. Visualizations of example 2D, 3D, and quasi-2D networks can be seen in \autoref{fig:network-voltages}. 

\subsection{Comparison of \acs{EMT} and \acs{PNM}}
\label{sec:methods-comparison}

To compare \ac{EMT} and \ac{PNM} results, a simple procedure was followed for a given network.
\begin{enumerate}
    \item Choose a \ac{PDF} to describe network conductances.
    \item Calculate the \ac{EMT} effective conductance, as described in \autoref{sec:methods-emt}. 
    \item Calculate the effective conductance with \ac{PNM} for a given $n$, using Monte Carlo sampling to average the results. 
    \item Repeat Step 3 on progressively larger networks.
    \item Calculate the \ac{RD} between \ac{EMT} and \ac{PNM} effective conductances as a function of network size $n$. 
    \item Calculate the \ac{RSD} of the \ac{PNM} conductances as a function of network size $n$. 
\end{enumerate}
These steps were repeated for a variety of different distributions $f(g)$ in both two and three dimensions. The quasi-2D case was only examined for a single \ac{PDF}. Initially, 100 Monte Carlo samples were taken per distribution at each $n$ while performing the \ac{PNM} calculations. In the interest of computational time required per simulation, this was subsequently reduced to 50 Monte Carlo samples, which proved to be a sufficient sample size. 

\ac{RD} for a given network was calculated according to 
\begin{equation}
    \label{eq:relative-difference}
    \varepsilon 
    =
    \frac{\lvert g_m - \overline{g}_{\mathrm{eff}} \rvert}
    {g_m} \times 100\%,
\end{equation}
where $g_m$ is the effective conductance according to \ac{EMT} and $\overline{g}_{\mathrm{eff}}$ is the averaged effective conductance from the Monte Carlo samplings of the \ac{PNM} calculations at a given $n$. Relative standard deviation was calculated as
\begin{equation}
    \label{eq:rsd}
    \sigma_{\mathrm{RSD}} 
    =
    \frac{\sigma_{g_{\mathrm{eff}}}}{\overline{g}_{\mathrm{eff}}} 
    \times 100\%,
\end{equation}
where $\sigma_{g_{\mathrm{eff}}}$ is the standard deviation of the Monte Carlo samplings.

The goal of the above procedure is to provide guidance for future computational research. The relationship between the \ac{RD} and the \ac{RSD}, and network size $n$ provides scaling laws governing the convergence of \ac{PNM} results. This can serve as a rule of thumb, providing an estimate for when the macroscopic limit is approached in \ac{PNM} simulations, and when \ac{EMT} can be applied instead of a numerical solution of the \ac{PNM}. 

\section{Results}
\label{sec:results}

The procedure described in \autoref{sec:methods-comparison} was carried out for nine statistical distributions of network conductance. These are summarized in \autoref{tab:pdf}. For illustration, only the plots from the \ac{PDF} referred to as ``Weibull 2'' will be shown in this section. Plots from every distribution in \autoref{tab:pdf} can be seen in \hyperref[app:plots]{Appendix~\ref*{app:plots}}. 

\begin{figure}[ht]
    \centering
    \includegraphics[width=.75\linewidth]{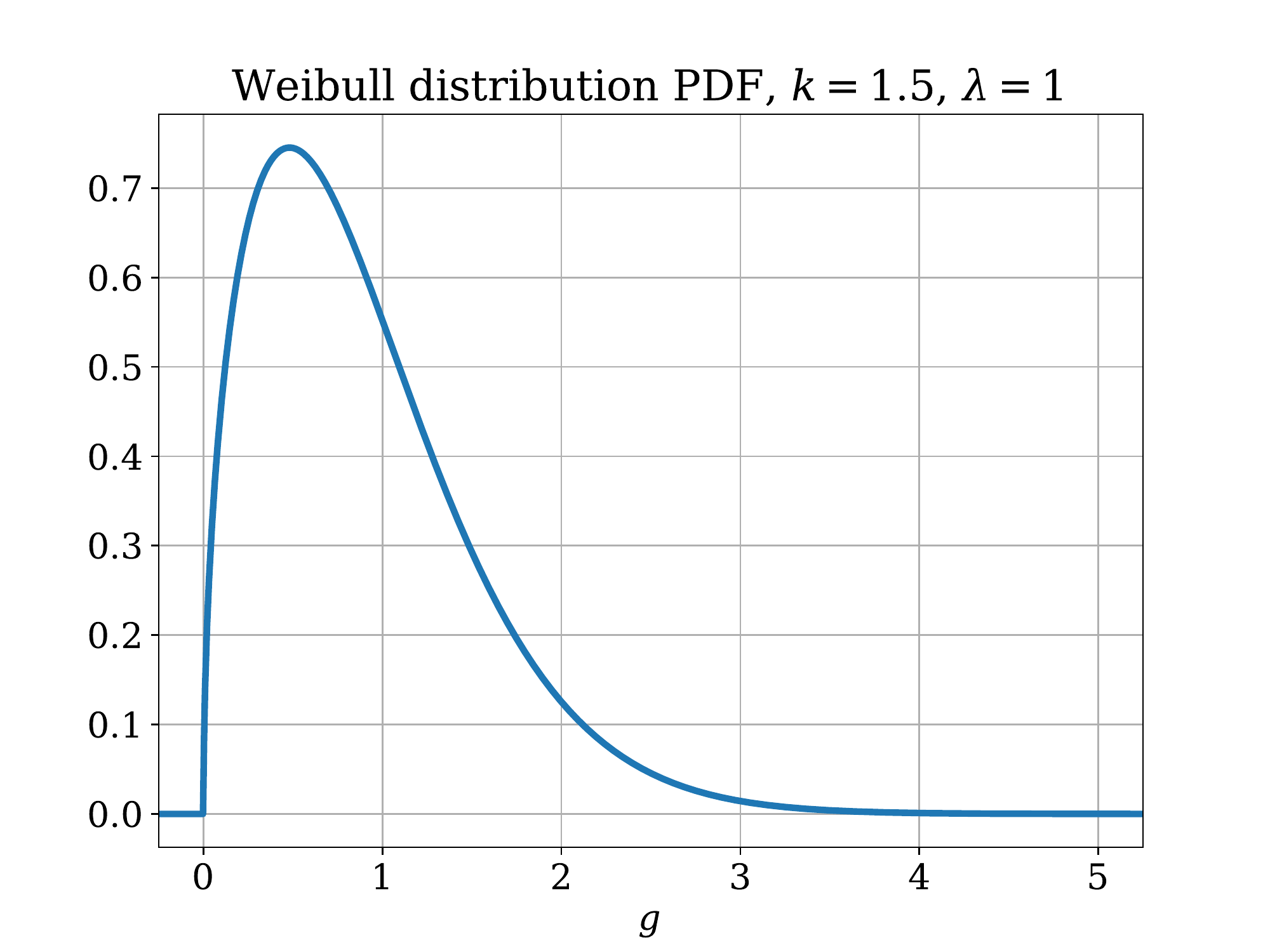}
    \caption{\acs{PDF} of Weibull 2 distribution with shape factor $k=1.5$.}
    \label{fig:weibull-1p5-pdf}
\end{figure}

The \ac{PDF}s were chosen to examine the convergence behaviour between \ac{EMT} and \ac{PNM} on a diverse array of network conductances. The arcsine distribution is continuous and bimodal, with two sharp peaks in conductance value. The bimodal sine distribution is also continuous and bimodal, but with less dramatic peaks. The binary and trinary distributions were chosen to examine \ac{PDF}s that are both discrete and have analytic \ac{EMT} solutions. The uniform distribution is to examine when each conductance within a certain range has an equal probability. The Weibull distributions were chosen to examine unimodal \ac{PDF}s with a peak at zero conductance (Weibull 1, equivalent to an exponential distribution), near zero (Weibull 2), and away from zero (Weibull 3). The \ac{PDF} for Weibull 2 is shown in \autoref{fig:weibull-1p5-pdf}. All other \ac{PDF}s are visible in \autoref{fig:app-pdf}. 

\renewcommand{\arraystretch}{1.5}
\begin{table}[t]
    \centering
    \caption{Summary of \aclp{PDF} examined.}
    \footnotesize
    \begin{tabular}{l|l} 
    \multicolumn{1}{l|}{\textit{Distribution}}&\multicolumn{1}{l}{\textit{\Acl{PDF}}, $f(g)$}  \\ \hline
    Arcsine & $\frac{1}{\pi \sqrt{g (1 - g)}},~g \in [0, 1]$ \\[0.4cm]
    Bimodal sine & 
    $\frac{\pi}{2} \begin{cases}
    \sin{(2 \pi g)}, & g \in [0, \frac{1}{2}) \\
    -\sin{(2 \pi g)}, & g \in [\frac{1}{2}, 1] \\
    0, & \text{otherwise}
    \end{cases}$ \\
    Binary 1 & $0.1 \cdot \delta(g - 0.3) + 0.9 \cdot \delta(g - 0.6)$ \\
    Binary 2 & $0.5 \cdot \delta(g - 0.3) + 0.5 \cdot \delta(g - 0.6)$ \\
    Trinary & $\frac{1}{3} \delta(g - 0.25) + \frac{1}{3} \delta(g - 0.5) + \frac{1}{3} \delta(g - 0.75)$ \\
    Uniform & 
    $\begin{cases} 
    1, & g \in [0, 1] \\ 
    0, & \text{otherwise} 
    \end{cases}$ \\
    Weibull 1 & $e^{-g}, ~g \in [0, \infty)$ \\
    Weibull 2 & $\frac{3}{2} g^{\frac{1}{2}} e^{-g^{\frac{3}{2}}}, ~g \in [0, \infty)$ \\
    Weibull 3 & $5 g^{4} e^{-g^{5}}, ~g \in [0, \infty)$ 
    \end{tabular}
    \label{tab:pdf}
\end{table}

\begin{figure*}[htbp]
    \centering
    \begin{subfigure}[b]{.4\linewidth}
        \includegraphics[width=\linewidth]{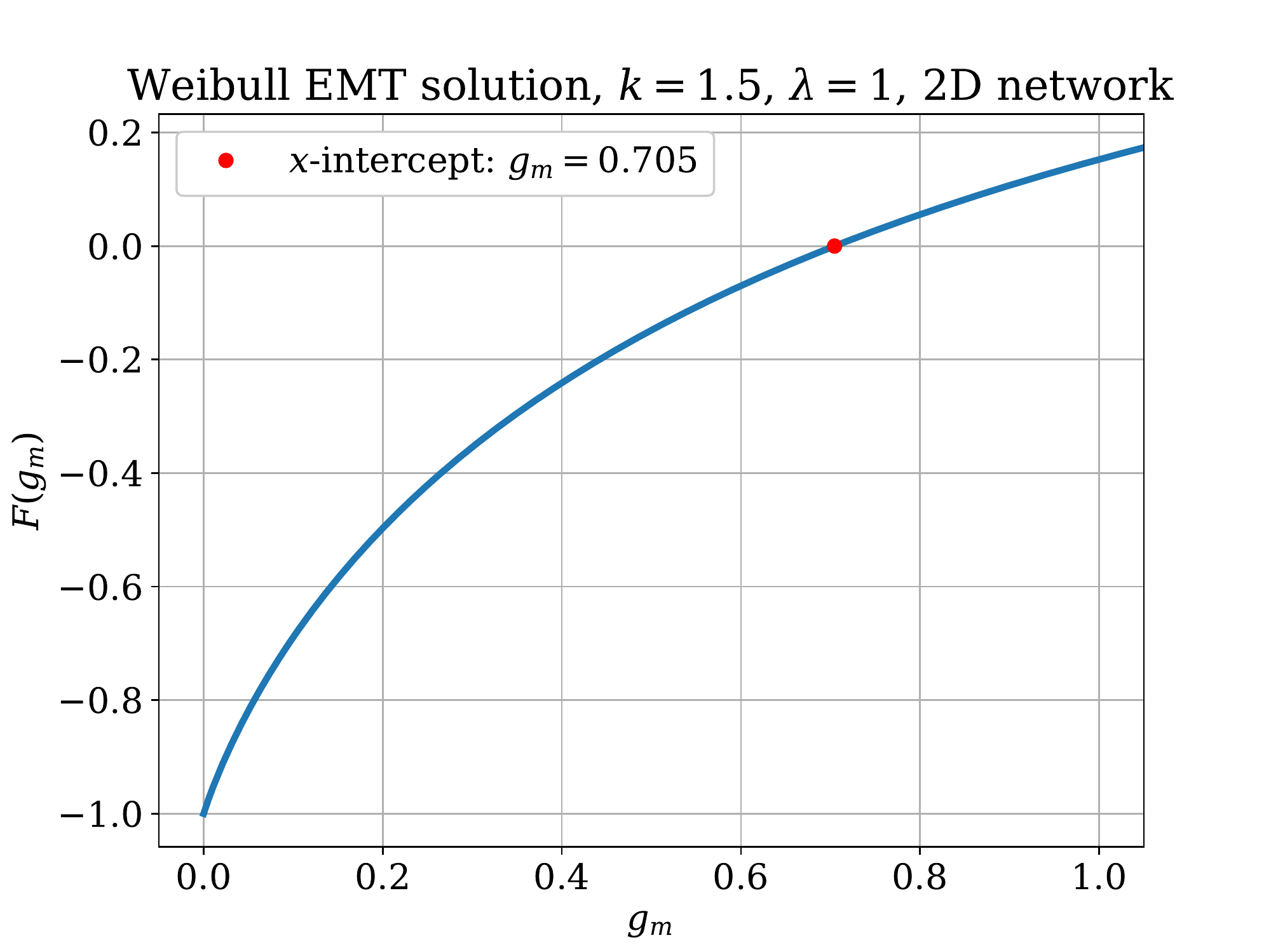}
        \caption{}
        \label{fig:weibull-1p5-2d-ki}
    \end{subfigure}
    \begin{subfigure}[b]{.4\linewidth}
        \includegraphics[width=\linewidth]{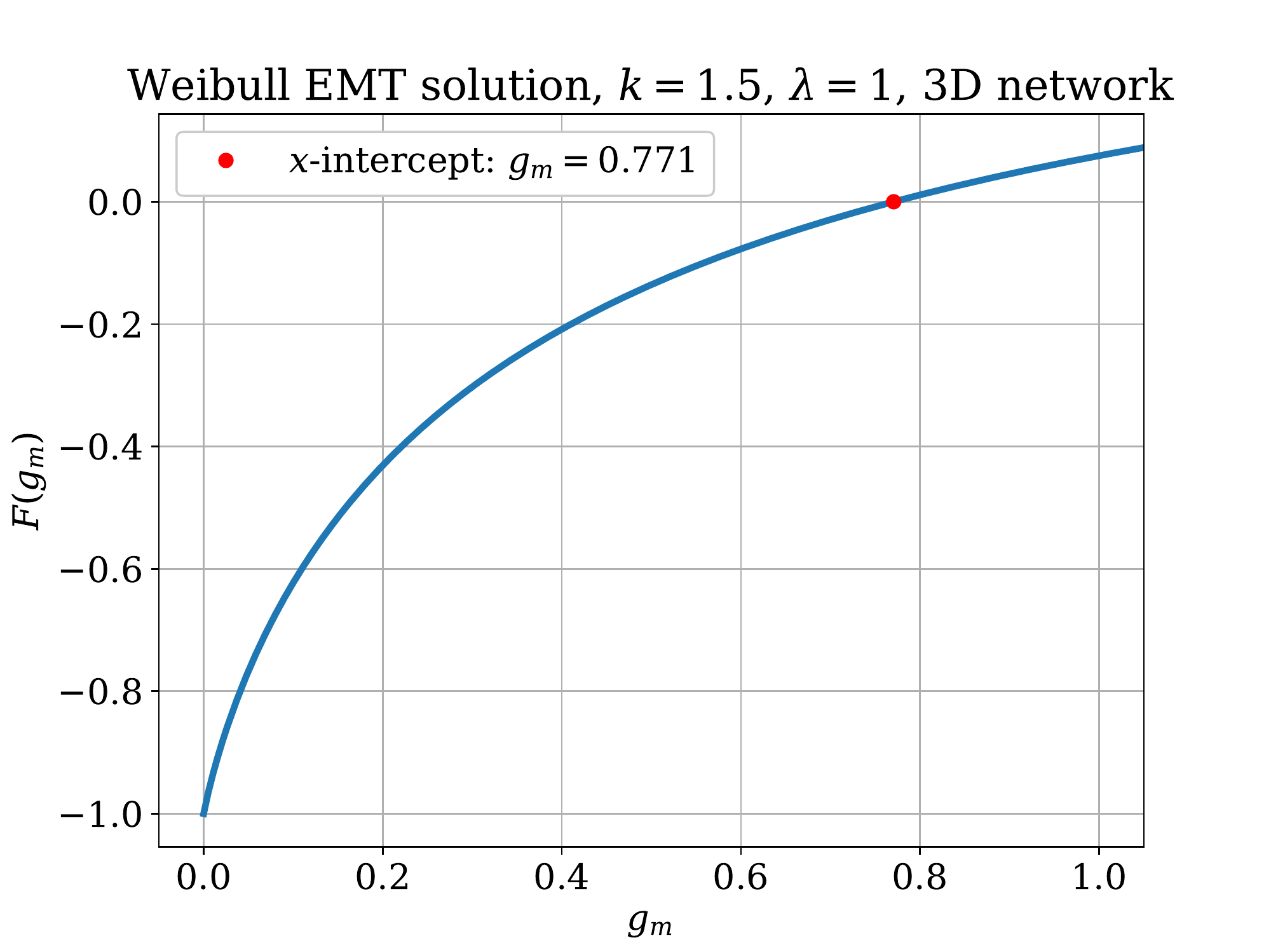}
        \caption{}
        \label{fig:weibull-1p5-3d-ki}
    \end{subfigure}
    \caption{\autoref{eq:kirkpatrick-vanish} solutions for Weibull 2 distribution.}
    \label{fig:weibull-1p5-ki}
\end{figure*}

The \ac{EMT} solutions to \autoref{eq:kirkpatrick-vanish} on both 2D and 3D networks are plotted in \autoref{fig:weibull-1p5-ki}. The intercept with the $g_m$ axis in the plots corresponds to the effective conductance $g_m$ that replaces the infinite network. Solutions for every \ac{PDF} can be seen in \autoref{fig:app-ki}. The solutions for every network are summarized in \autoref{tab:soln-ki}.

\begin{table}[t]
    \centering
    \caption{\acs{EMT} solutions to \autoref{eq:kirkpatrick-vanish}.}
    \footnotesize
    \begin{tabular}{l|r|r}
    \multicolumn{1}{l|}{\textit{Distribution}}&\multicolumn{1}{l|}{\textit{2D network}}&\multicolumn{1}{l}{\textit{3D network}}  \\
    \hline
    Arcsine & 0.333 & 0.400 \\
    Bimodal sine &  0.414 & 0.445 \\
    Binary 1 & 0.561 & 0.564 \\
    Binary 2 & 0.424 & 0.433 \\
    Trinary & 0.455 & 0.475 \\
    Uniform & 0.398 & 0.437 \\
    Weibull 1 &  0.610 & 0.727 \\
    Weibull 2 & 0.705 & 0.771 \\
    Weibull 3 & 0.892 & 0.901 \\
    \end{tabular}
    \label{tab:soln-ki}
\end{table}

\begin{table*}[tp]
  \centering
  \caption{Least-square power law fits for \acs{RD} and \acs{RSD}, 2D and 3D networks.}
  \footnotesize
    \begin{tabular}{l|r|r|r|r|r|r|r|r}
    \textit{Distribution} & \multicolumn{1}{l|}{2D $a_{\mathrm{RD}}$} & \multicolumn{1}{l|}{2D $b_{\mathrm{RD}}$} & \multicolumn{1}{l|}{2D $a_{\mathrm{RSD}}$} & \multicolumn{1}{l|}{2D $b_{\mathrm{RSD}}$}& \multicolumn{1}{l|}{3D $a_{\mathrm{RD}}$} & \multicolumn{1}{l|}{3D $b_{\mathrm{RD}}$} & \multicolumn{1}{l|}{3D $a_{\mathrm{RSD}}$} & \multicolumn{1}{l}{3D $b_{\mathrm{RSD}}$} \\
    \hline 
    Arcsine & $\mathbf{-2.638}$ & $\mathbf{17947.503}$ & $-1.035$ & $154.325$ & $-1.088$ & $136.064$ & $-1.576$ & $132.446$ \\
    Bimodal sine & $-1.108$ & $149.310$ & $-0.946$ & $56.447$ & $-1.076$ & $134.850$ & $-1.472$ & $60.106$ \\
    Binary 1 & $-0.992$ & $99.080$ & $-1.014$ & $20.701$ & $-1.031$ & $114.829$ & $-1.248$ & $7.825$ \\
    Binary 2 & $-1.010$ & $107.622$ & $-0.911$ & $23.143$ & $-1.035$ & $117.497$ & $-1.793$ & $94.133$ \\
    Trinary & $-1.016$ & $109.052$ & $-0.997$ & $46.446$ & $\mathbf{-1.330}$ & $\mathbf{233.389}$ & $-1.625$ & $70.957$ \\
    Uniform & $-1.092$ & $129.761$ & $-1.020$ & $97.236$ & $-1.014$ & $106.608$ & $-1.571$ & $90.009$ \\
    Weibull 1 & $-0.961$ & $74.099$ & $-1.097$ & $172.700$ & $-1.000$ & $108.749$ & $-1.459$ & $89.196$ \\
    Weibull 2 & $-1.061$ & $129.554$ & $-1.041$ & $99.519$ & $-1.035$ & $117.506$ & $-1.508$ & $68.338$ \\
    Weibull 3 & $-1.038$ & $119.236$ & $-1.046$ & $31.159$ & $-1.043$ & $120.107$ & $-1.621$ & $36.443$ \\
    \end{tabular}%
  \label{tab:fits}%
\end{table*}%

\begin{figure*}[htbp]
    \centering
    \begin{subfigure}[b]{.4\linewidth}
        \includegraphics[width=\linewidth]{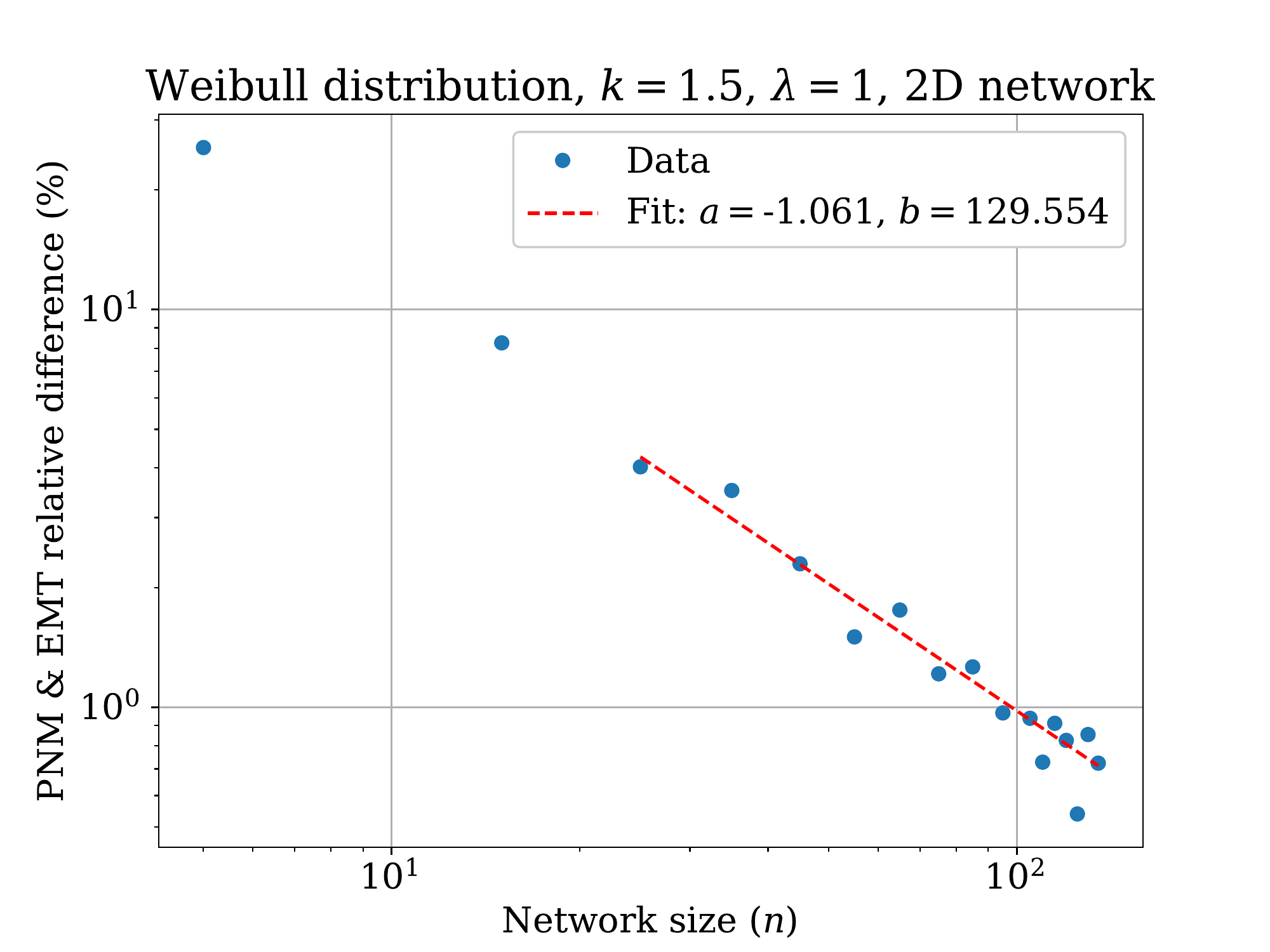}
        \caption{\acs{PNM} and \acs{EMT} \acs{RD}.}
        \label{fig:weibull-1p5-2d-err}
    \end{subfigure}
    \begin{subfigure}[b]{.4\linewidth}
        \includegraphics[width=\linewidth]{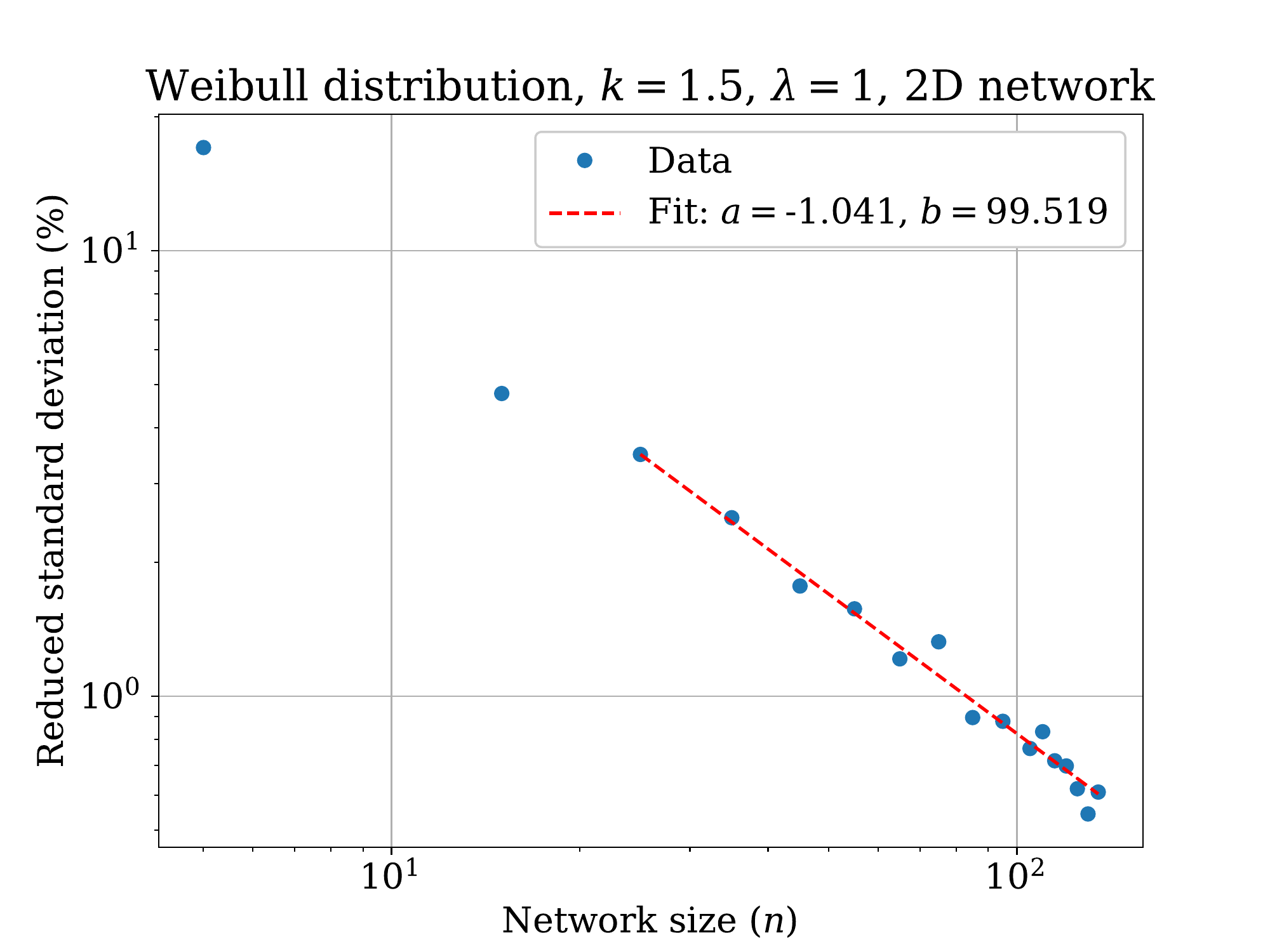}
        \caption{\acs{PNM} \acs{RSD}.}
        \label{fig:weibull-1p5-2d-rsd}
    \end{subfigure}
    
    \begin{subfigure}[b]{.4\linewidth}
        \includegraphics[width=\linewidth]{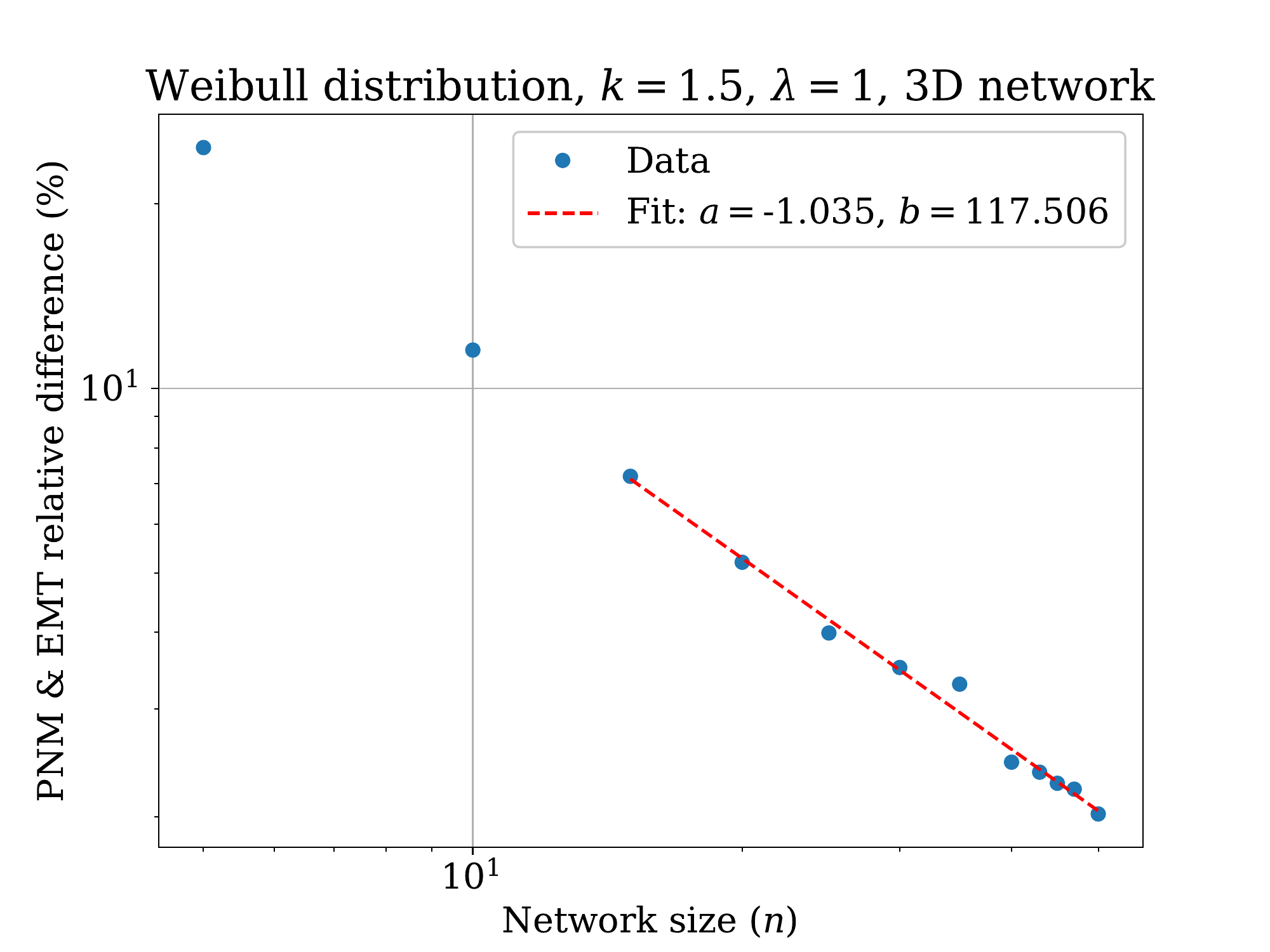}
        \caption{\acs{PNM} and \acs{EMT} \acs{RD}.}
        \label{fig:weibull-1p5-3d-err}
    \end{subfigure}
    \begin{subfigure}[b]{.4\linewidth}
        \includegraphics[width=\linewidth]{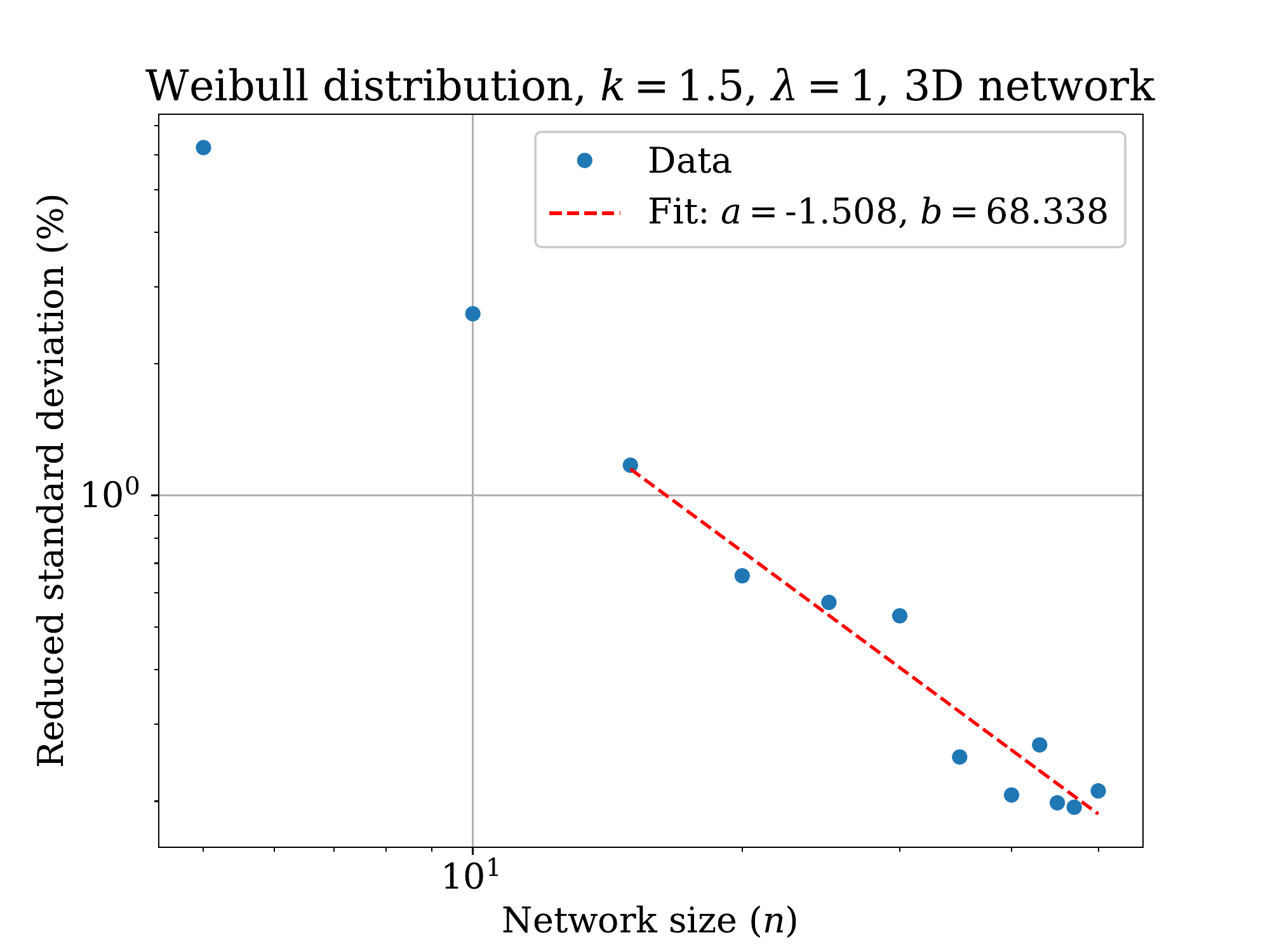}
        \caption{\acs{PNM} \acs{RSD}.}
        \label{fig:weibull-1p5-3d-rsd}
    \end{subfigure}
    \caption{Power law fits to \acs{RD} and \acs{RSD}.}
    \label{fig:weibull-1p5-fits}
\end{figure*}

Least-squares fits of the \ac{RD} between \ac{EMT} and \ac{PNM} (\autoref{eq:relative-difference}) and the \ac{PNM} Monte Carlo sampling \ac{RSD} (\autoref{eq:rsd}) were made using Python's native \texttt{scipy.\hspace{0pt}optimize.\hspace{0pt}curve\_\hspace{0pt}fit} function. Fits were made to a power law,
\begin{equation}
    \label{eq:power-law}
    y(n) = b n^{a},
\end{equation}
where $n$ is the size of the network, and $b$ and $a$ are the parameters of the fit. The fits for the Weibull 2 distribution are in \autoref{fig:weibull-1p5-fits}. All other fits can be seen in \autoref{fig:app-fits}. Fit parameter values are summarized in \autoref{tab:fits} for 2D and 3D networks. Errors in the fit parameters are presented in \hyperref[app:errors]{Appendix~\ref*{app:errors}}.

\begin{figure*}[htbp]
    \centering
    \begin{subfigure}[b]{.4\linewidth}
        \includegraphics[width=\linewidth]{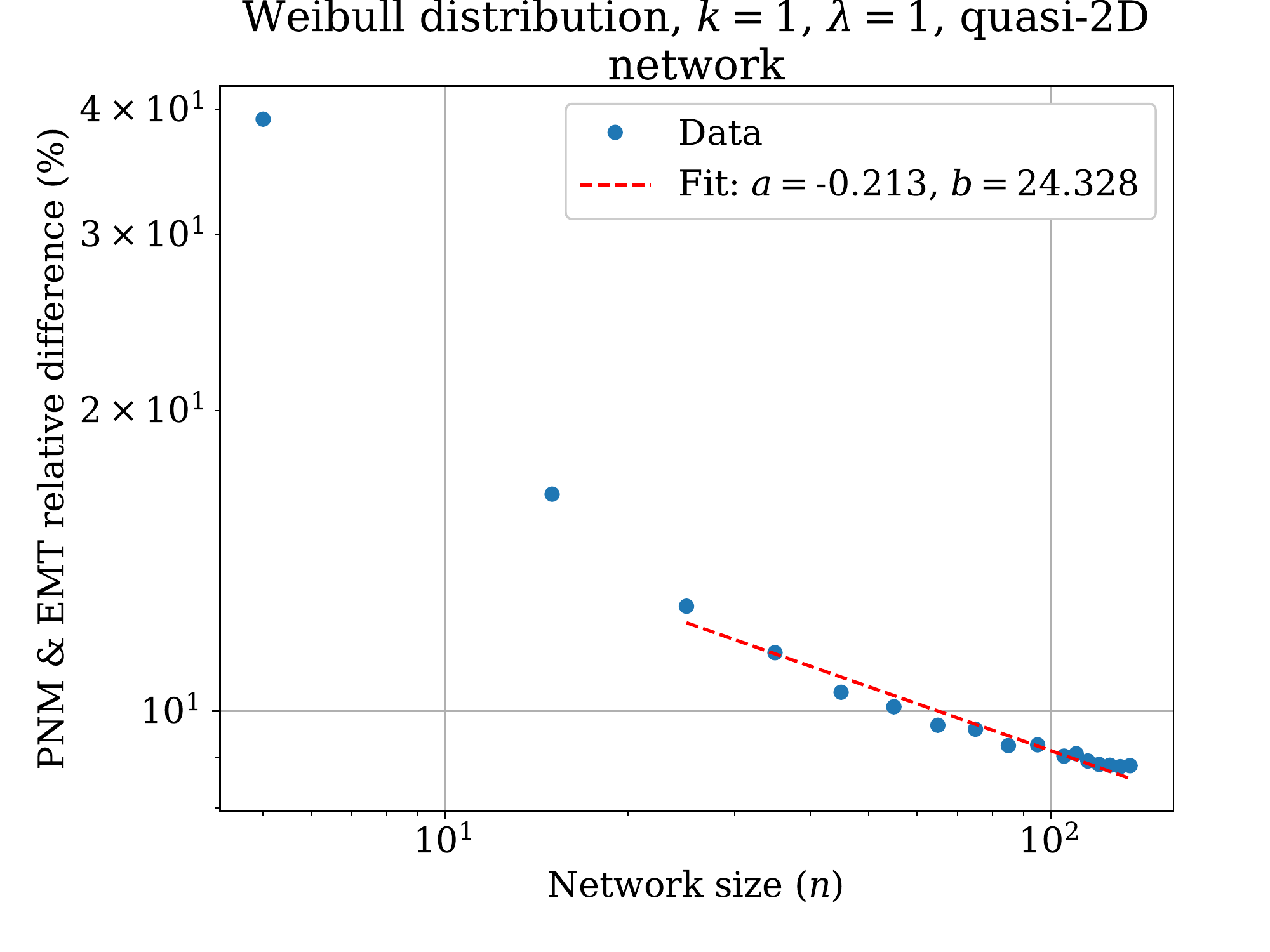}
        \caption{\acs{RD} between 2D and quasi-2D.}
        \label{fig:weibull-1p5-quasi2d-2d-err}
    \end{subfigure}
    \begin{subfigure}[b]{.4\linewidth}
        \includegraphics[width=\linewidth]{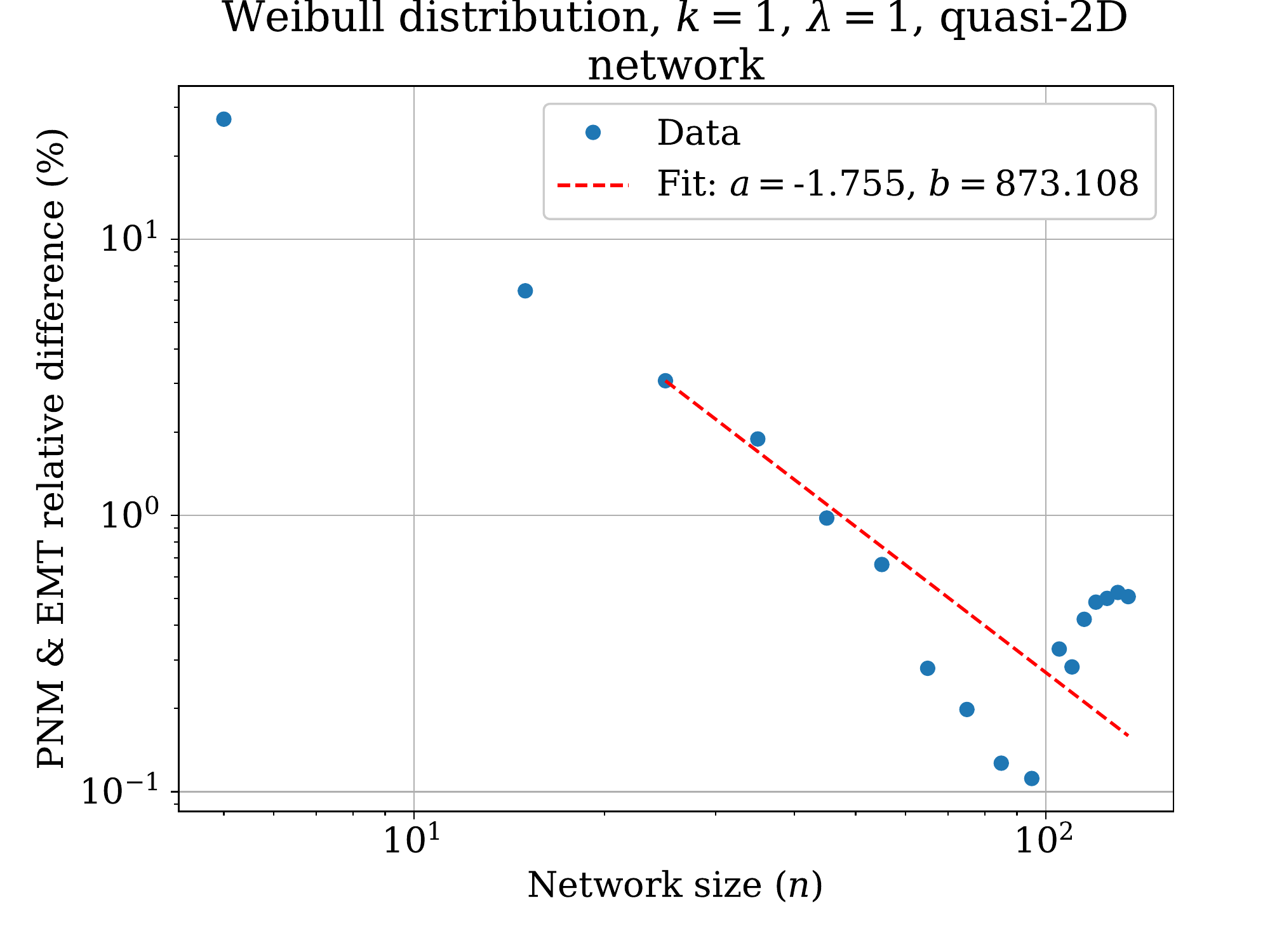}
        \caption{\acs{RD} between 3D and quasi-2D.}
        \label{fig:weibull-1p5-quasi2d-3d-err}
    \end{subfigure}
    \begin{subfigure}[b]{.4\linewidth}
        \includegraphics[width=\linewidth]{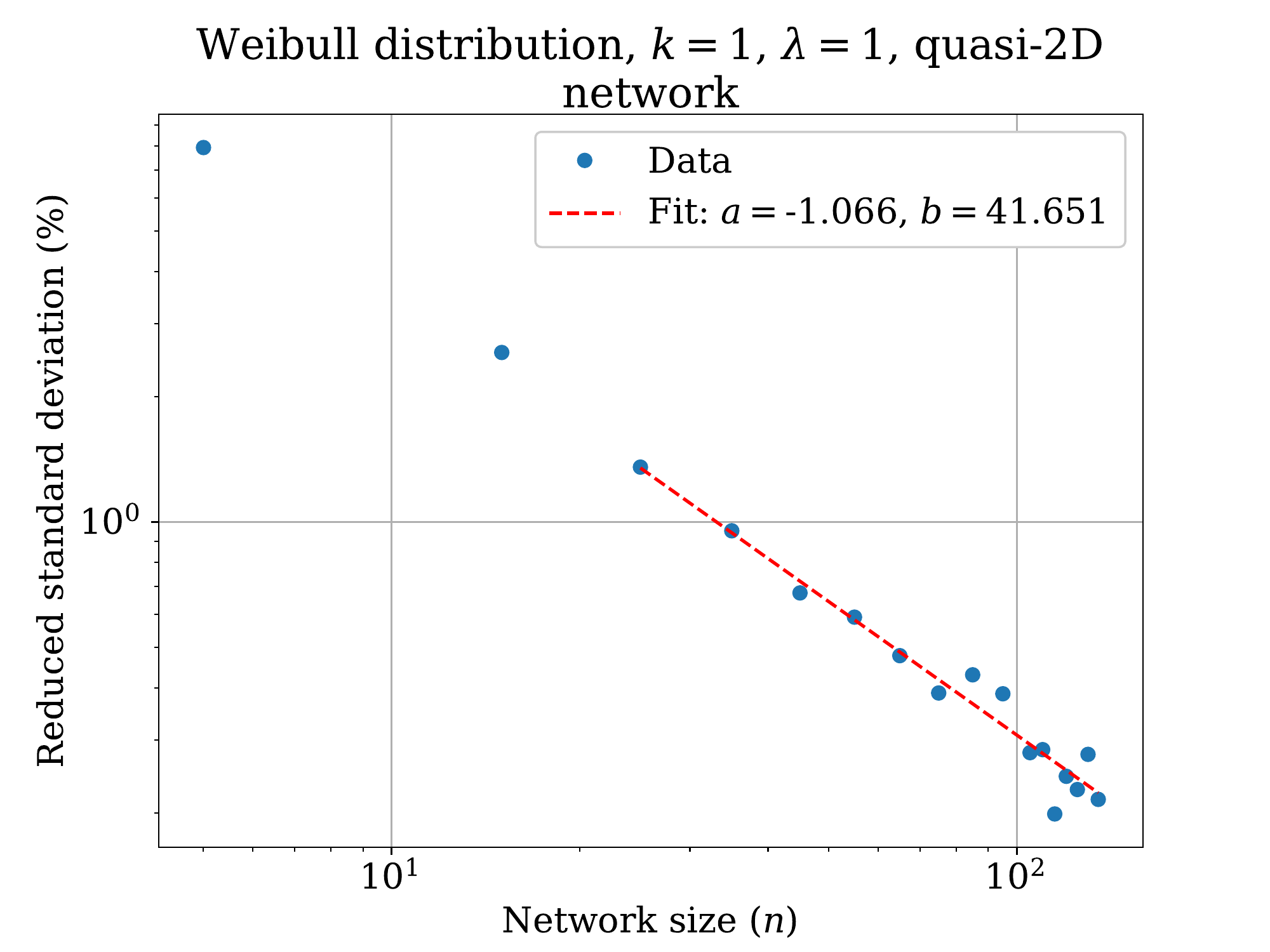}
        \caption{\acs{RSD} of quasi-2D network.}
        \label{fig:weibull-1p5-quasi2d-rsd}
    \end{subfigure}
    \caption{Power law fits to quasi-2d \aclp{RD} and \acs{RSD}, in-plane.}
    \label{fig:weibull-1p5-quasi2d-fits}
\end{figure*}

In addition to these comparisons between \ac{EMT} and \ac{PNM} on 2D and 3D networks, the ``quasi-2D'' case was examined for the Weibull 2 distribution. Effective conductance was calculated for transport along one of the long axes of $n$ nodes length, known as in-plane transport, and along the five-node axis, known as through-plane transport. Both are scenarios outside of the scope considered by Kirkpatrick's \ac{EMT}, which only considers infinite 2D or 3D networks. The idealized quasi-2D case can be thought of as a 3D network that is infinite in two dimensions, but is finite in the third dimension. Consequently, there is no \ac{EMT} solution to which the \ac{PNM} results can be directly compared. However, the \ac{RD} between the \ac{PNM} solution and both the 2D and 3D \ac{EMT} solutions was still calculated, in addition to the \ac{RSD}. These can be seen in \autoref{fig:weibull-1p5-quasi2d-fits}. The fit parameters for the in-plane transport are summarized in \autoref{tab:fits-quasi-2d}. Through-plane transport reached a constant effective conductivity value after $n\sim20$, so fits were not performed on those data. Little information could be gained from the latter analysis.

\begin{table*}[t]
    \centering
    \caption{Quasi-2D network fit parameters, in-plane transport.}
    \footnotesize
    \begin{tabular}{l|r|r|r|r|r|r}
    \textit{Distribution} & \multicolumn{1}{l|}{$a_{\mathrm{RD}}$, 2D \acs{EMT}} & \multicolumn{1}{l|}{$b_{\mathrm{RD}}$, 2D \acs{EMT}} & \multicolumn{1}{l|}{$a_{\mathrm{RD}}$, 3D \acs{EMT}} & \multicolumn{1}{l|}{$b_{\mathrm{RD}}$, 3D \acs{EMT}} & \multicolumn{1}{l|}{$a_{\mathrm{RSD}}$} & \multicolumn{1}{l}{$b_{\mathrm{RSD}}$} \\
    \hline
    Weibull 2 & $-0.213$ & $24.328$ & $-1.755$ & $873.108$ & $-1.066$ & $41.651$
    \end{tabular}
    \label{tab:fits-quasi-2d}
\end{table*}

\section{Discussion}
\label{sec:discussion}

\subsection{Relative difference}
\label{sec:disc-rd}

The problem posited in \autoref{sec:problem} was to examine the relationship between \ac{EMT} and \ac{PNM} in modelling transport phenomena within composite materials and porous media. Namely, what is the convergence behaviour between the two techniques as the size of the system being examined grows? This was investigated for the conductance distributions in \autoref{tab:pdf}, to see the effect of different distributions on this behaviour. Two and three-dimensional networks were examined. As evidenced from the fit parameters in \autoref{tab:fits}, it appears that the distribution of conductances has little effect on the convergence behaviour of \ac{RD} values between \ac{EMT} and \ac{PNM}. Indeed, the dimension of the network also seems to have little effect. 

For most of the \ac{PDF}s examined, on both 2D and 3D cubic networks, the power law convergence behaviour for \ac{RD} follows the general relationship
\begin{equation}
    \label{eq:thumb-rd-convergence}
    \varepsilon(n) \propto n^{-1}.
\end{equation}
There is much agreement in the value of the exponential fit parameter $a$ in \autoref{eq:power-law}. There is less agreement in the scaling fit parameter $b$, though the scaling parameter ultimately is less informative as to the underlying trends being exhibited.

The \ac{RD} is inversely proportional to the network size. This provides a handy rule of thumb. If one would like their \ac{PNM} solutions to exhibit a certain \ac{RD} from the theoretically exact \ac{EMT} value for infinitely large systems, one can use \autoref{eq:thumb-rd-convergence} to determine approximately how large the network must be. For example, if one desires an \ac{RD} of $2\,\%$, then for most conductivity distributions a network size, $n$, of
\begin{equation}
    \label{eq:thumb-network-size}
    n \approx \left( \frac{2\,\%}{100\,\%} \right)^{-1} = 50
\end{equation}
would be approximately sufficient, where dividing by $100\,\%$ accounts for the $b$ scaling parameter, which is of order $100$ for almost all simulations. This rule of thumb is an aid in determining appropriate domain sizes for direct \ac{PNM} calculations for continuum macroscale behaviour to emerge. Though these calculations were specifically performed for electrical conductance, transport behaviours of a variety of phenomena behave in analogous ways, as discussed in \autoref{sec:lit-review}. It can be expected that similar trends will hold for other transport problems. 

There are two exceptions to this trend in our simulations, whose values are bolded in \autoref{tab:fits}. The most notable is the arcsine distribution on a 2D network. Its exponent fit parameter is $a=-2.638$ and scaling fit parameter is $b=17947.504$, very different from every other distribution. This distribution has a large number of bonds in the network having a conductance at or very near to zero. It is analogous to a softened binary distribution at zero and one with equal probability. Such a binary distribution is exactly the percolation threshold on 2D networks, where transport through the network ceases entirely. The reduced connectivity in a square lattice (four nearest-neighbour bonds, as opposed to six in a cubic lattice) places the arcsine distribution very close to this threshold, which likely leads to atypical transport behaviour within the network. As mentioned above, Kirkpatrick's method breaks down near the percolation method and the arcsine example supports that notion.

The arcsine distribution is an unusual \ac{PDF} and is not often used in the study of composite media. The fact that it does not follow the general trends of other, more conventional \ac{PDF}s, demonstrates that great care should be taken when applying such atypical models. 

The other exception is in the trinary distribution on a 3D network. Its exponent fit parameter is $a=-1.330$. This is not so outlandish as the arcsine power law, but it remains significantly different from the other distributions examined. Its scaling fit parameter is $b=233.389$, within an order of magnitude of the other distributions, but still roughly twice as large. It is unclear why this particular distribution differs from the overall trend. It is seldom physically accurate to apply discrete \ac{PDF}s to the properties of composite media, so this discrepancy can perhaps be overlooked. Nevertheless, the behaviour of this particular distribution remains an open question in this analysis. Further studies into the convergence behaviour between \ac{PNM} and \ac{EMT} on a greater variety of discrete \ac{PDF}s could provide further insight.

\subsection{Reduced standard deviation}
\label{sec:disc-rsd}

Much like the \ac{RD} convergence discussed in \autoref{sec:disc-rd}, the exponential fit parameter of the \ac{RSD} values of the \ac{PNM} Monte Carlo sampled conductances appeared to be independent of the \ac{PDF} used to distribute the conductances. However, there was greater discrepancy between the results from 2D and 3D networks, with 3D networks reducing their \ac{RSD} values more quickly. This, however, is to be expected. For a given network size $n$, a 3D network will contain more bonds and, hence, a larger sample volume. Again, these results can be applied more generally to other transport phenomena within porous media. Additionally, there was greater discrepancy in the fit parameter $b$ between different distributions.  

On 2D networks, the exponential fit parameter $a$ of the power law governing the relationship between \ac{RSD} and network size was similar to \autoref{eq:thumb-rd-convergence}; inversely proportional. Consequently, the rule of thumb used to approximate the network size required to achieve a desired \ac{RSD} is the same as \autoref{eq:thumb-network-size}. 

Discrete or narrow distributions (such as Weibull 3) have smaller $b$ fit parameters, whereas wider distributions have larger $b$ fit parameters. This indicates that the expected \ac{RSD} is smaller for narrower distributions or ones with more limited available conductance values, and it is larger for wide and continuous distributions.

On 3D networks, the approximate power law is
\begin{equation}
    \label{eq:thumb-rsd-3d}
    \sigma_{\mathrm{RSD}} \propto n^{-1.5},
\end{equation}
which provides a good rule of thumb to estimate required network size for a desired \ac{RSD}. This simplified relationship neglects the influence of the $b$ parameter. Unless the distribution has been examined ahead of time, the $b$ parameter cannot be easily taken into account. However, the general power-law behaviour of the networks as described by the $a$ parameter will still hold. 

Again, let us take the example of $2\,\%$ as the desired value and a value of $b$ of order $100$. This would provide an estimate for an appropriate network size of
\begin{equation}
    \label{eq:thumb-rsd-network-size}
    n \approx \left( \frac{2\,\%}{100\,\%} \right)^{-0.667} \sim 14.
\end{equation}
This is of the right magnitude but somewhat smaller than results found from \ac{PNM} simulations undertaken by \citet{berg-2021-pnm}. There, it was determined that a 3D network size of $n=24$ was needed to get the \ac{RSD} values below $2\,\%$. However, the \ac{PNM} treatment in that study was significantly more complex than undertaken here, which would be reflected by a value of $b$ that deviates from $100$. The general trends found in this study can still provide a useful starting point when determining the appropriate domain sizes for simulation. 

\subsection{Quasi-two-dimensional case}
\label{sec:disc-quasi-2d}

The final case examined is the quasi-2D network. As stated in \autoref{sec:methods-pnm}, this is defined as a network with dimensions of $n \times n \times 5$, imitating thin porous media such as ionomer membranes. Only the Weibull 2 distribution was examined on this network. The \ac{RSD} behaviour on this network for in-plane and through-plane transport was very similar to that displayed by true 2D networks, following an approximately inverse relationship between $\sigma_{\mathrm{RSD}}$ and $n$. 

However, the results from \ac{PNM} for this network did not conform with the expected \ac{EMT} results for either 2D or 3D lattices, as can be seen in Figures~\hyperref[fig:weibull-1p5-quasi2d-2d-err]{\ref*{fig:weibull-1p5-quasi2d-2d-err}} and \hyperref[fig:weibull-1p5-quasi2d-3d-err]{\ref*{fig:weibull-1p5-quasi2d-3d-err}} for in-plane transport. Instead, as the network size grew, the $\overline{g}_{\mathrm{eff}}$ approached a value distinct from either solution (approximately 0.767 for in-plane and 0.995 for through-plane). This underscores a very important detail when it comes to applying \ac{EMT}. These theories are developed to describe idealized effective media. They only strictly apply in the continuum limit, when microscopic properties are negligible and the domains are infinite. To use these results on real materials, it must be assumed that the domains being examined are large enough to be effectively infinite. Indeed, this is a reasonable approximation for many fields of research, and \ac{EMT} can be applied reliably.

However, as the methods of \ac{EMT} and \ac{PNM} begin to be applied to model phenomena on micro- and nanoscales, these approximations can no longer be reasonably held. Porous thin films used in nanotechnologies are three-dimensional. However, they are significantly smaller in one dimension than in the other two, but not to the extent that they can be treated as simply two-dimensional. The utmost care must be used when attempting to apply \ac{EMT} techniques in such fields. Often, \ac{EMT} will not be suitable.

\section{Conclusion}
\label{sec:conclusion}

The goal of this project is to provide insight into the relationship between the macroscopic approaches of \ac{EMT} and the discretized approaches of \ac{PNM} in regards to transport through composite and porous media. The specific case examined in this treatment is electrical conductance, though the trends discovered here can be generalized to other transport phenomena. 

To that end, comparisons are made between predicted effective conductances from \ac{EMT} and directly calculated effective conductances from \ac{PNM} on 2D and 3D networks. Ultimately, from the analysis of the relative difference (\ac{RD}) between these values on progressively larger networks, useful rules of thumb were developed governing this relationship. From the power law fits made to the resulting data, it was determined that the \ac{RD} is approximately inversely proportional to the size of the network being examined independent of the chosen \ac{PDF} for conductance. This relationship held for both 2D and 3D networks. This can be used as a guide for future research into these areas, to determine what domain size is sufficient in order for macroscopic \ac{EMT} models to accurately replace \ac{PNM} techniques. It can also be used to determine how large of a network is required in order for \ac{PNM} to model bulk transport properties. All that is necessary is to decide what is an adequate threshold value for \ac{RD} for the system under investigation. 

There are two caveats, however, to this rule of thumb. It breaks down when examining the arcsine distribution on 2D networks and -- to a lesser extent -- the discrete trinary distribution on 3D networks. This underscores that, ultimately, the relationship is most useful as a guide but does not guarantee a particular result. Caution is still required, especially when applying unorthodox distributions to a particular medium. Further insight could be gained by examining discrete distributions in more detail. Additionally, as the \ac{EMT} being examined was only applicable to square and cubic lattices, these results will not necessarily apply to more complex network geometries. Also, nonlinear transport properties (i.e. conductances) may have an impact on the scaling law, which could be a potential subject of future research. Nonetheless, the underlying convergence behaviour remains a useful tool. 

The reduced standard deviation (\ac{RSD}) of the Monte Carlo sampled, effective conductance values from \ac{PNM} were also examined on 2D and 3D networks, as a function of network size. Again, it was found that a power law fit described the convergence behaviour, though dimensionality did affect the result in this case. On 2D networks, \ac{RSD} was found to be inversely proportional to network size. On 3D networks, the power law was closer to $-1.5$; 3D networks converge more quickly as network size increases. The faster convergence behaviour of 3D networks was not unexpected, and the results were in-line with \ac{RSD} values from another \ac{PNM} study of transport within porous media~\cite{berg-2021-pnm}. 

When these techniques were applied to the quasi-2D case, however, the clean convergence behaviour between \ac{EMT} and \ac{PNM} broke down. \ac{EMT} applies to a continuous, infinite medium, whether that medium be in two or three dimensions. The quasi-2D case, however, is analogous to a three-dimensional network that is infinite in two dimensions, but finite in the third. This can be thought of as a proxy for thin porous media. As the network size grew larger, the effective conductance did not converge to either the 2D or 3D predictions from \ac{EMT}. Assumptions that continuum methods can be applied to such a material do not necessarily hold. This underscores that much caution must be used if one wishes to utilize \ac{EMT} techniques in analyzing the in-plane and through-plane behaviour of thin porous media. A closer analysis of the convergence scaling laws for quasi-2D systems forms another subject for future research.

\bibliographystyle{myabbrvnat}
\bibliography{biblio}

\newpage

\onecolumngrid

\appendix

\section{Code}
\label{app:code}

\subsection{Kirkpatrick \acs{EMT} and derivative}
\label{app:code-emt}

The following code was written to calculate the value of the \ac{EMT} function $F(g_m)$ (\autoref{eq:kirkpatrick-integral}) and its derivative (\autoref{eq:kirkpatrick-derivative}), for numerical root-finding.

\begin{minted}[
frame=lines,
framesep=2mm,
baselinestretch=1.2,
fontsize=\footnotesize,
linenos
]{Python}
import numpy as np
import scipy.integrate as integrate

def kirkpatrick_integral(f, gm, z):
    """
    Numerically calculates the integral described in Kirkpatrick (1973)

    Parameters
    ----------
    f : FUNCTION
        Probability density function describing the distribution of 
            conductances within the network.
    gm : FLOAT
        The conductivity of the effective medium being considered.
    z : INT
        Describes the dimensions of the network. 4 corresponds to a 2D square
            lattice, each node having 4 nearest neighbours. 6 corresponds to a
            3D cubic lattice, each node having 6 nearest neighbours.

    Returns
    -------
    F : FLOAT
        The numerically integrated value returned by the integral.

    """
    integrand = lambda g: (f(g)*(gm - g))/(g + (0.5*z - 1)*gm)
    
    F, err = integrate.quad(integrand, 0, np.inf)
    return F 

def kirkpatrick_derivative(f, gm, z):
    """
    Numerically calculates the derivative of the integral equation described
        in Kirkpatrick (1973) wrt to d/d(gm)

    Parameters
    ----------
    f : FUNCTION
        Probability density function describing the distribution of 
            conductances within the network.
    gm : FLOAT
        The conductivity of the effective medium being considered.
    z : INT
        Describes the dimensions of the network. 4 corresponds to a 2D square
            lattice, each node having 4 nearest neighbours. 6 corresponds to a
            3D cubic lattice, each node having 6 nearest neighbours.

    Returns
    -------
    Fprime : FLOAT
        The numerically integrated value of the derivative.

    """
    integrand = lambda g: (f(g)*0.5*z*g)/(g + (0.5*z - 1)*gm)**2
    
    Fprime, err = integrate.quad(integrand, 0, np.inf)
    return Fprime
\end{minted}

\subsection{Root finders}
\label{app:code-root}

The following functions were used to numerically calculate the root of the \ac{EMT} function $F(g_m)$, using bisection or Newton methods.

\begin{minted}[
frame=lines,
framesep=2mm,
baselinestretch=1.2,
fontsize=\footnotesize,
linenos
]{Python}
def bisection(f,a,b,N):
    '''Approximate solution of f(x)=0 on interval [a,b] by bisection method.

    Parameters
    ----------
    f : function
        The function for which we are trying to approximate a solution f(x)=0.
    a,b : numbers
        The interval in which to search for a solution. The function returns
        None if f(a)*f(b) >= 0 since a solution is not guaranteed.
    N : (positive) integer
        The number of iterations to implement.

    Returns
    -------
    x_N : number
        The midpoint of the Nth interval computed by the bisection method. The
        initial interval [a_0,b_0] is given by [a,b]. If f(m_n) == 0 for some
        midpoint m_n = (a_n + b_n)/2, then the function returns this solution.
        If all signs of values f(a_n), f(b_n) and f(m_n) are the same at any
        iteration, the bisection method fails and return None.

    Examples
    --------
    >>> f = lambda x: x**2 - x - 1
    >>> bisection(f,1,2,25)
    1.618033990263939
    >>> f = lambda x: (2*x - 1)*(x - 3)
    >>> bisection(f,0,1,10)
    0.5
    '''
    if f(a)*f(b) >= 0:
        print("Bisection method fails.")
        return None
    a_n = a
    b_n = b
    for n in range(1,N+1):
        m_n = (a_n + b_n)/2
        f_m_n = f(m_n)
        if f(a_n)*f_m_n < 0:
            a_n = a_n
            b_n = m_n
        elif f(b_n)*f_m_n < 0:
            a_n = m_n
            b_n = b_n
        elif f_m_n == 0:
            print("Found exact solution.")
            return m_n
        else:
            print("Bisection method fails.")
            return None
    return (a_n + b_n)/2

def newton(f,Df,x0,epsilon,max_iter):
    '''Approximate solution of f(x)=0 by Newton's method.

    Parameters
    ----------
    f : function
        Function for which we are searching for a solution f(x)=0.
    Df : function
        Derivative of f(x).
    x0 : number
        Initial guess for a solution f(x)=0.
    epsilon : number
        Stopping criteria is abs(f(x)) < epsilon.
    max_iter : integer
        Maximum number of iterations of Newton's method.

    Returns
    -------
    xn : number
        Implement Newton's method: compute the linear approximation
        of f(x) at xn and find x intercept by the formula
            x = xn - f(xn)/Df(xn)
        Continue until abs(f(xn)) < epsilon and return xn.
        If Df(xn) == 0, return None. If the number of iterations
        exceeds max_iter, then return None.
    '''
    xn = x0
    for n in range(0,max_iter):
        fxn = f(xn)
        if abs(fxn) < epsilon:
            print('Found solution after',n,'iterations.')
            return xn
        Dfxn = Df(xn)
        if Dfxn == 0:
            print('Zero derivative. No solution found.')
            return None
        xn = xn - fxn/Dfxn
    print('Exceeded maximum iterations. No solution found.')
    return None
\end{minted}

\pagebreak

\subsection{\acs{PNM} Monte Carlo sampler}
\label{app:code-pnm}

The following is an example code used to carry out Monte Carlo sampling using \ac{PNM} techniques. It is currently configured for examining 3D networks while taking samples from a binary distribution. Open\ac{PNM} contains several native statistical distributions. The binary distribution is an example of a custom-written distribution.

\begin{minted}[
frame=lines,
framesep=2mm,
baselinestretch=1.2,
fontsize=\footnotesize,
linenos
]{Python}
import numpy as np
import openpnm as op
import openpnm.models as mods
import custom_models as cm

# Array of network sizes
network_sizes = np.array((5, 10, 15, 20, 25, 30, 35, 40, 43, 45))
# Arrays to store the mean and rsd values of geff
mean_effective_conductivity = np.zeros(len(network_sizes))
rsd_effective_conductivity  = np.zeros(len(network_sizes))
 
# Number of Monte Carlo samples to take
M = 50
# Connectivity of the cubic lattice
connectivity = 6

for j in range(0, len(network_sizes)):
    # The size of the network
    n = network_sizes[j]
    x = n
    y = n
    # 2D or 3D network; 1 if 2D, n if 3D
    z = n
    
    # Lattice constant of the network; set to 1, then conductivity and 
    #   conductance have the same numerical value
    Lc = 1
    
    # Array to store the effective conductivities of calculated on the random
    #   networks
    effective_conductivity = np.zeros(M)
    
    # Setup the network
    pn = op.network.Cubic(shape=[x,y,z], spacing=Lc, connectivity=connectivity)
    Ps = pn.pores()
    Ts = pn.throats()
    geo = op.geometry.GenericGeometry(network=pn, pores=Ps, throats=Ts)
    
    for i in range(0, M):
        geo.add_model(propname='throat.seed', 
                      model=mods.geometry.throat_seed.random)
        geo.add_model(propname='pore.diameter',
                      model=mods.misc.constant,
                      value=0)
        geo.add_model(propname='throat.diameter',
                      model=mods.misc.constant,
                      value=Lc)
        geo.add_model(propname='pore.area', 
                      model=mods.geometry.pore_area.cube)
        geo.add_model(propname='throat.area',
                      model=mods.geometry.throat_area.cuboid)
        geo.add_model(propname='throat.endpoints',
                      model=mods.geometry.throat_endpoints.cubic_pores)
        geo.add_model(propname='throat.conduit_lengths', 
                      model=mods.geometry.throat_length.conduit_lengths)
        
        # Create phase
        phase = op.phases.GenericPhase(network=pn)
        phase['throat.seed'] = geo['throat.seed']
        
        # Assign conductivities according to statistical distribution
        phase.add_model(propname='throat.electrical_conductivity',
                        model = cm.binary_distribution, seeds='throat.seed', 
                        p=0.1, alpha=0.3, beta=0.6)
        # Pores need a value, though it does not affect the geff calculation
        phase['pore.electrical_conductivity'] = np.zeros(len(Ps))
        
        # Add physics objects
        phys = op.physics.GenericPhysics(network=pn, phase=phase, geometry=geo)
        phys.add_model(propname='throat.electrical_conductance',
                       model=mods.physics.electrical_conductance.series_resistors)
        
        # Add algorithm for calculated net conductance
        ohm = op.algorithms.OhmicConduction(network=pn)
        ohm.setup(phase=phase)
        ohm.set_value_BC(pores=pn.pores('left'), values=0)
        ohm.set_value_BC(pores=pn.pores('right'), values=100)
        ohm.run()
        geff = ohm.calc_effective_conductivity(inlets=pn.pores('right'),
                                                outlets=pn.pores('left'),
                                                domain_area=x*z,
                                                domain_length=y)    
        effective_conductivity[i] = geff
        
    mean_geff = np.mean(effective_conductivity)
    mean_effective_conductivity[j] = mean_geff
    std_geff  = np.std(effective_conductivity)
    rsd_effective_conductivity[j]  = std_geff*100/mean_geff
    
np.savetxt('binary_p0p1_3d_mean_g.csv', 
           np.array((network_sizes, mean_effective_conductivity)), 
           delimiter=',')
np.savetxt('binary_p0p1_3d_rsd_g.csv', 
           np.array((network_sizes, rsd_effective_conductivity)), 
           delimiter=',')
\end{minted}

\newpage

\section{Additional plots}
\label{app:plots}

\newcommand\width{.45}

\subsection{Probability density functions}
\label{app:plots-pdf}

\begin{figure}[H]
    \centering
    \begin{subfigure}[b]{\width\linewidth}
        \includegraphics[width=\linewidth]{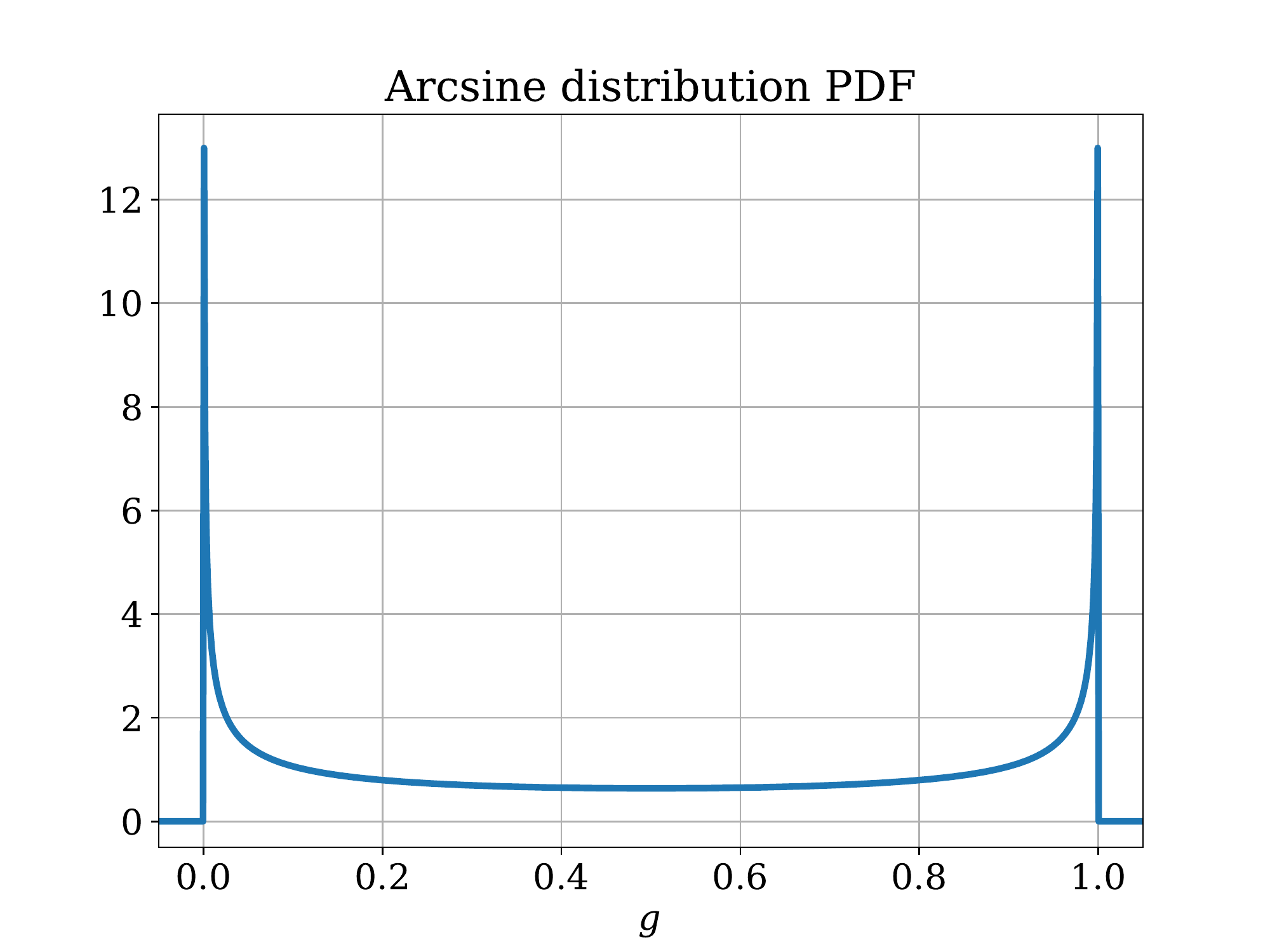}
        \caption{Extreme bimodal values.}
        \label{fig:app-arcsine-pdf}
    \end{subfigure}
    \begin{subfigure}[b]{\width\linewidth}
        \includegraphics[width=\linewidth]{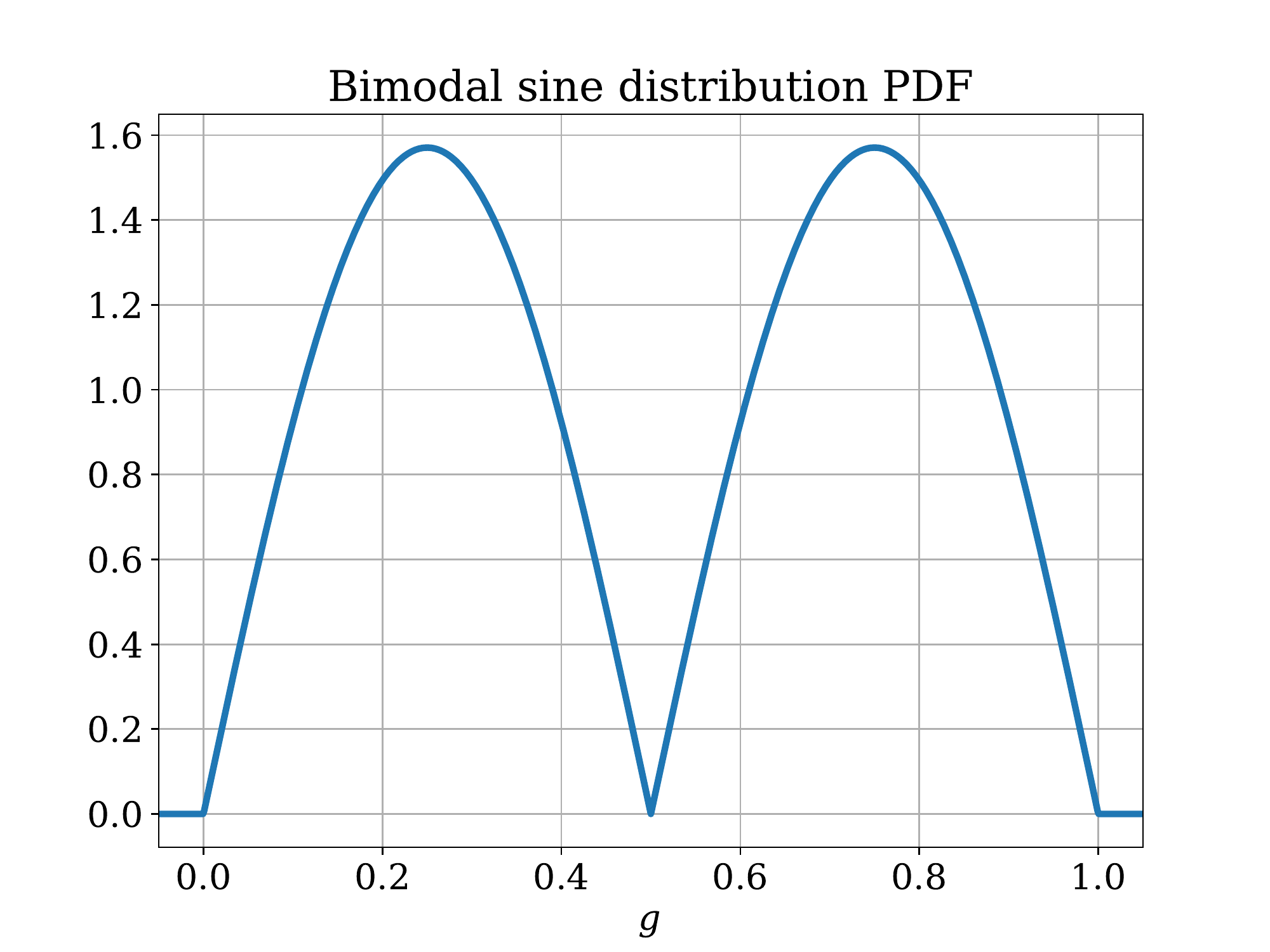}
        \caption{Continuous bimodal values.}
        \label{fig:app-bimodal-pdf}
    \end{subfigure}
    
    \begin{subfigure}[b]{\width\linewidth}
        \includegraphics[width=\linewidth]{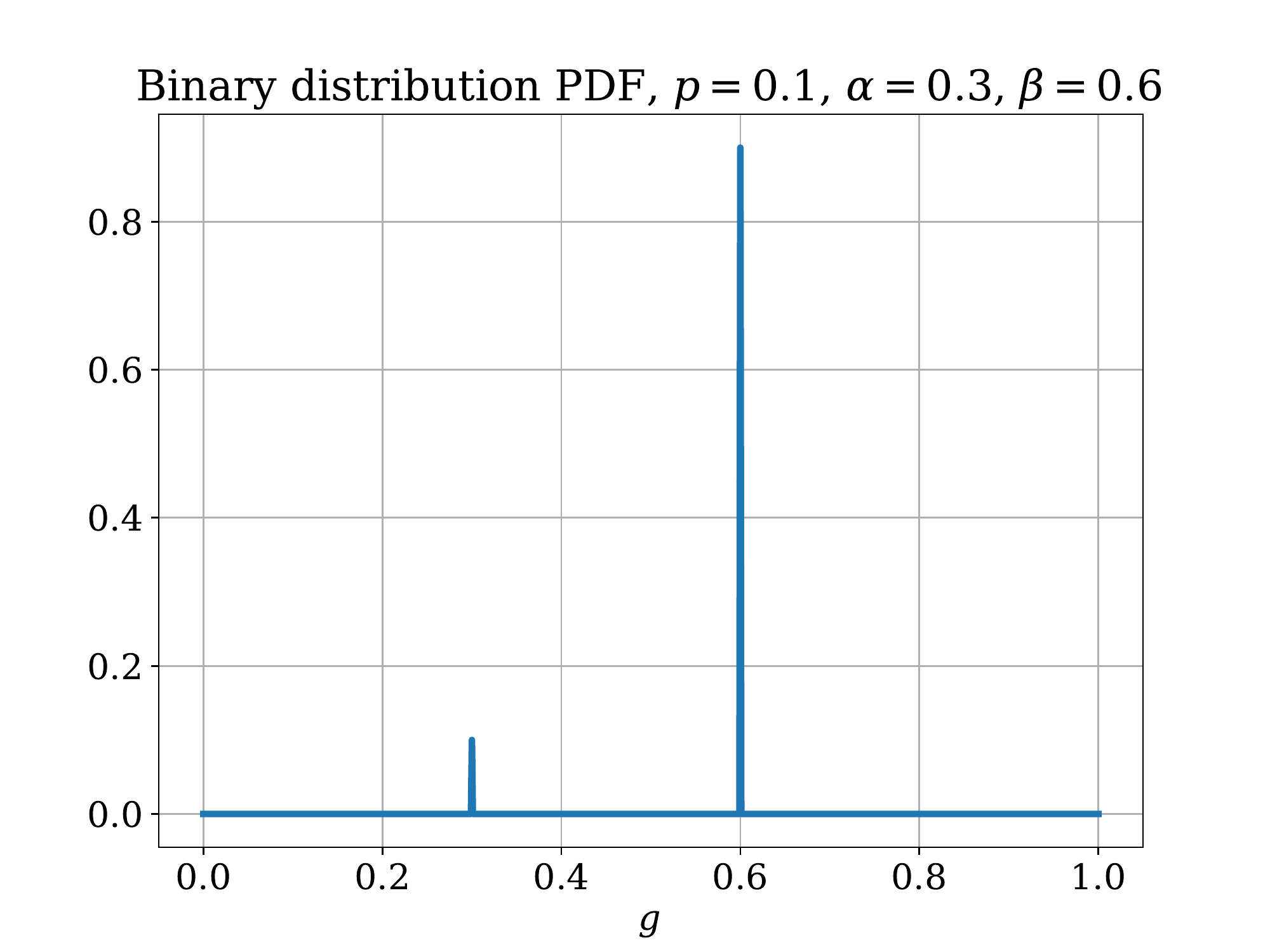}
        \caption{0.1--0.9 split between values.}
        \label{fig:app-binary-0p1-pdf}
    \end{subfigure}
    \begin{subfigure}[b]{\width\linewidth}
        \includegraphics[width=\linewidth]{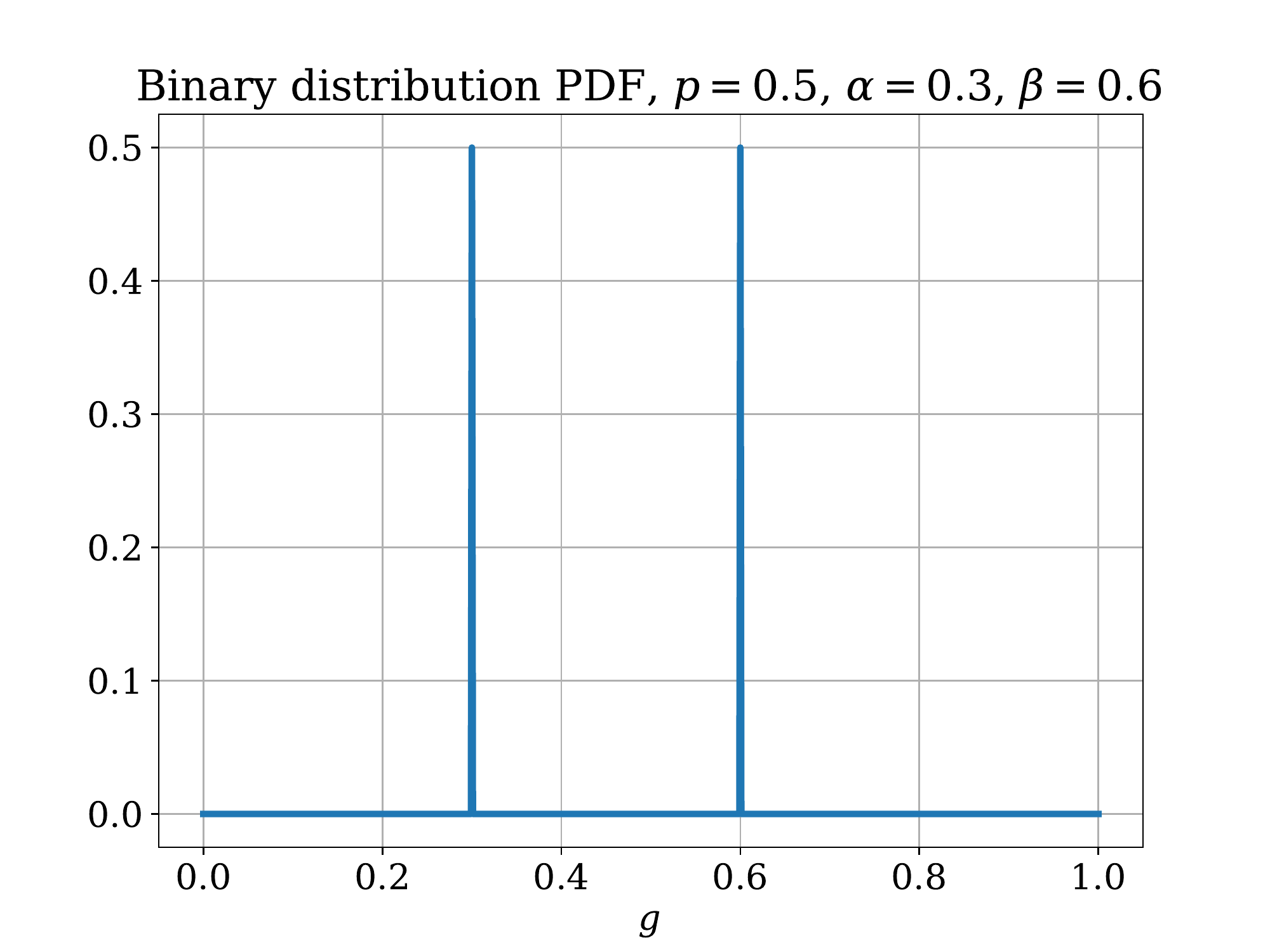}
        \caption{Even split between values.}
        \label{fig:app-binary-0p5-pdf}
    \end{subfigure}

    \begin{subfigure}[b]{\width\linewidth}
        \includegraphics[width=\linewidth]{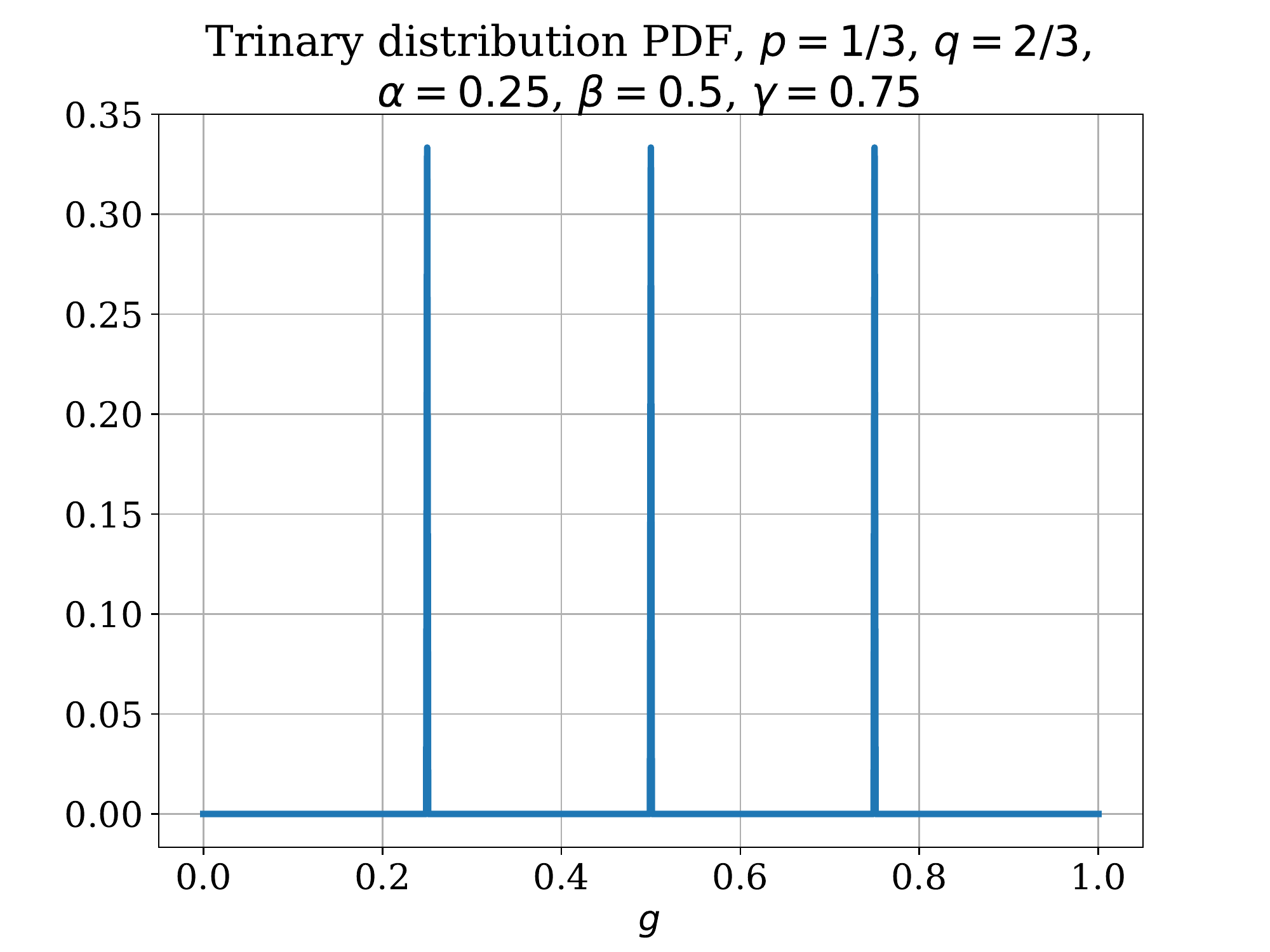}
        \caption{Even split between values.}
        \label{fig:app-trinary-pdf}
    \end{subfigure}
    \begin{subfigure}[b]{\width\linewidth}
        \includegraphics[width=\linewidth]{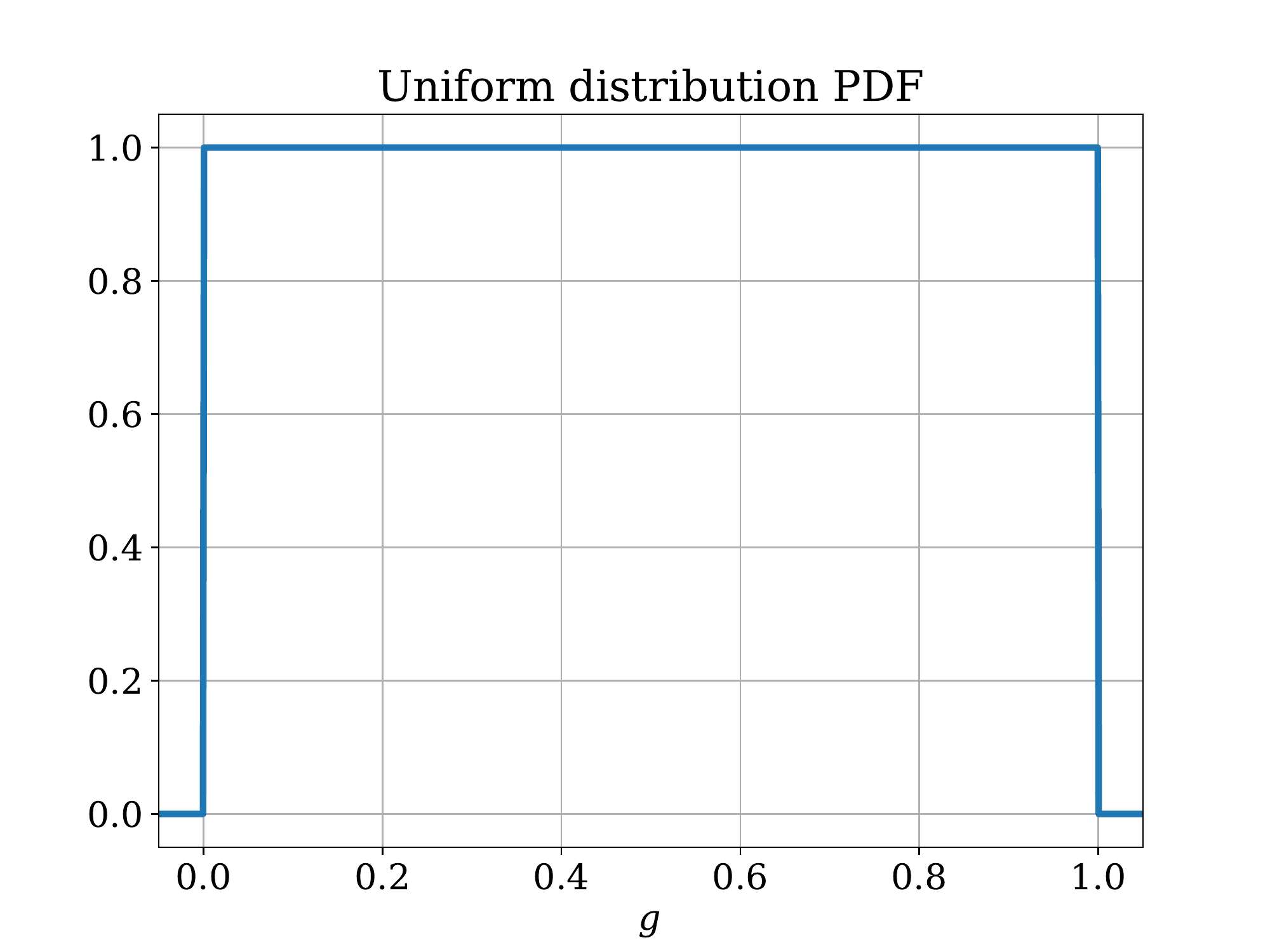}
        \caption{Uniform distribution.}
        \label{fig:app-uniform-pdf}
    \end{subfigure}    
    \caption{Probability density functions (\acs{PDF}s), $f(g)$, examined in our analysis.}
    \label{fig:app-pdf}
\end{figure}
    
\begin{figure}[H]\ContinuedFloat
    \centering
    \begin{subfigure}[b]{\width\linewidth}
        \includegraphics[width=\linewidth]{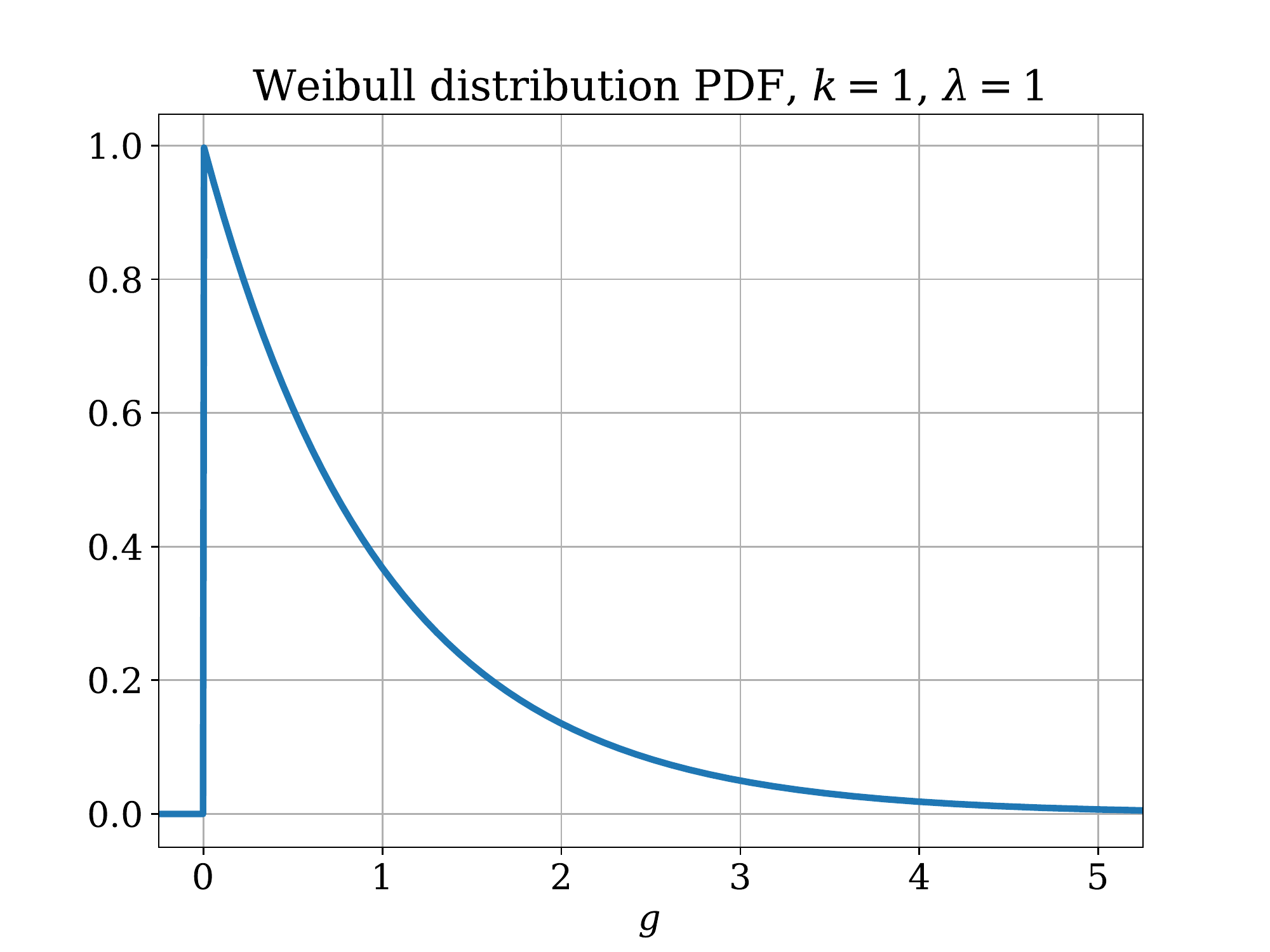}
        \caption{Unimodal peak near zero-conductance.}
        \label{fig:app-weibull-1-pdf}
    \end{subfigure}
    \begin{subfigure}[b]{\width\linewidth}
        \includegraphics[width=\linewidth]{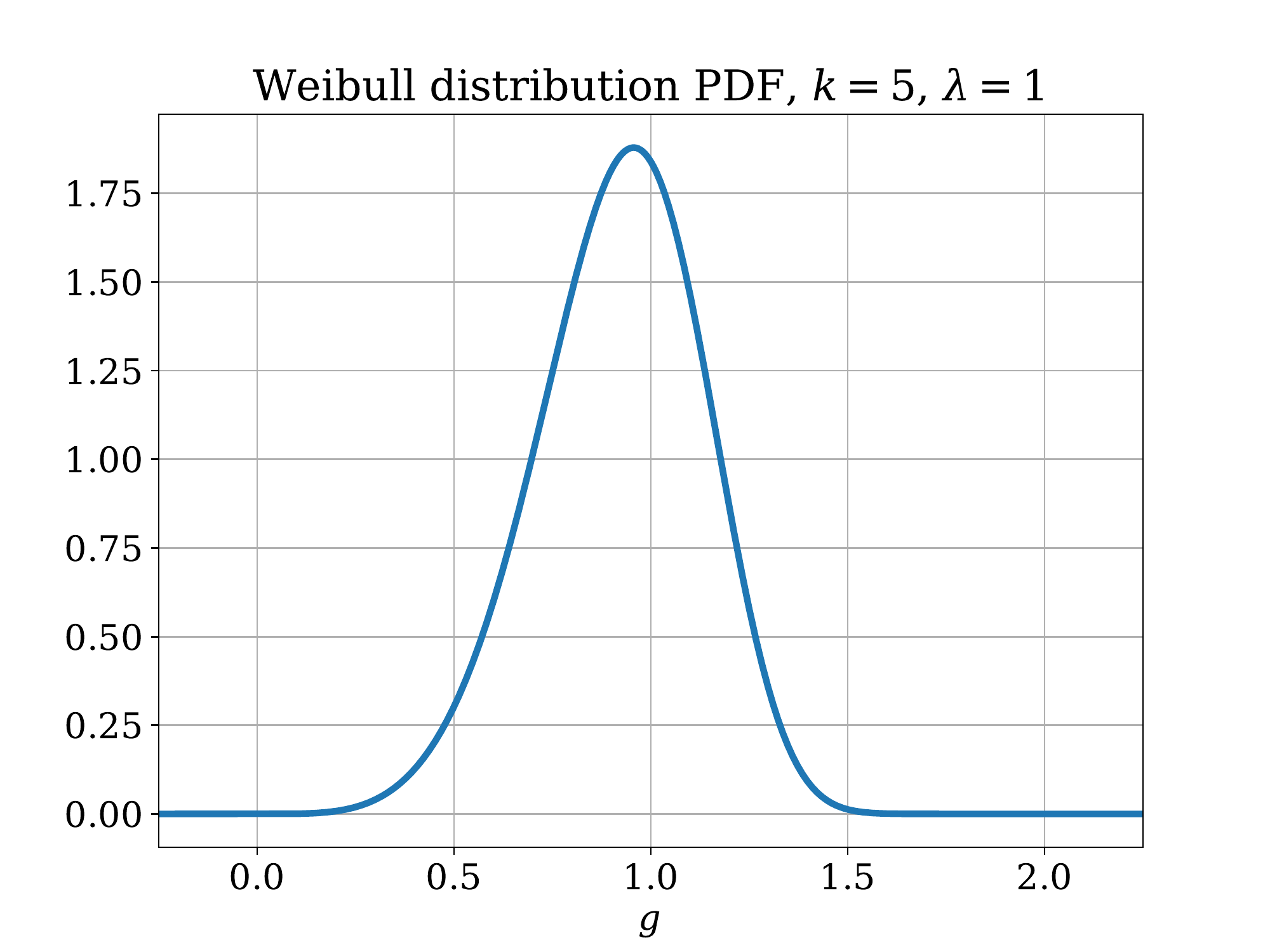}
        \caption{Unimodal peak away from zero.}
        \label{fig:app-weibull-5-pdf}
    \end{subfigure}
    
    \caption[]{Probability density functions (\acs{PDF}s), $f(g)$, examined in our analysis.}
\end{figure}

\subsection{\acs{EMT} plots}
\label{app:plots-emt}

\begin{figure}[H]
    \centering
    \begin{subfigure}[b]{\width\linewidth}
        \includegraphics[width=\linewidth]{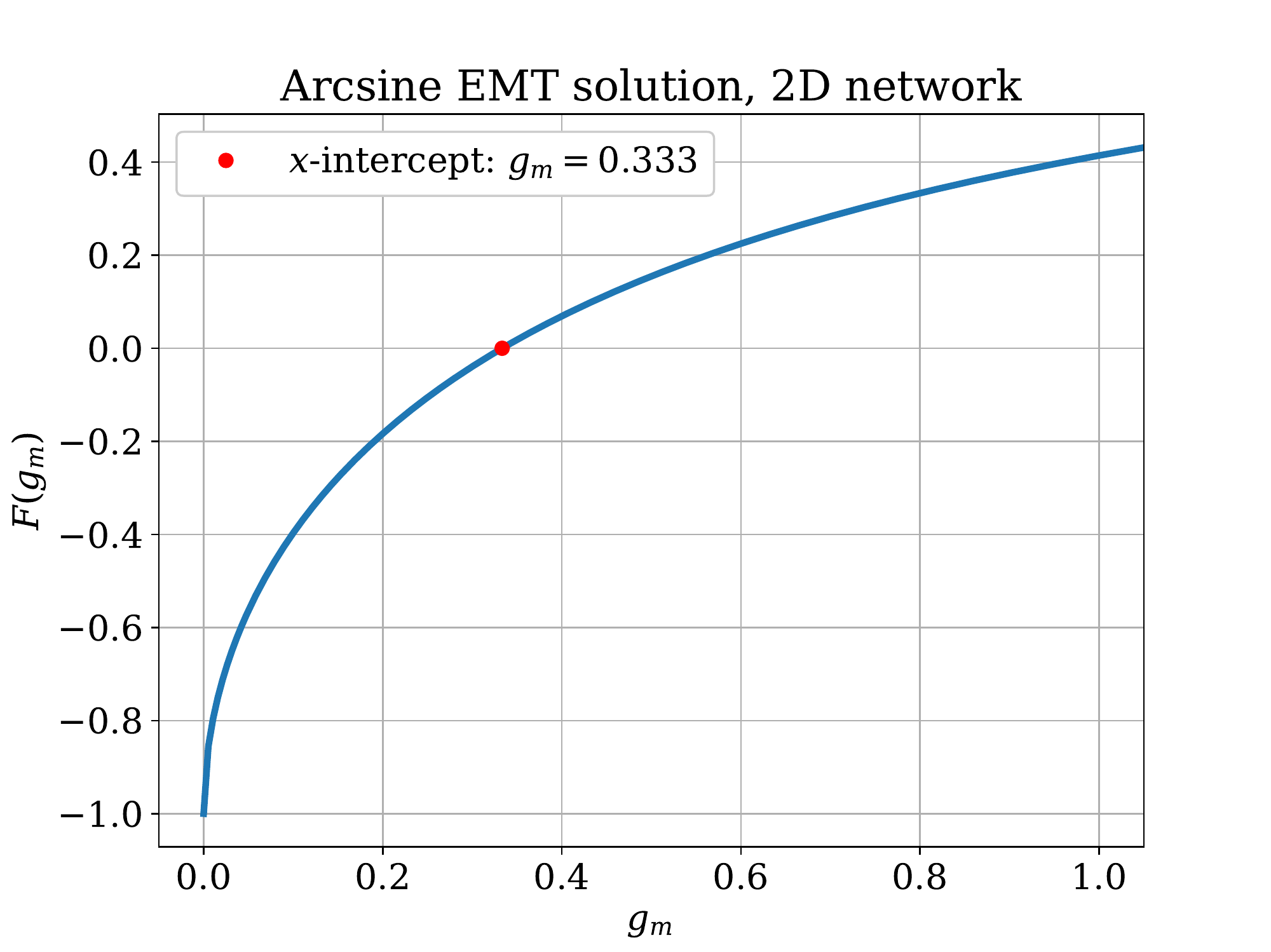}
        \caption{}
        \label{fig:app-arcsine-2d-ki}
    \end{subfigure}
    \begin{subfigure}[b]{\width\linewidth}
        \includegraphics[width=\linewidth]{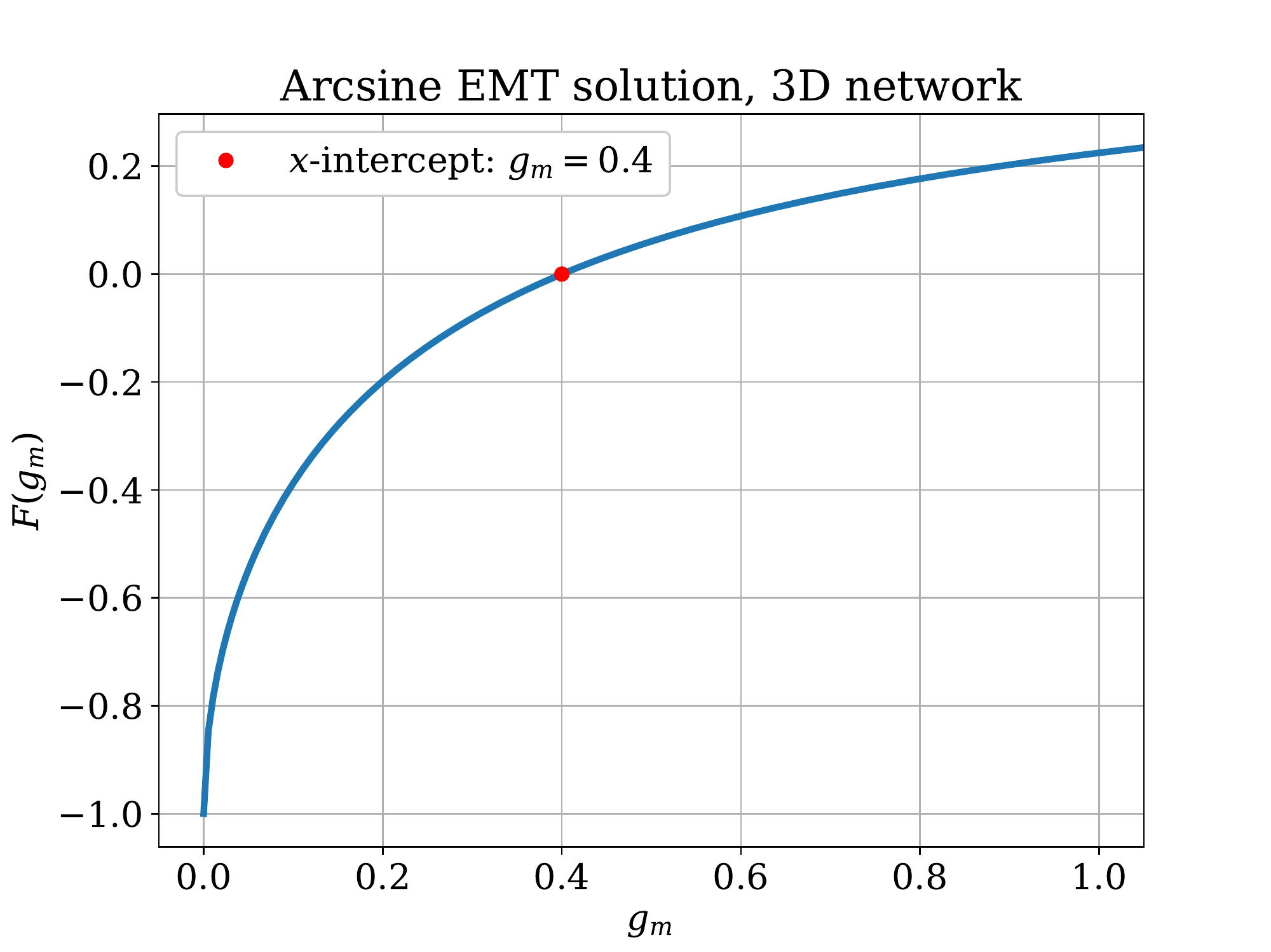}
        \caption{}
        \label{fig:app-arcsine-3d-ki}
    \end{subfigure}
    
    \begin{subfigure}[b]{\width\linewidth}
        \includegraphics[width=\linewidth]{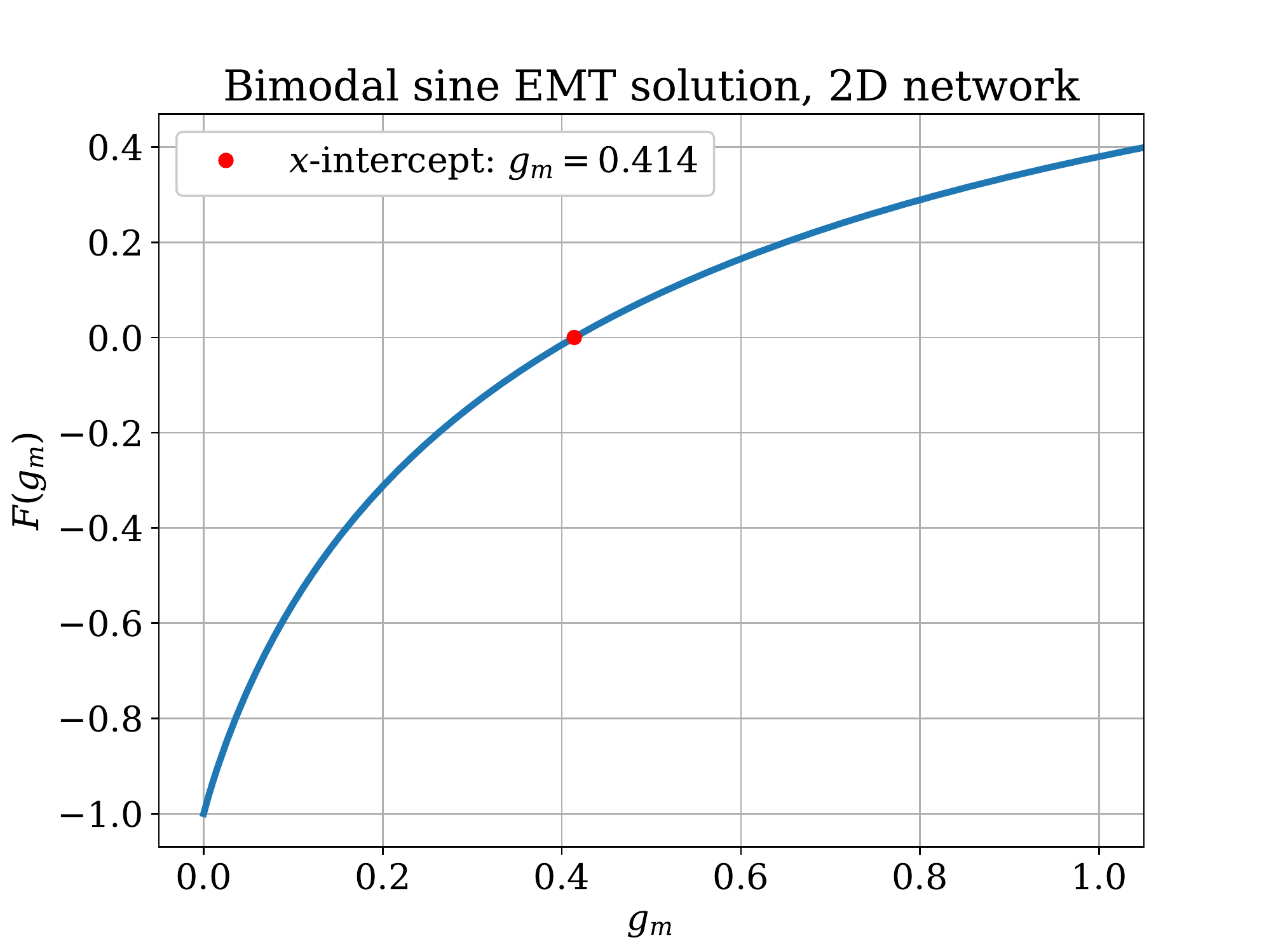}
        \caption{}
        \label{fig:app-bimodal-sine-2d-ki}
    \end{subfigure}
    \begin{subfigure}[b]{\width\linewidth}
        \includegraphics[width=\linewidth]{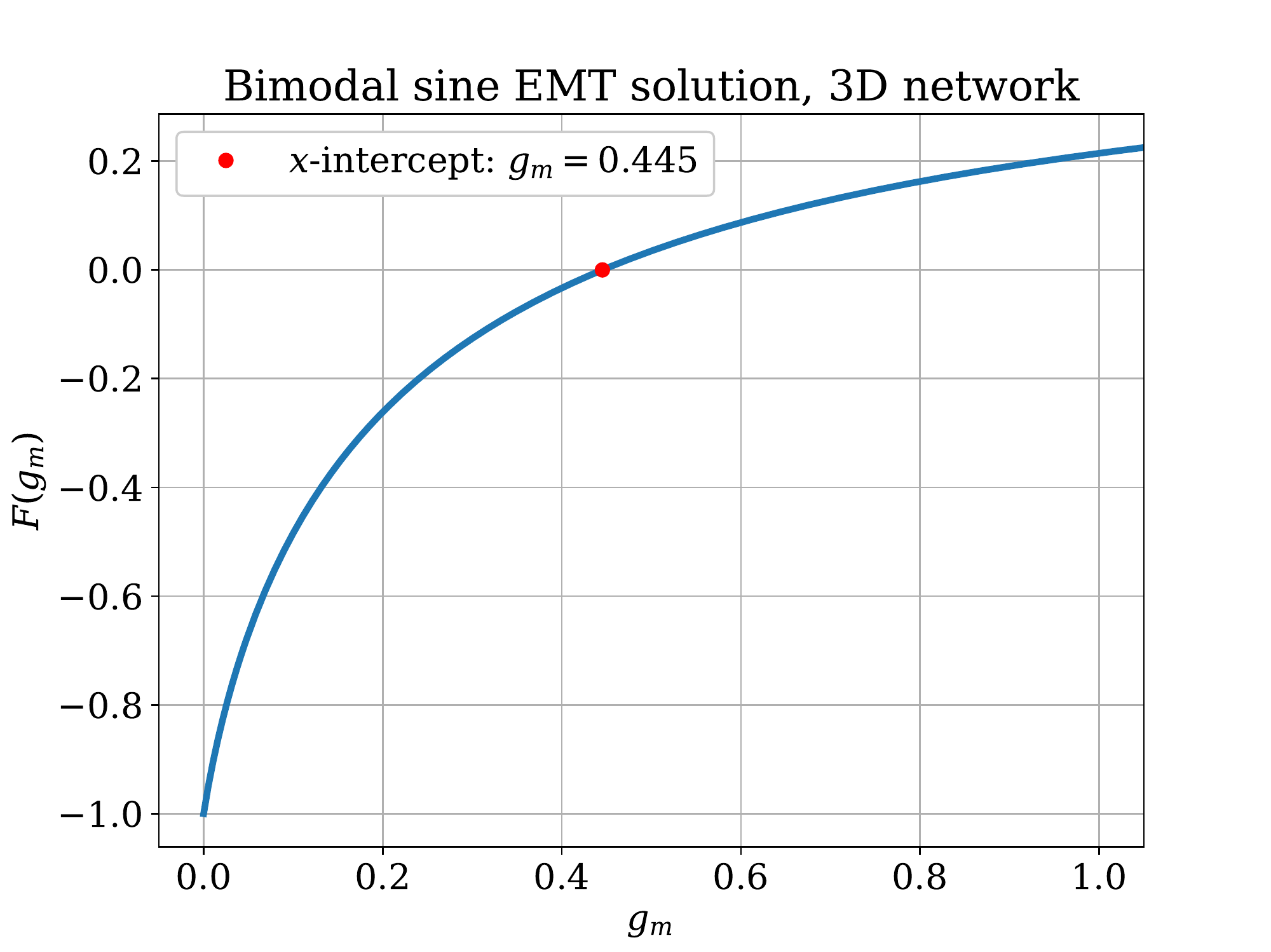}
        \caption{}
        \label{fig:app-bimodal-sine-3d-ki}
    \end{subfigure}
    
    \begin{subfigure}[b]{\width\linewidth}
        \includegraphics[width=\linewidth]{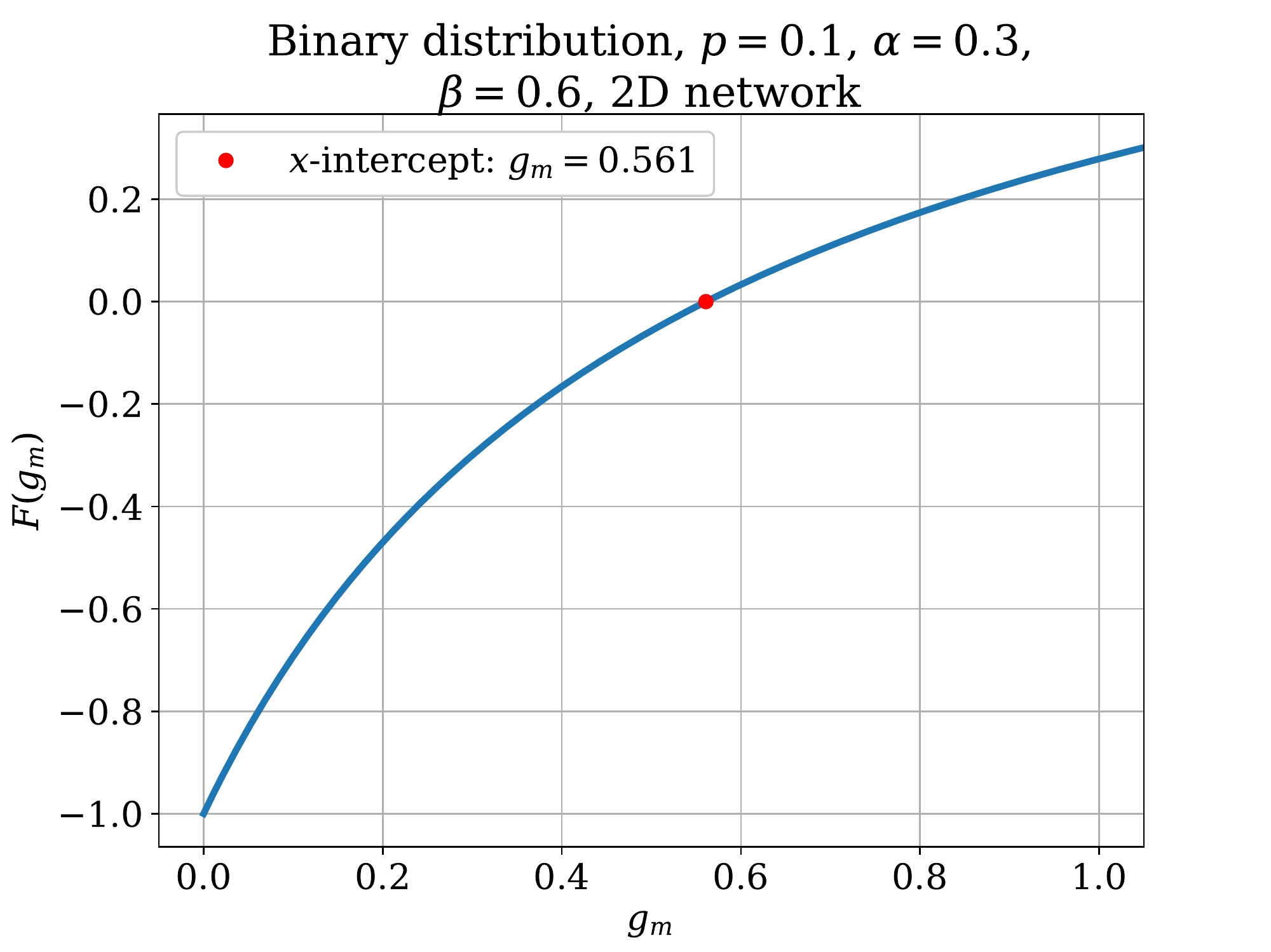}
        \caption{}
        \label{fig:app-binary-0p1-2d-ki}
    \end{subfigure}
    \begin{subfigure}[b]{\width\linewidth}
        \includegraphics[width=\linewidth]{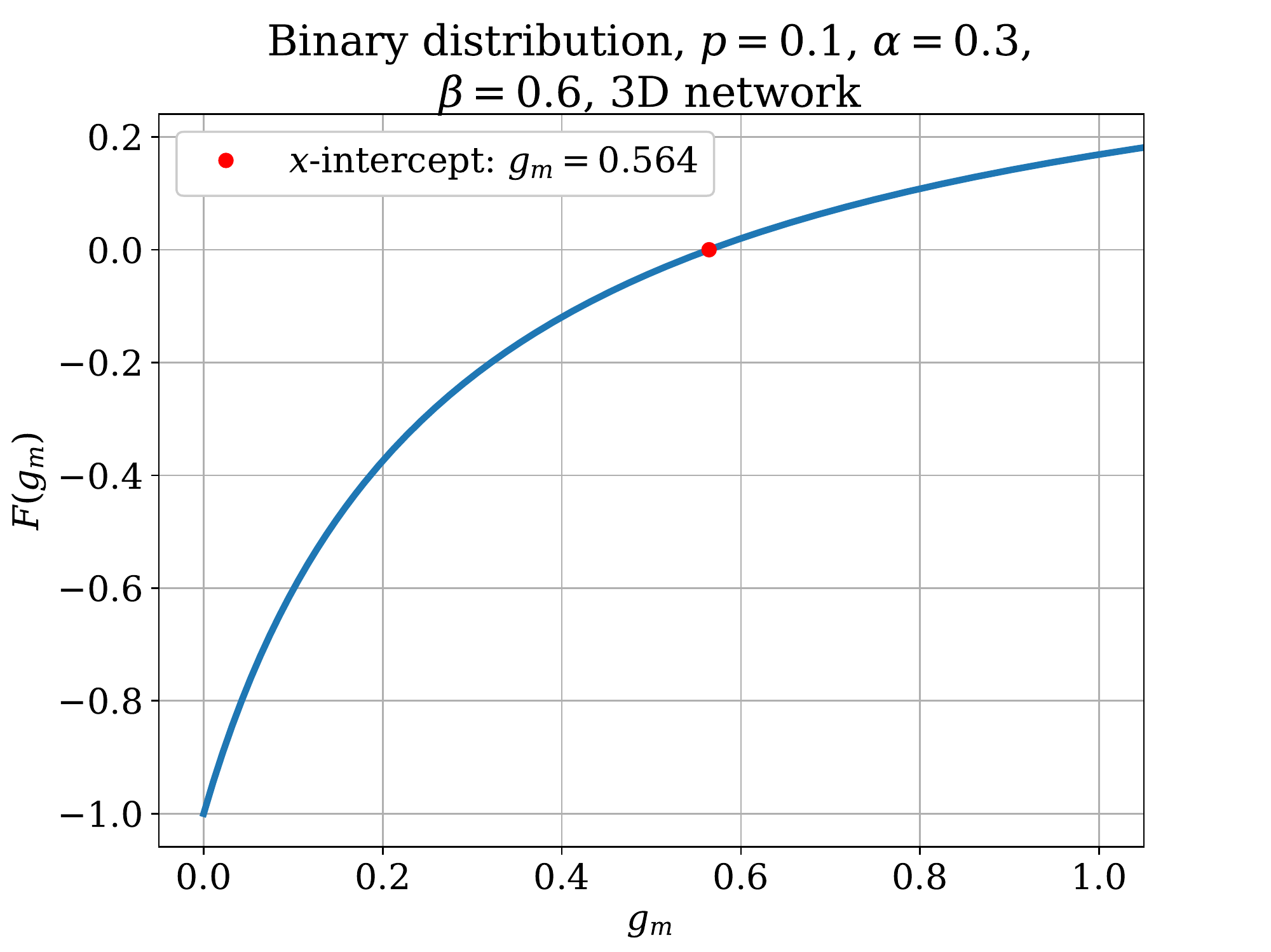}
        \caption{}
        \label{fig:app-binary-0p1-3d-ki}
    \end{subfigure}
    \caption{Solutions of \autoref{eq:kirkpatrick-vanish} for \acs{PDF}s listed in \autoref{tab:pdf}.}
    \label{fig:app-ki}
\end{figure}

\begin{figure}[H]\ContinuedFloat
    \centering
    \begin{subfigure}[b]{\width\linewidth}
        \includegraphics[width=\linewidth]{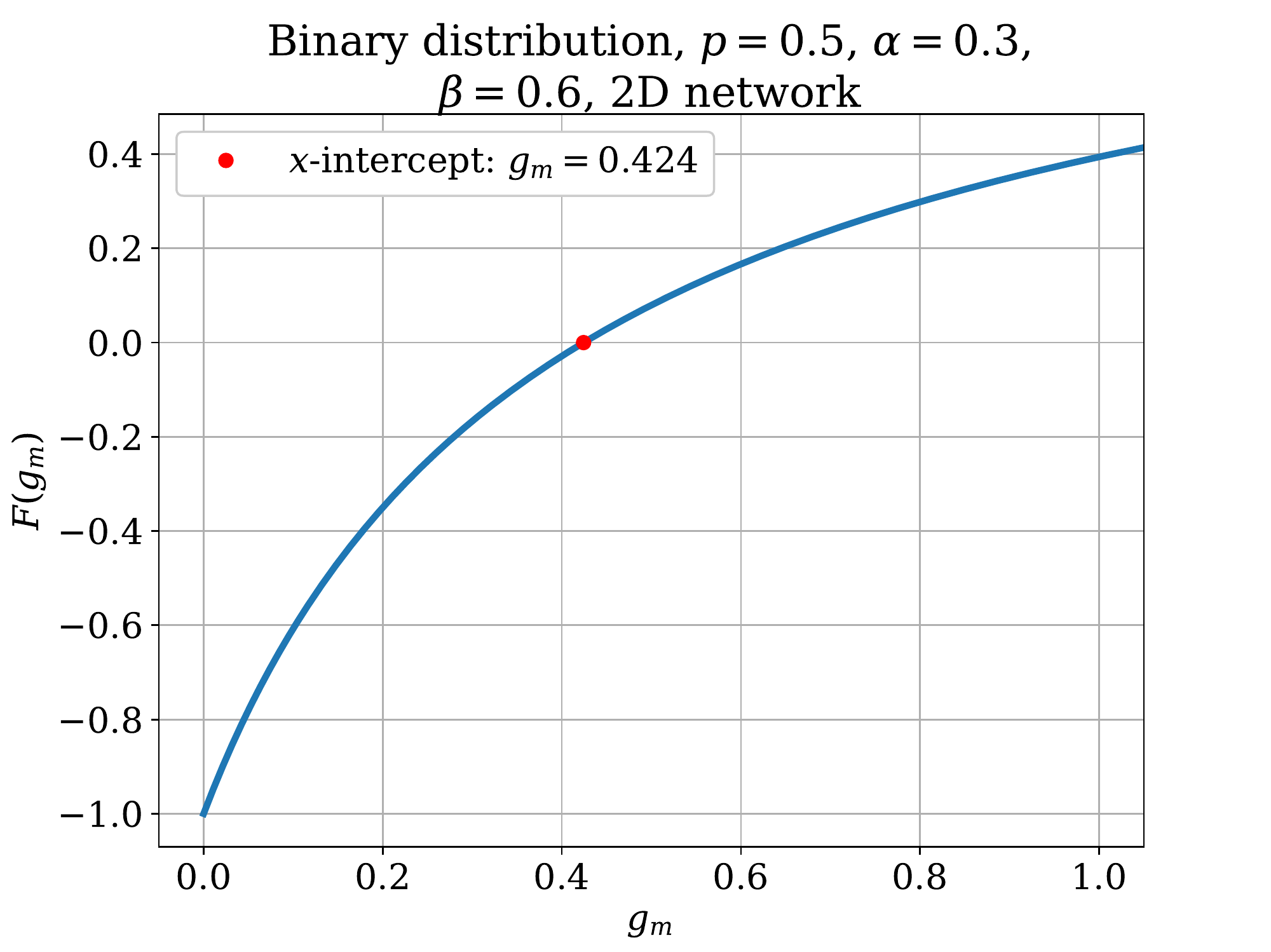}
        \caption{}
        \label{fig:app-binary-0p5-2d-ki}
    \end{subfigure}
    \begin{subfigure}[b]{\width\linewidth}
        \includegraphics[width=\linewidth]{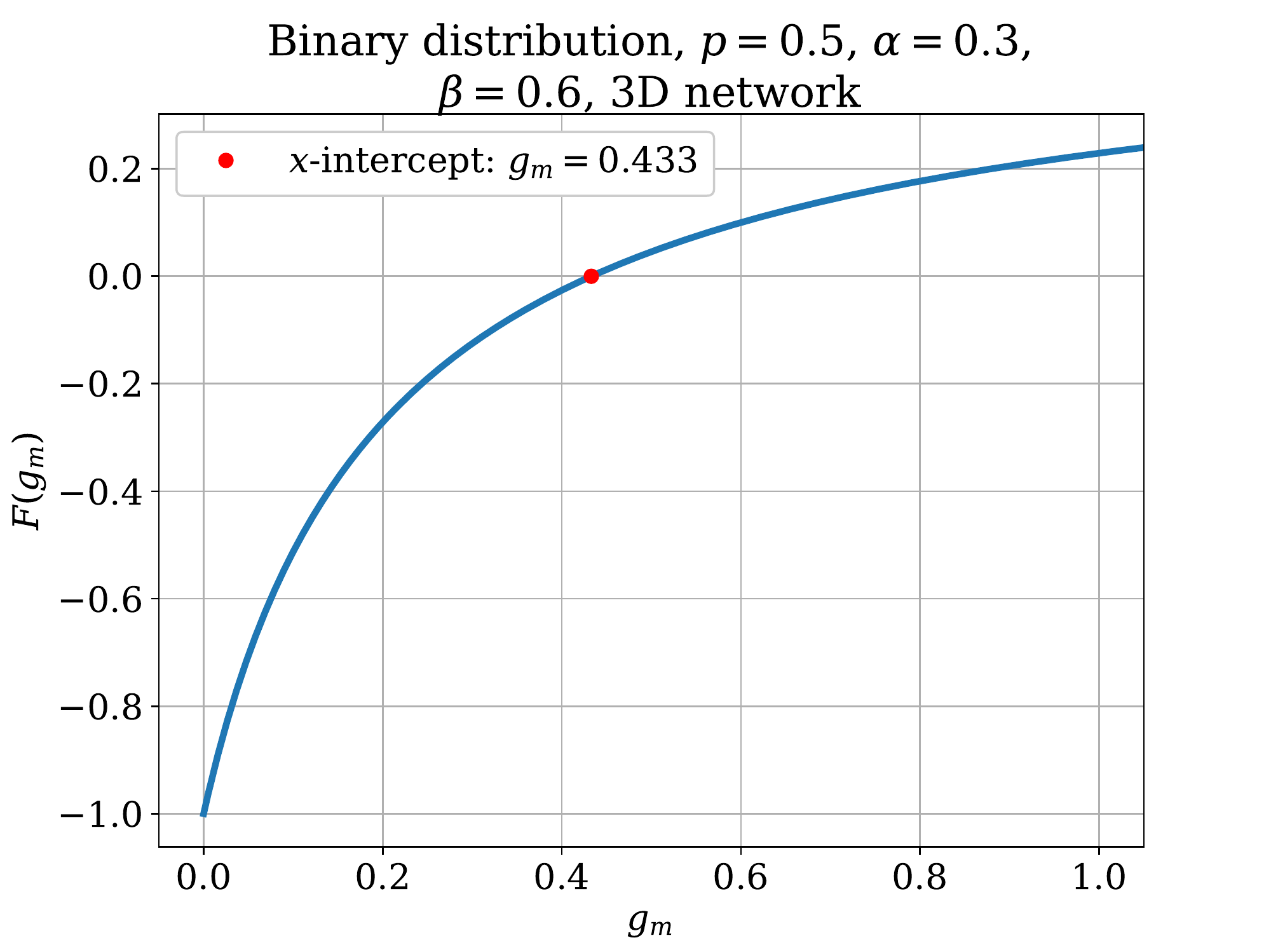}
        \caption{}
        \label{fig:app-binary-0p5-3d-ki}
    \end{subfigure}
    
    \begin{subfigure}[b]{\width\linewidth}
        \includegraphics[width=\linewidth]{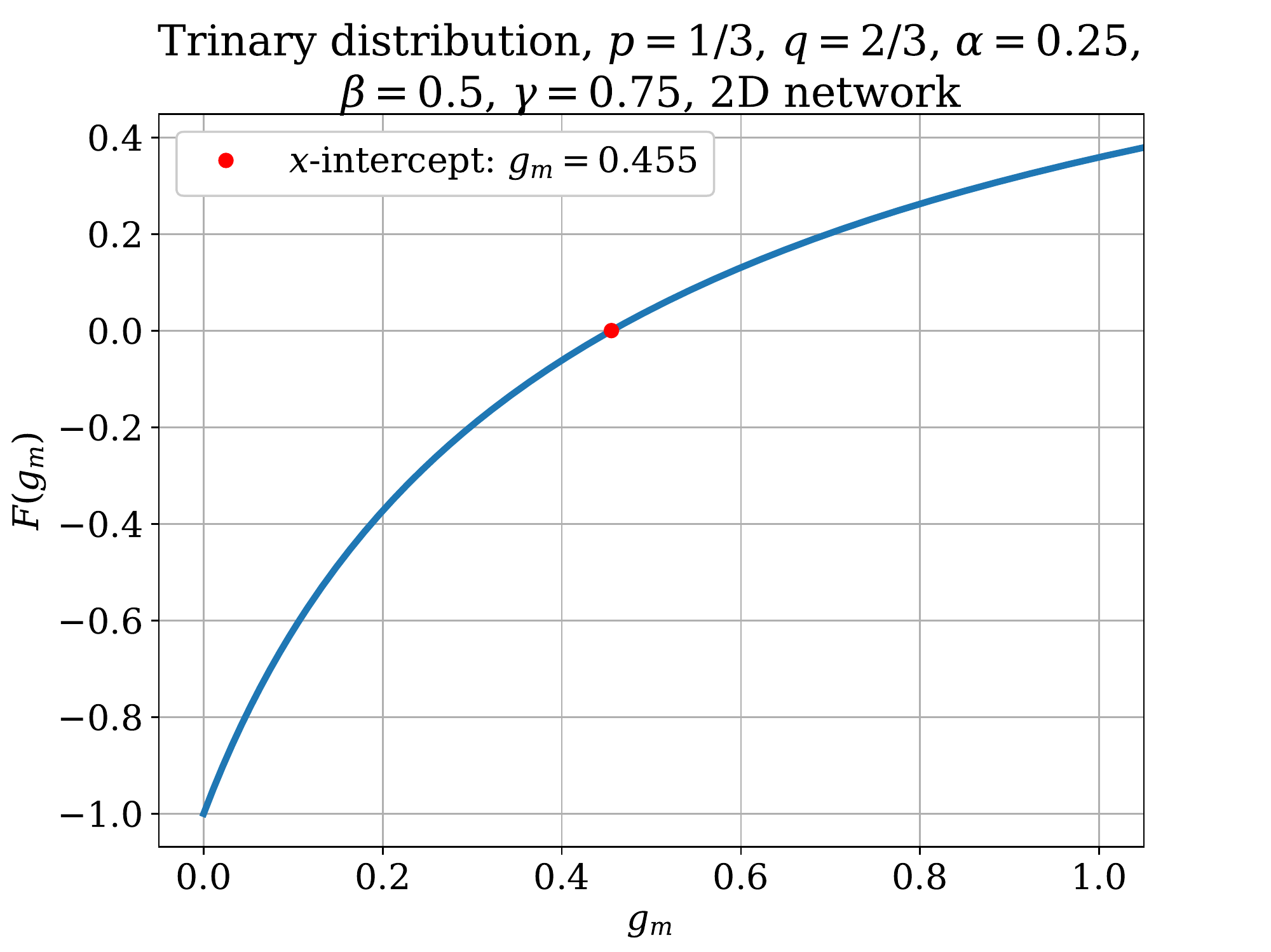}
        \caption{}
        \label{fig:app-trinary-2d-ki}
    \end{subfigure}
    \begin{subfigure}[b]{\width\linewidth}
        \includegraphics[width=\linewidth]{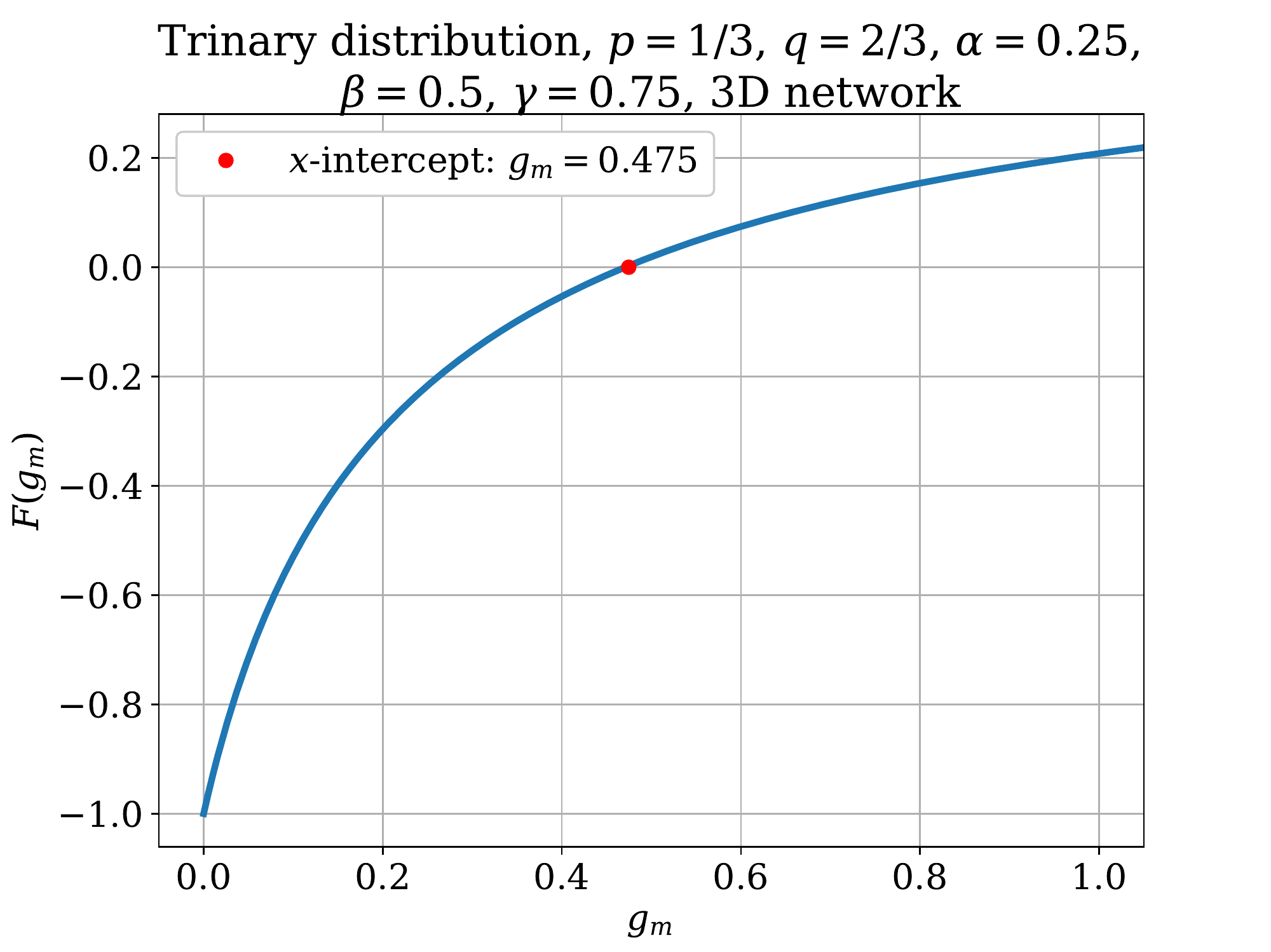}
        \caption{}
        \label{fig:app-trinary-3d-ki}
    \end{subfigure}
    
    \begin{subfigure}[b]{\width\linewidth}
        \includegraphics[width=\linewidth]{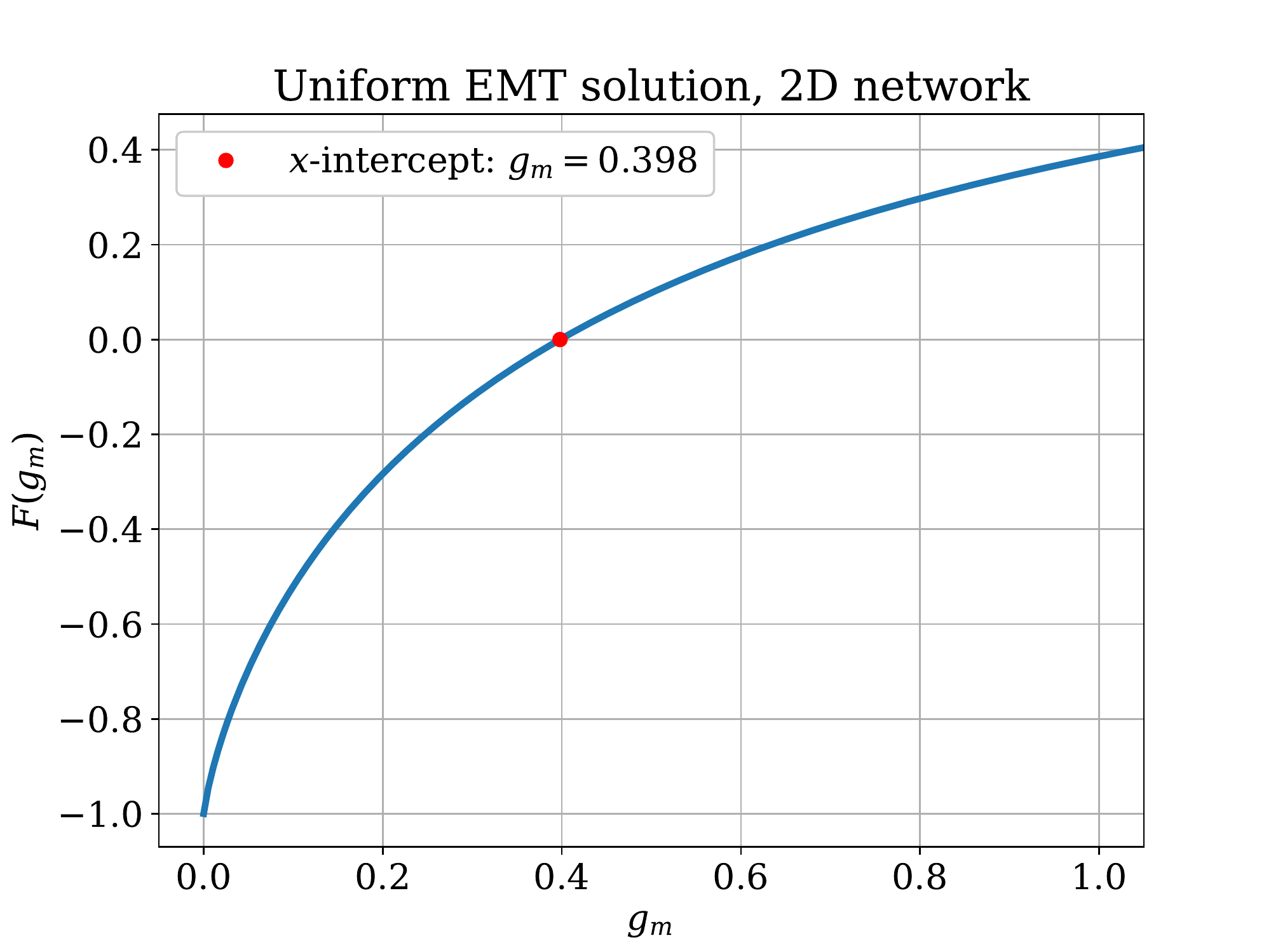}
        \caption{}
        \label{fig:app-uniform-2d-ki}
    \end{subfigure}
    \begin{subfigure}[b]{\width\linewidth}
        \includegraphics[width=\linewidth]{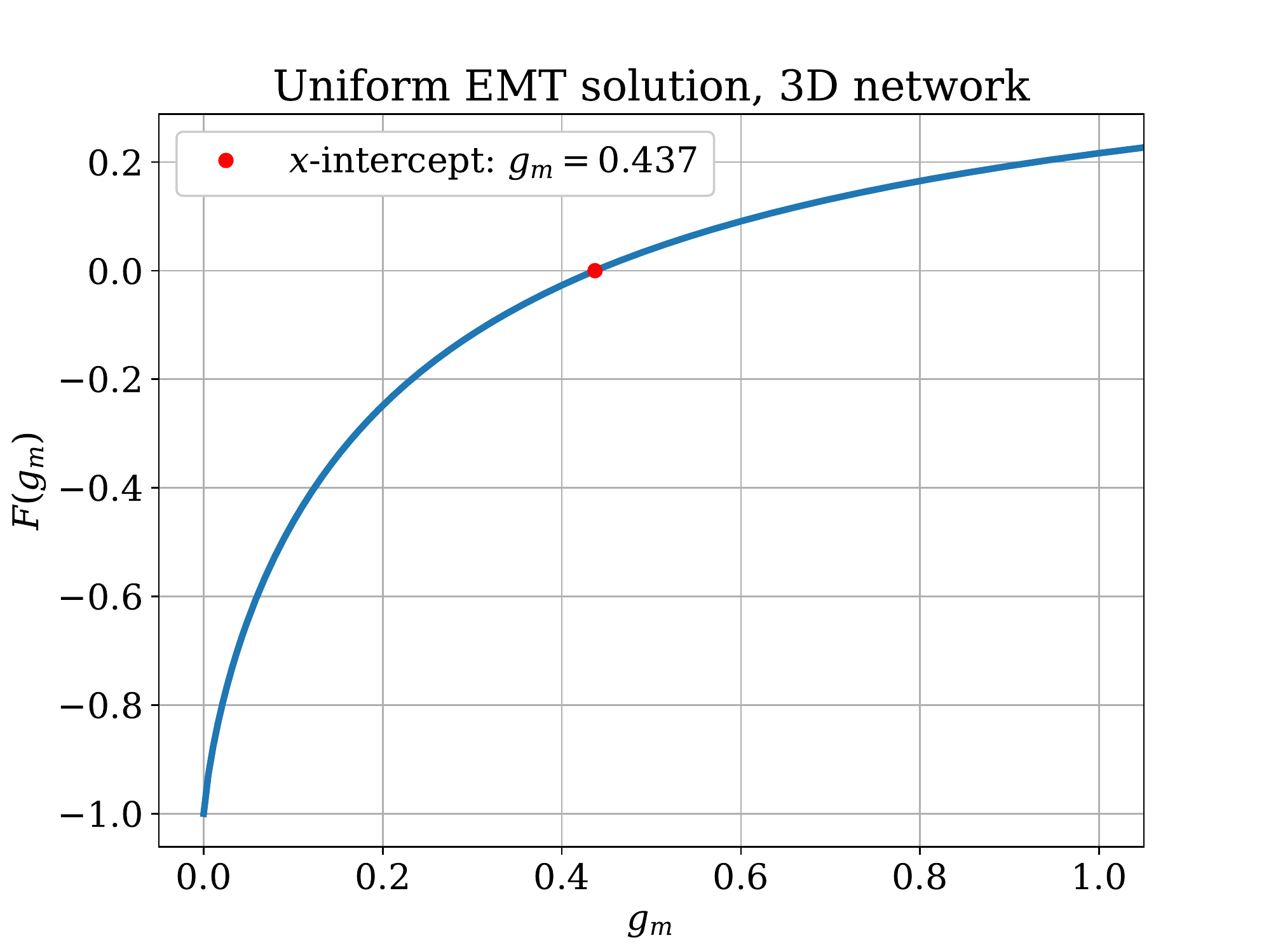}
        \caption{}
        \label{fig:app-uniform-3d-ki}
    \end{subfigure}
    \caption[]{Solutions of \autoref{eq:kirkpatrick-vanish} for \acs{PDF}s listed in \autoref{tab:pdf}.}
\end{figure}

\begin{figure}[H]\ContinuedFloat
    \centering    
    \begin{subfigure}[b]{\width\linewidth}
        \includegraphics[width=\linewidth]{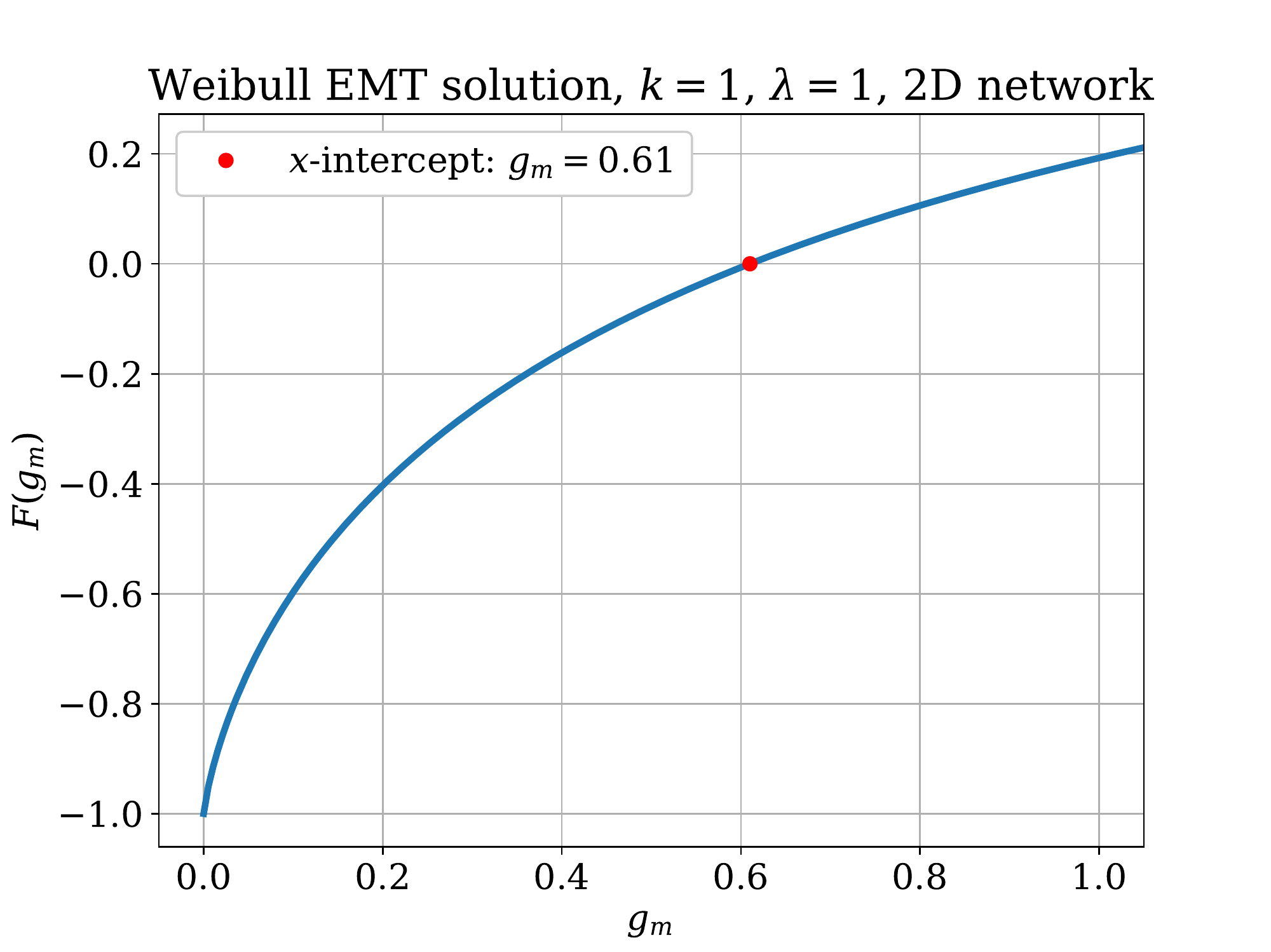}
        \caption{}
        \label{fig:app-weibull-1-2d-ki}
    \end{subfigure}
    \begin{subfigure}[b]{\width\linewidth}
        \includegraphics[width=\linewidth]{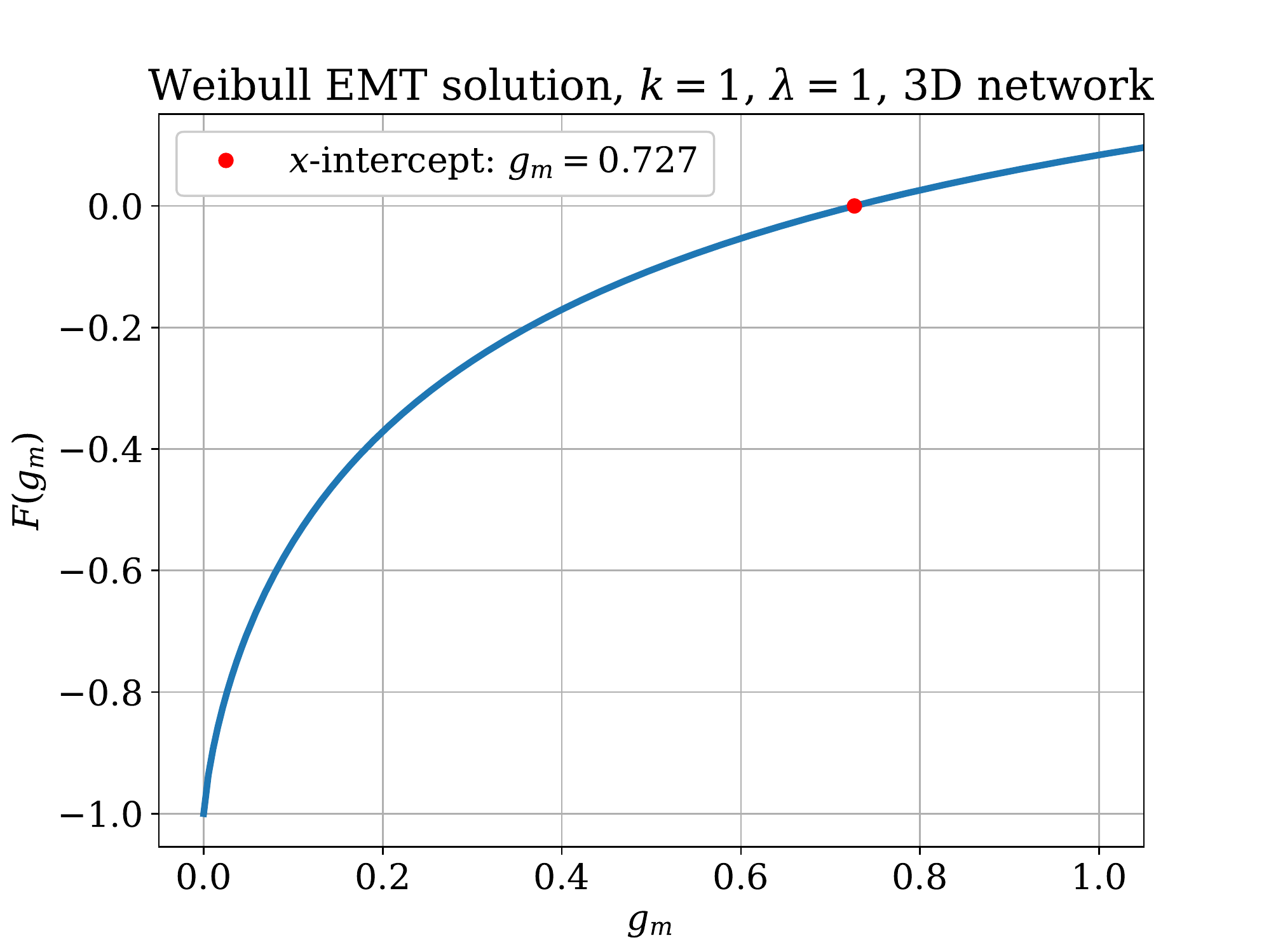}
        \caption{}
        \label{fig:app-weibull-1-3d-ki}
    \end{subfigure}
    
    \begin{subfigure}[b]{\width\linewidth}
        \includegraphics[width=\linewidth]{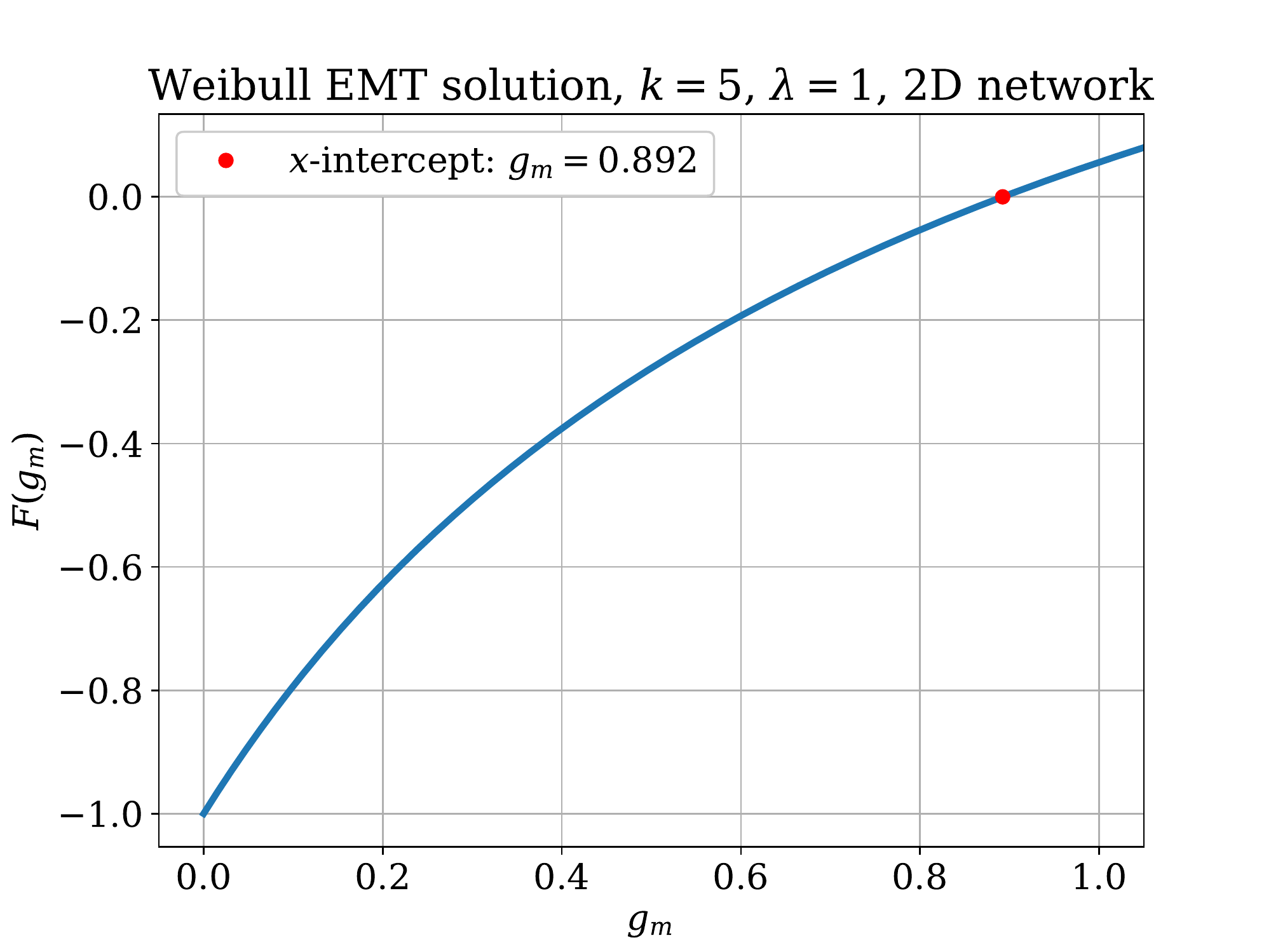}
        \caption{}
        \label{fig:app-weibull-5-2d-ki}
    \end{subfigure}
    \begin{subfigure}[b]{\width\linewidth}
        \includegraphics[width=\linewidth]{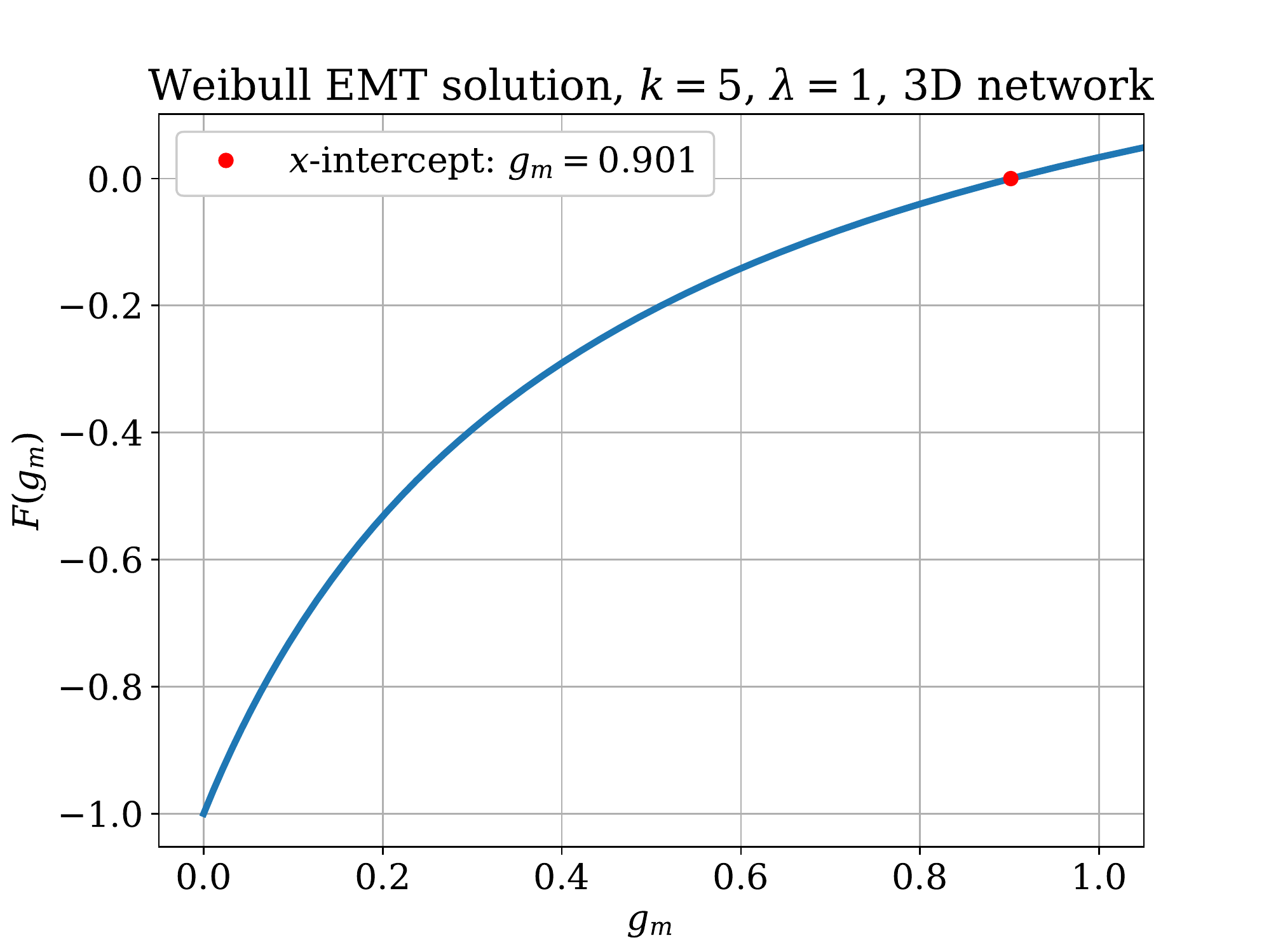}
        \caption{}
        \label{fig:app-weibull-5-3d-ki}
    \end{subfigure}
    
    \caption[]{Solutions of \autoref{eq:kirkpatrick-vanish} for \acs{PDF}s listed in \autoref{tab:pdf}.}
\end{figure}

\subsection{Power law fit plots: \acs{RD} and \acs{RSD}}
\label{app:plots-fits}

\begin{figure}[H]
    \centering
    
    \begin{subfigure}[b]{\width\linewidth}
        \includegraphics[width=\linewidth]{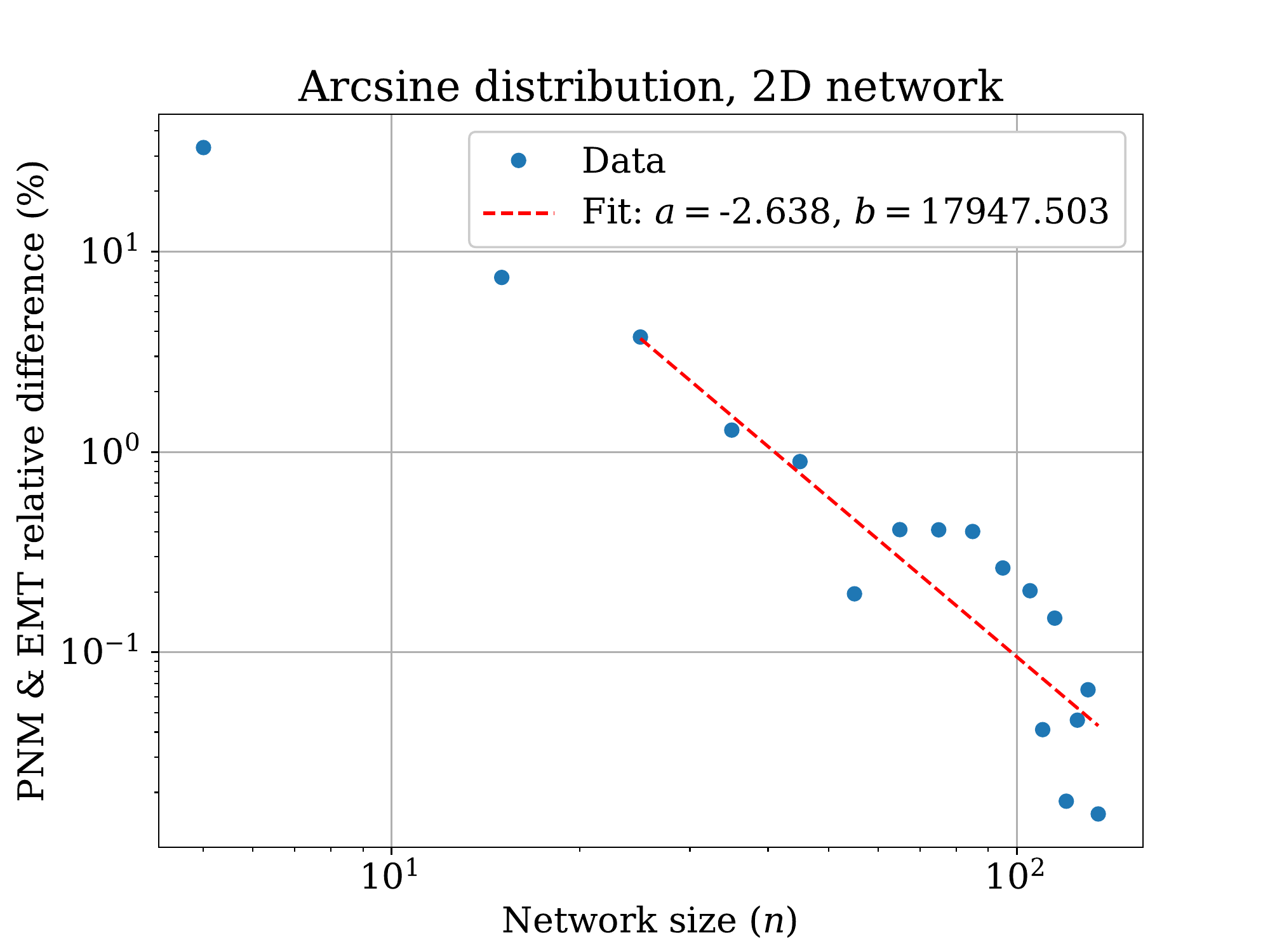}
        \caption{}
        \label{fig:app-arcsine-2d-err}
    \end{subfigure}
    \begin{subfigure}[b]{\width\linewidth}
        \includegraphics[width=\linewidth]{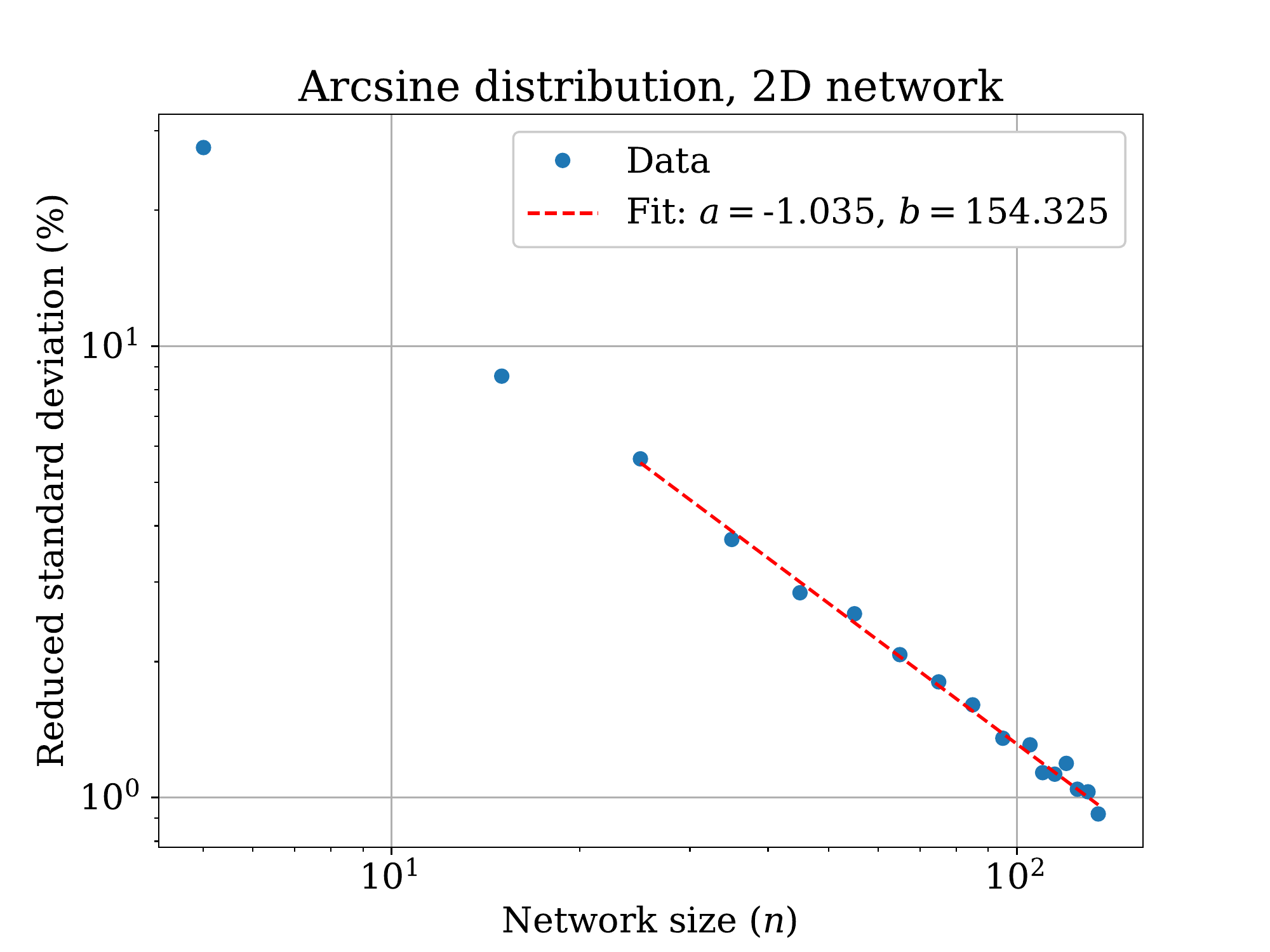}
        \caption{}
        \label{fig:app-arcsine-2d-rsd}
    \end{subfigure}
    
    \begin{subfigure}[b]{\width\linewidth}
        \includegraphics[width=\linewidth]{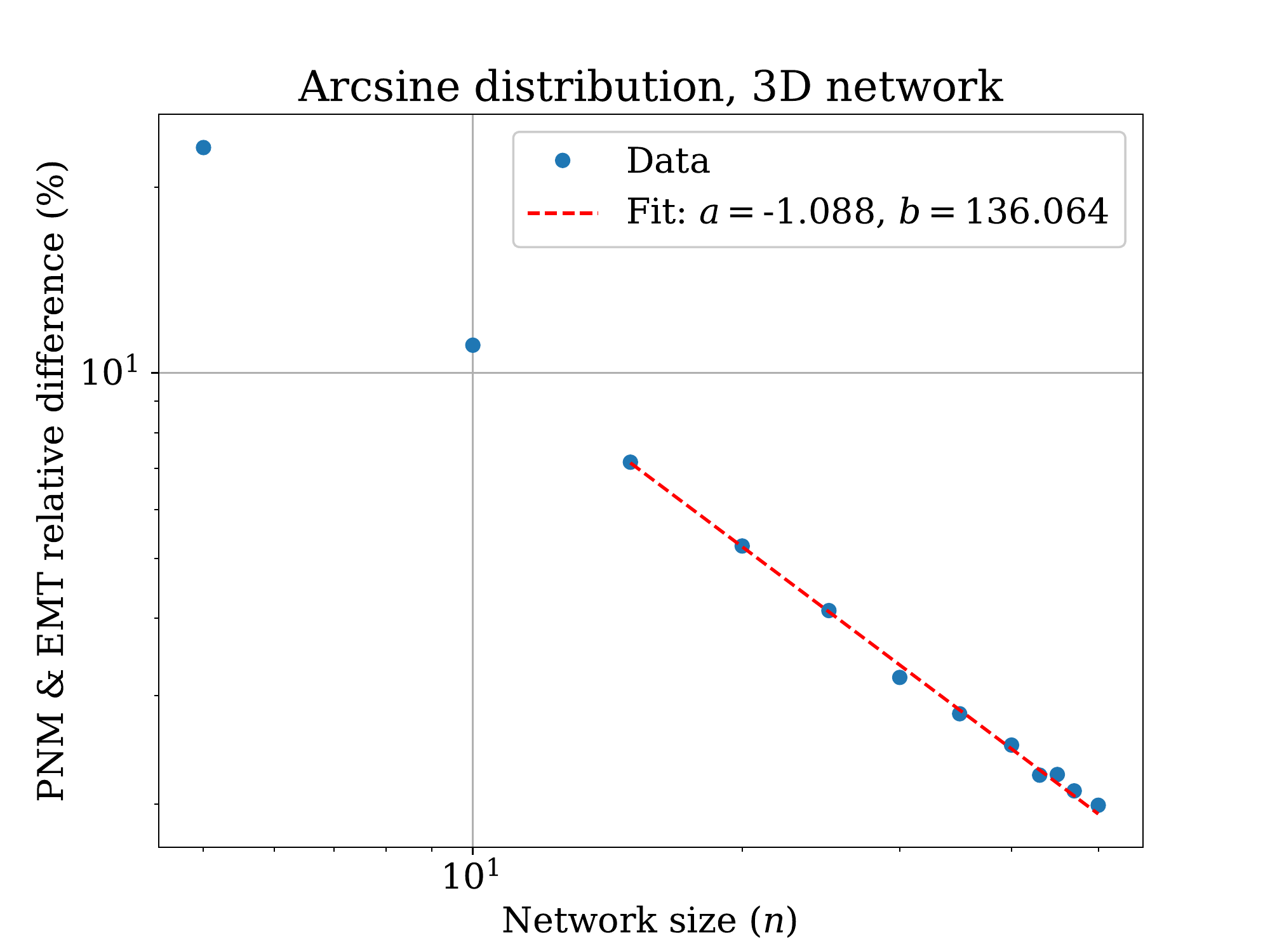}
        \caption{}
        \label{fig:app-arcsine-3d-err}
    \end{subfigure}
    \begin{subfigure}[b]{\width\linewidth}
        \includegraphics[width=\linewidth]{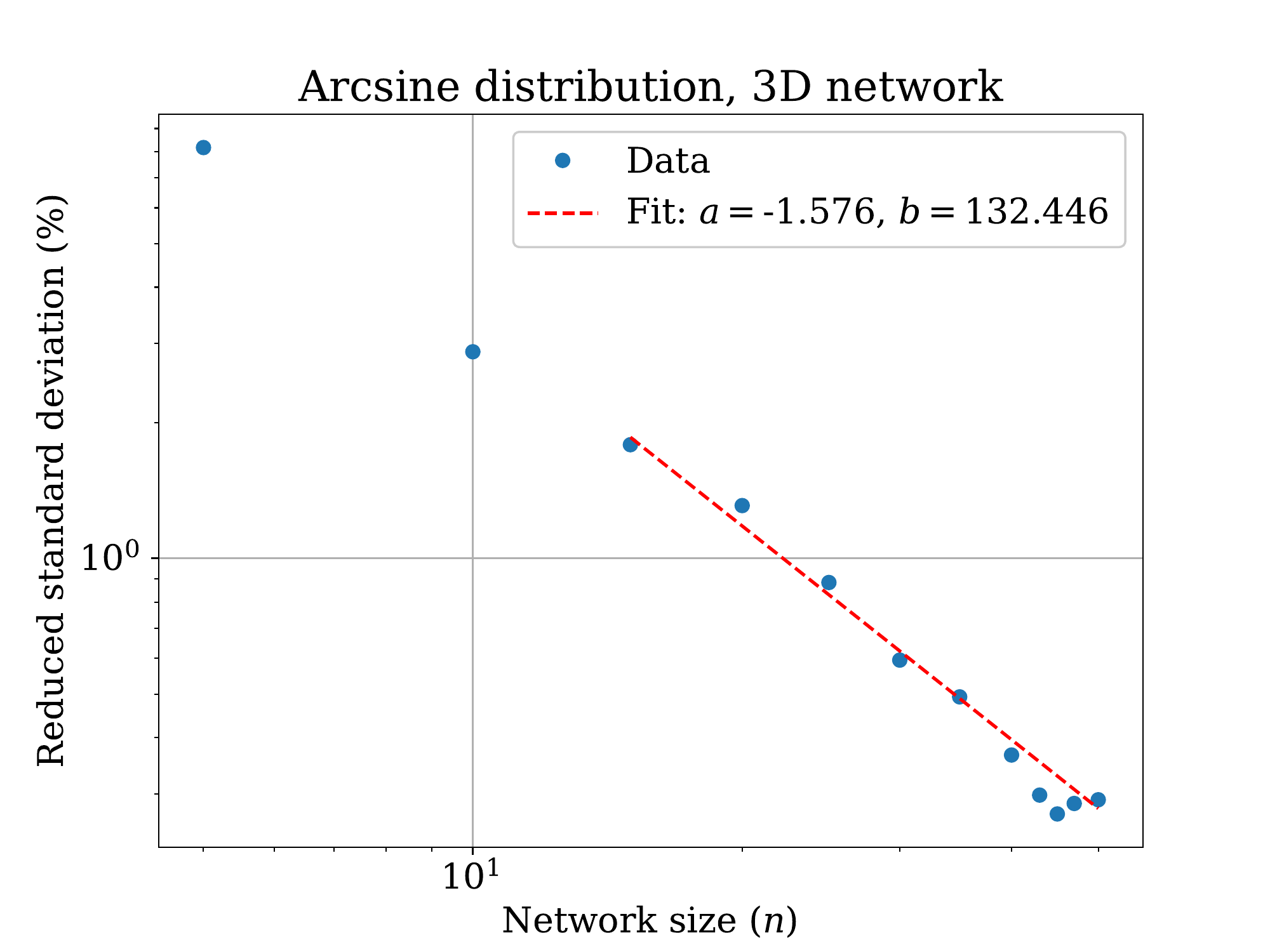}
        \caption{}
        \label{fig:app-arcsine-3d-rsd}
    \end{subfigure}
    
    \begin{subfigure}[b]{\width\linewidth}
        \includegraphics[width=\linewidth]{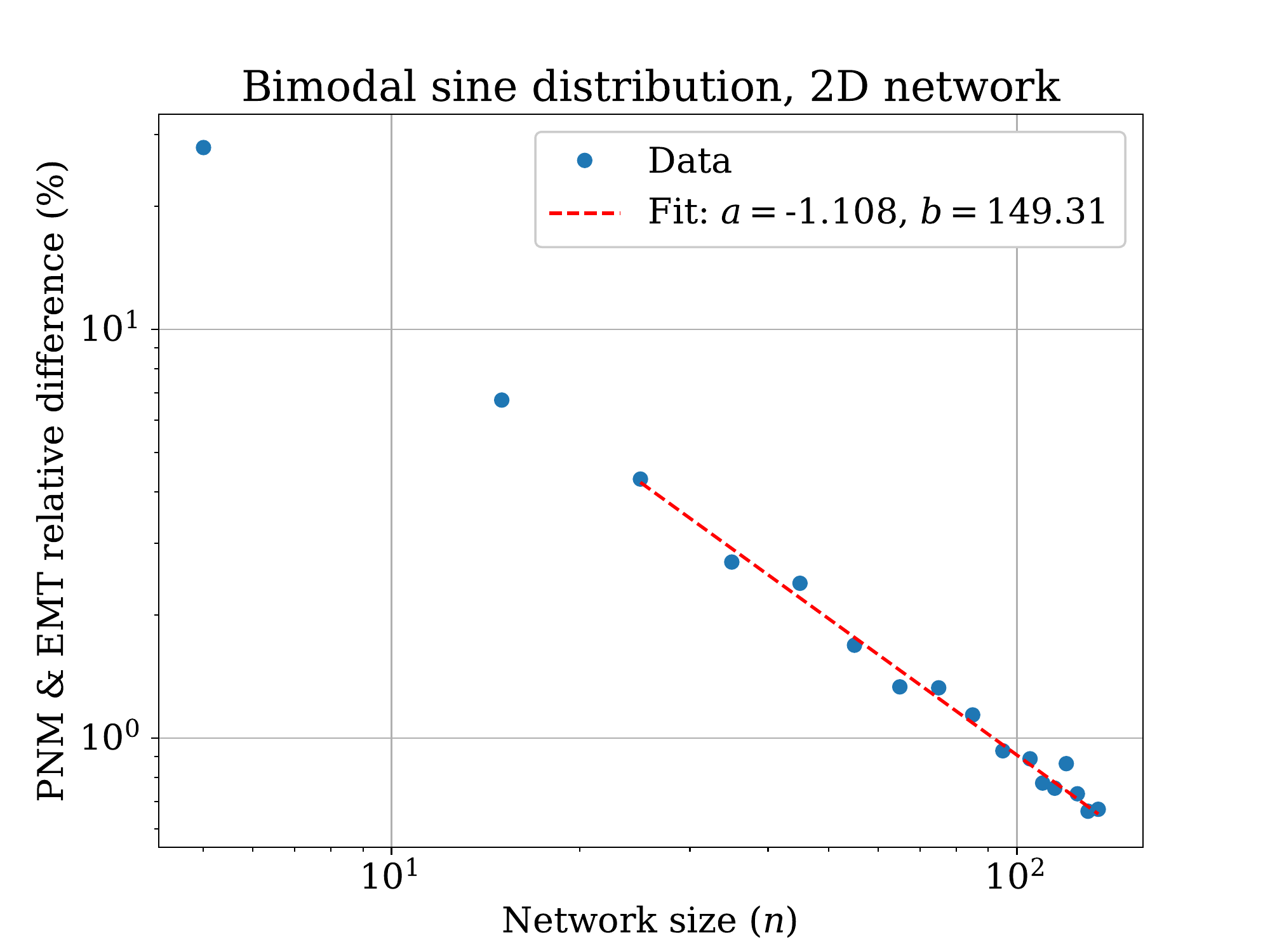}
        \caption{}
        \label{fig:app-bimodal-sine-2d-err}
    \end{subfigure}
    \begin{subfigure}[b]{\width\linewidth}
        \includegraphics[width=\linewidth]{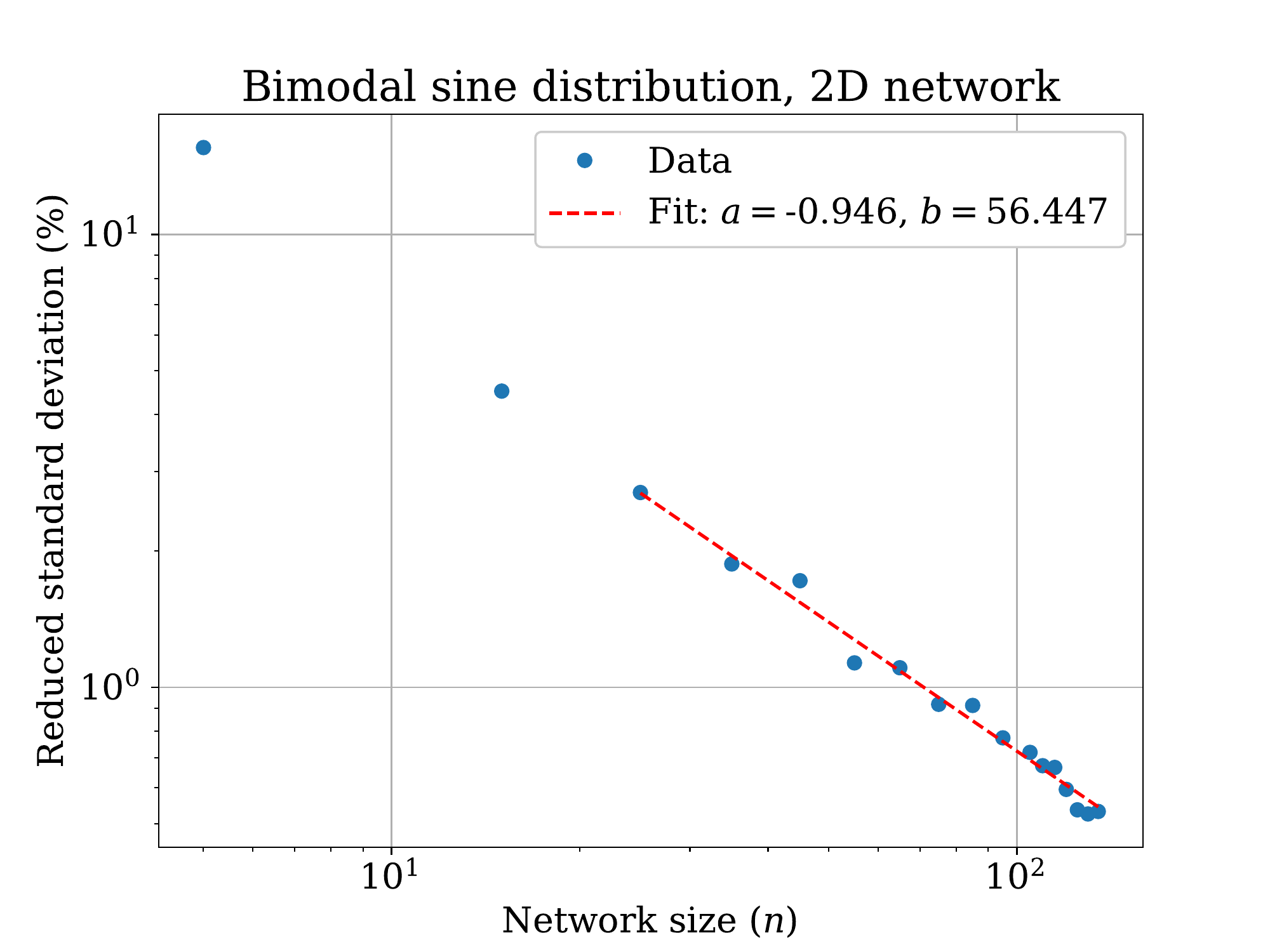}
        \caption{}
        \label{fig:app-bimodal-sine-2d-rsd}
    \end{subfigure}
    
    \caption{Power law fits to relative difference (\acs{RD}) and reduced standard deviation (\acs{RSD}) of \acs{PNM} solution, as a function of $n$.}
    \label{fig:app-fits}
\end{figure}

\begin{figure}[H]\ContinuedFloat
    \centering
    
    \begin{subfigure}[b]{\width\linewidth}
        \includegraphics[width=\linewidth]{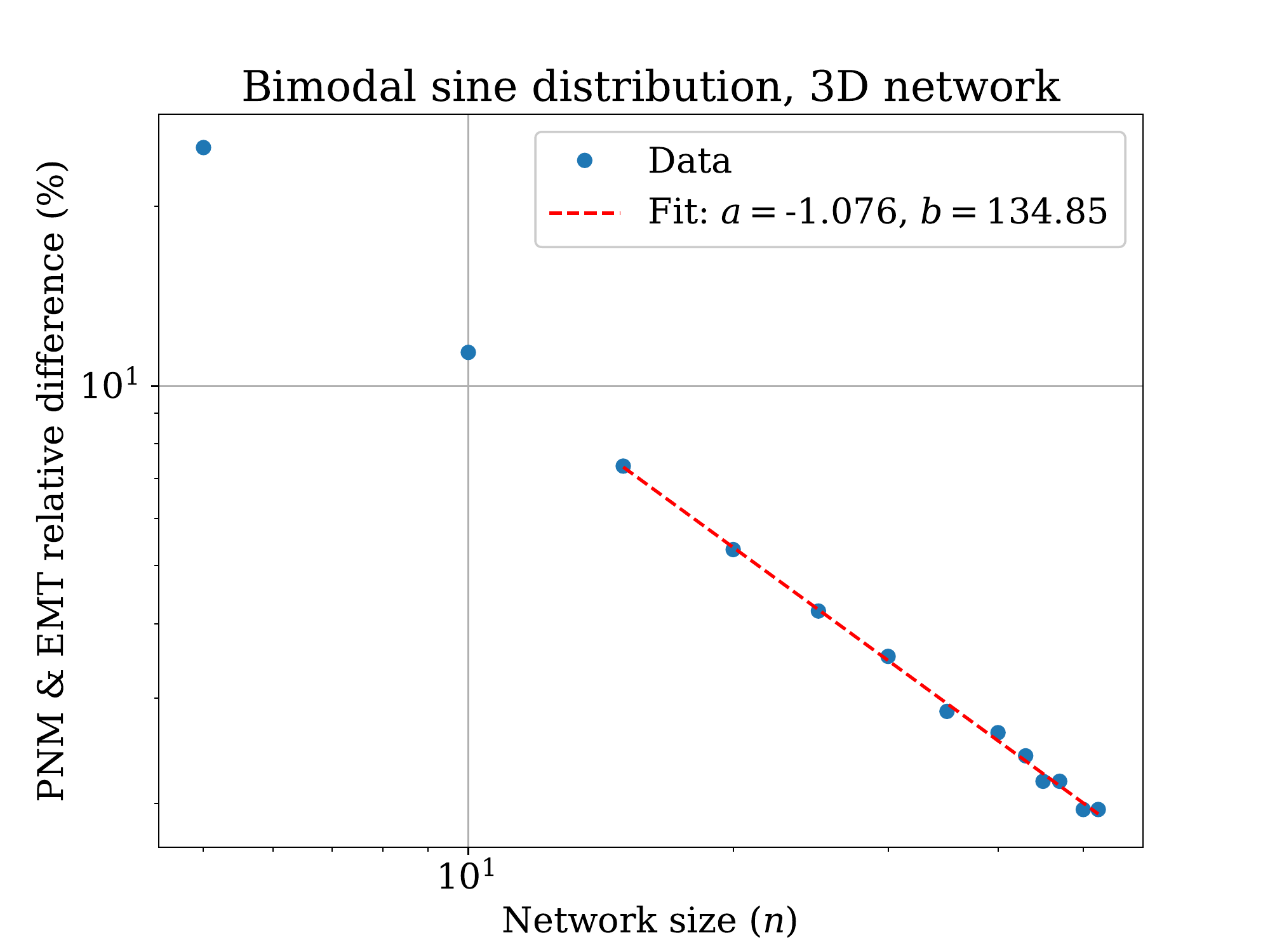}
        \caption{}
        \label{fig:app-bimodal-sine-3d-err}
    \end{subfigure}
    \begin{subfigure}[b]{\width\linewidth}
        \includegraphics[width=\linewidth]{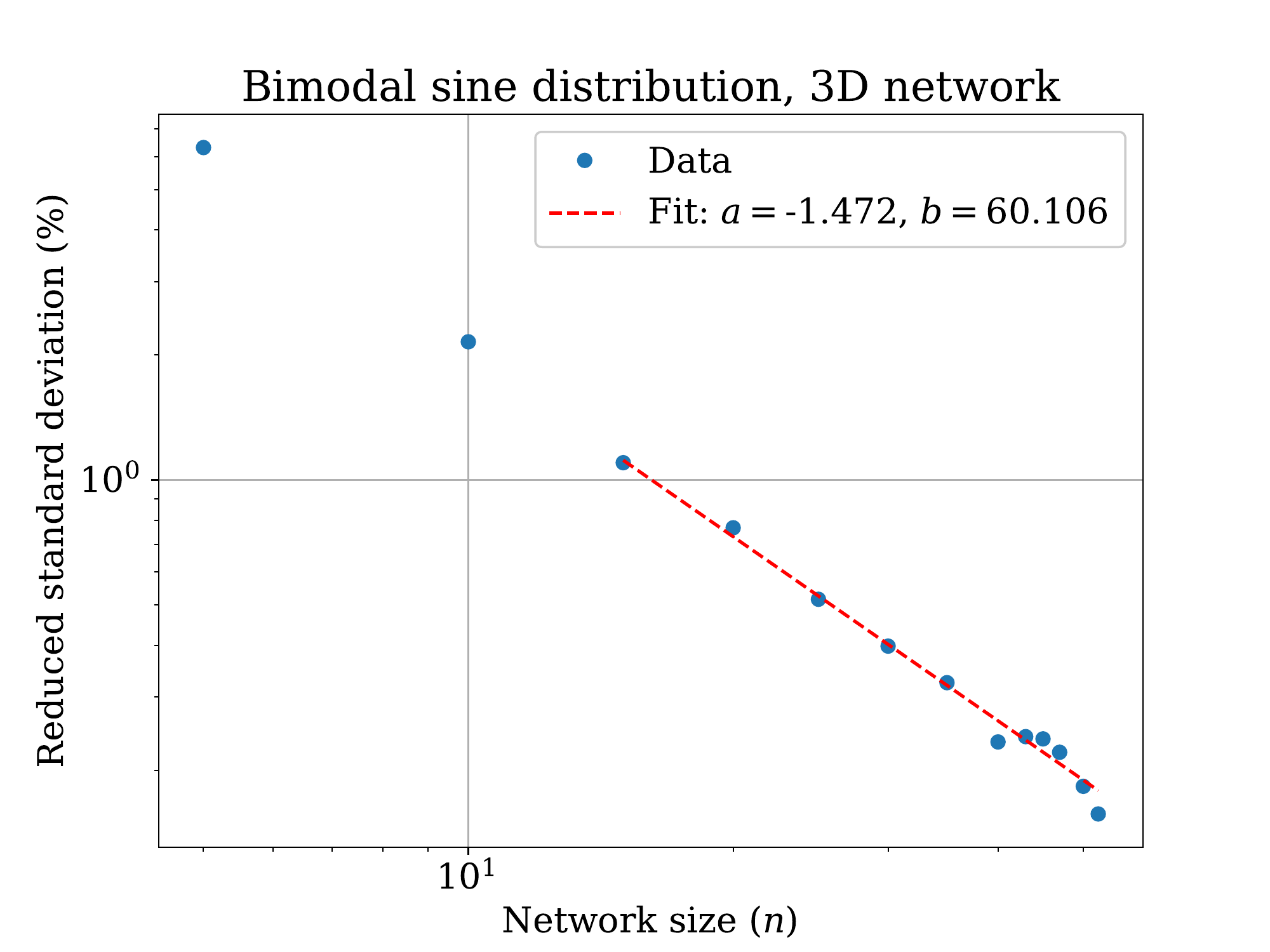}
        \caption{}
        \label{fig:app-bimodal-sine-3d-rsd}
    \end{subfigure}
    
    \begin{subfigure}[b]{\width\linewidth}
        \includegraphics[width=\linewidth]{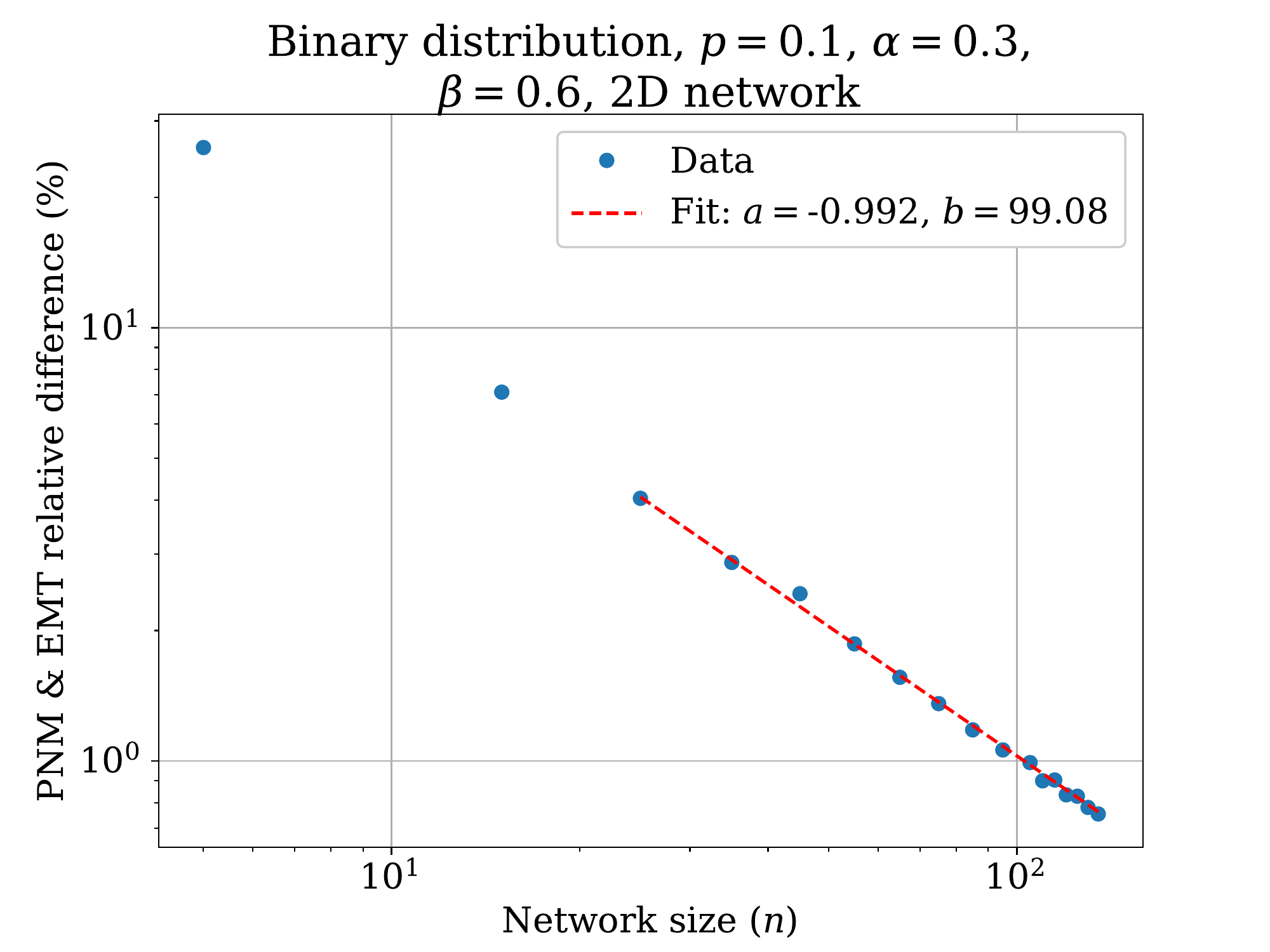}
        \caption{}
        \label{fig:app-binary-p0p1-2d-err}
    \end{subfigure}
    \begin{subfigure}[b]{\width\linewidth}
        \includegraphics[width=\linewidth]{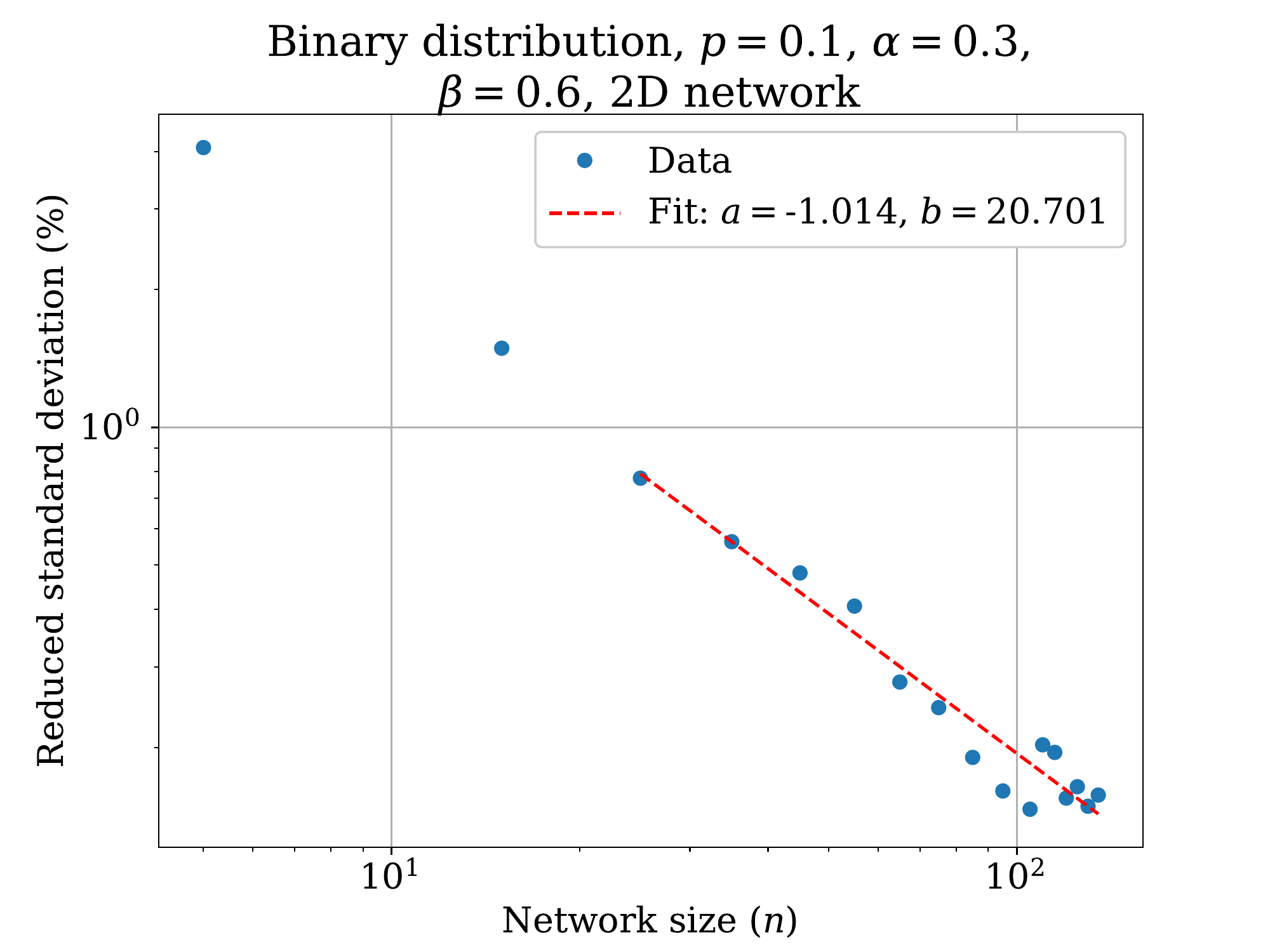}
        \caption{}
        \label{fig:app-binary-p0p1-2d-rsd}
    \end{subfigure}
    
    \begin{subfigure}[b]{\width\linewidth}
        \includegraphics[width=\linewidth]{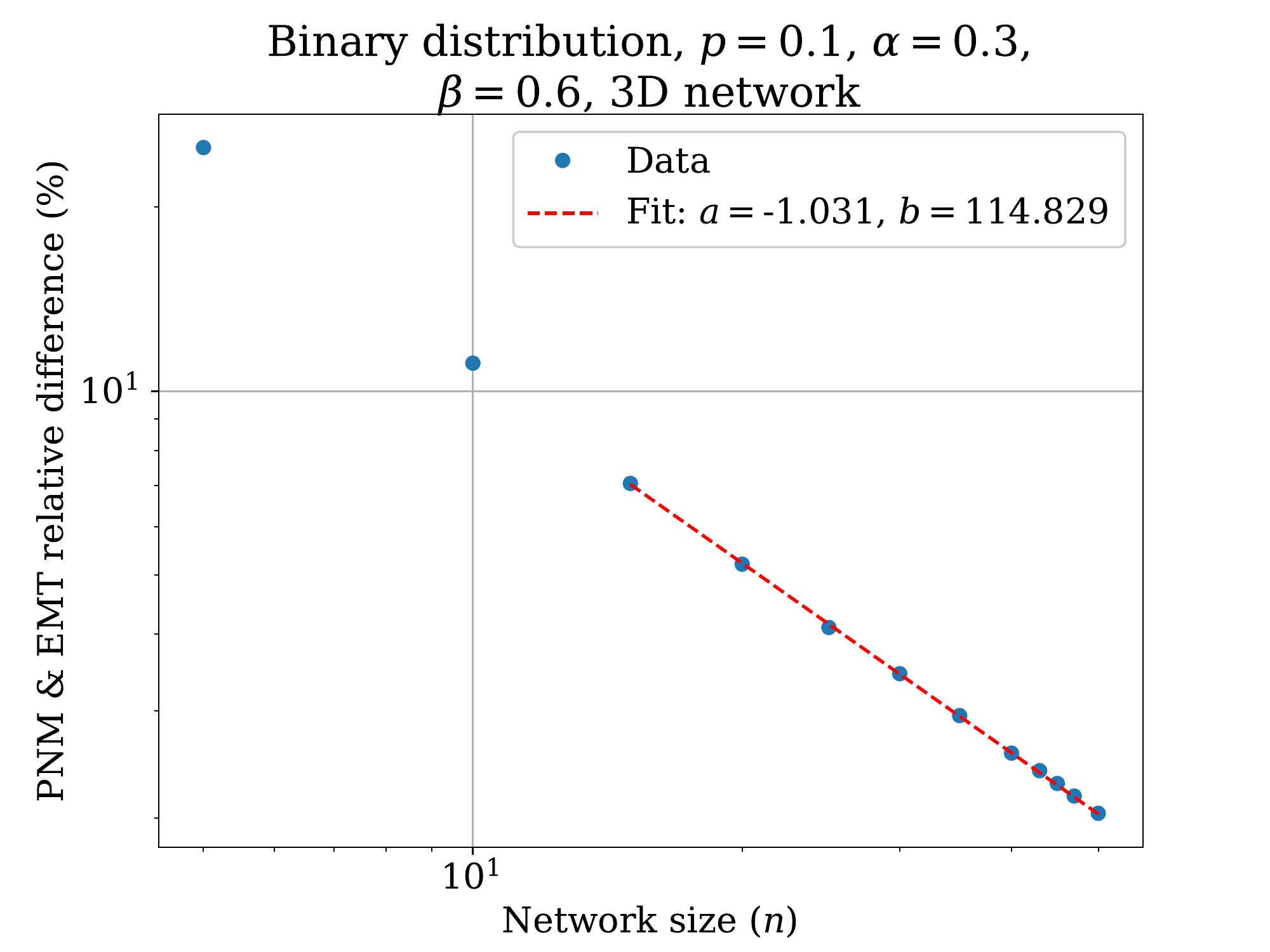}
        \caption{}
        \label{fig:app-binary-p0p1-3d-err}
    \end{subfigure}
    \begin{subfigure}[b]{\width\linewidth}
        \includegraphics[width=\linewidth]{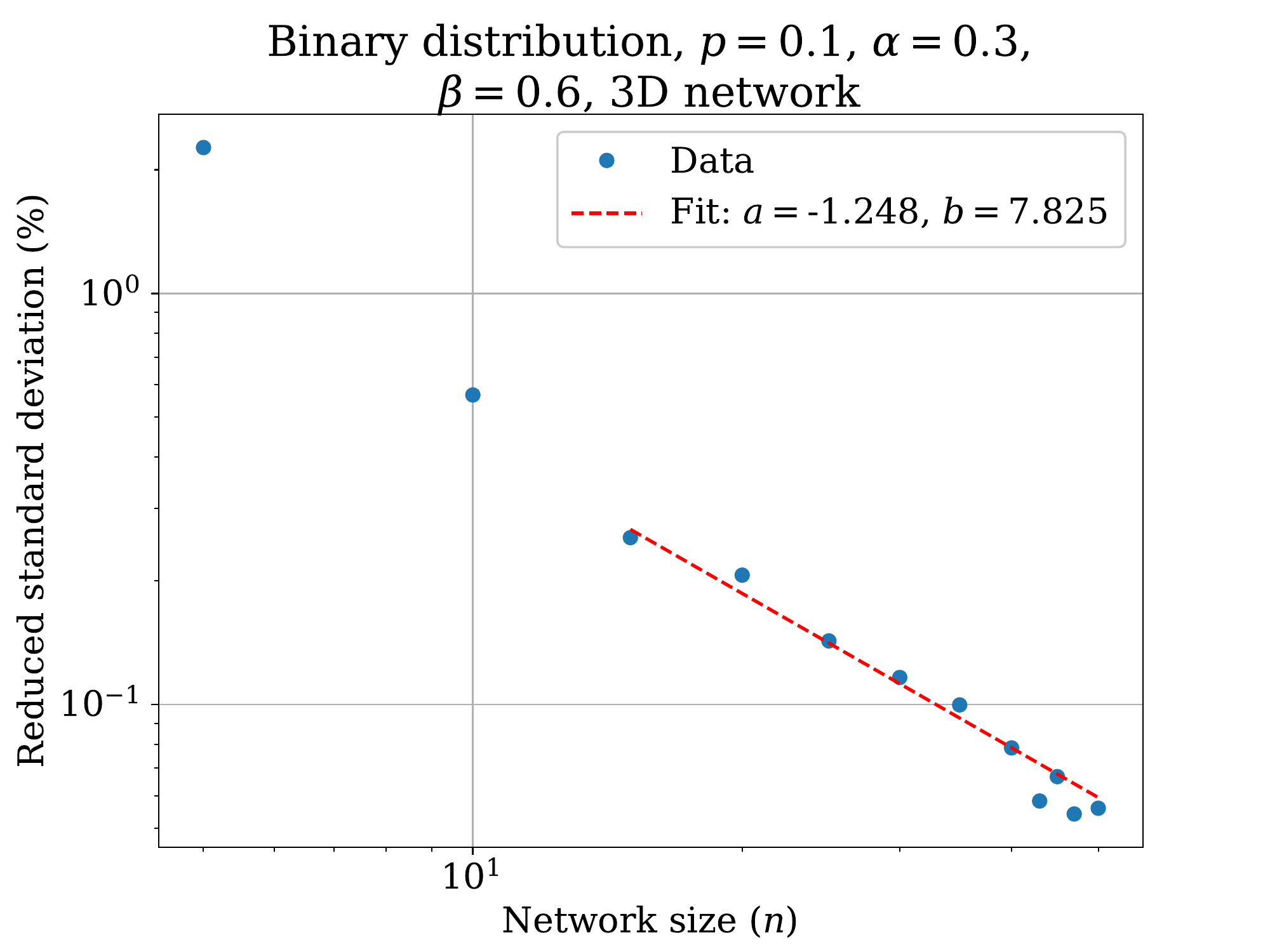}
        \caption{}
        \label{fig:app-binary-p0p1-3d-rsd}
    \end{subfigure}
    
\caption[]{Power law fits to relative difference (\acs{RD}) and reduced standard deviation (\acs{RSD}) of \acs{PNM} solution, as a function of $n$.}
\end{figure}

\begin{figure}[H]\ContinuedFloat
    \centering
    
    \begin{subfigure}[b]{\width\linewidth}
        \includegraphics[width=\linewidth]{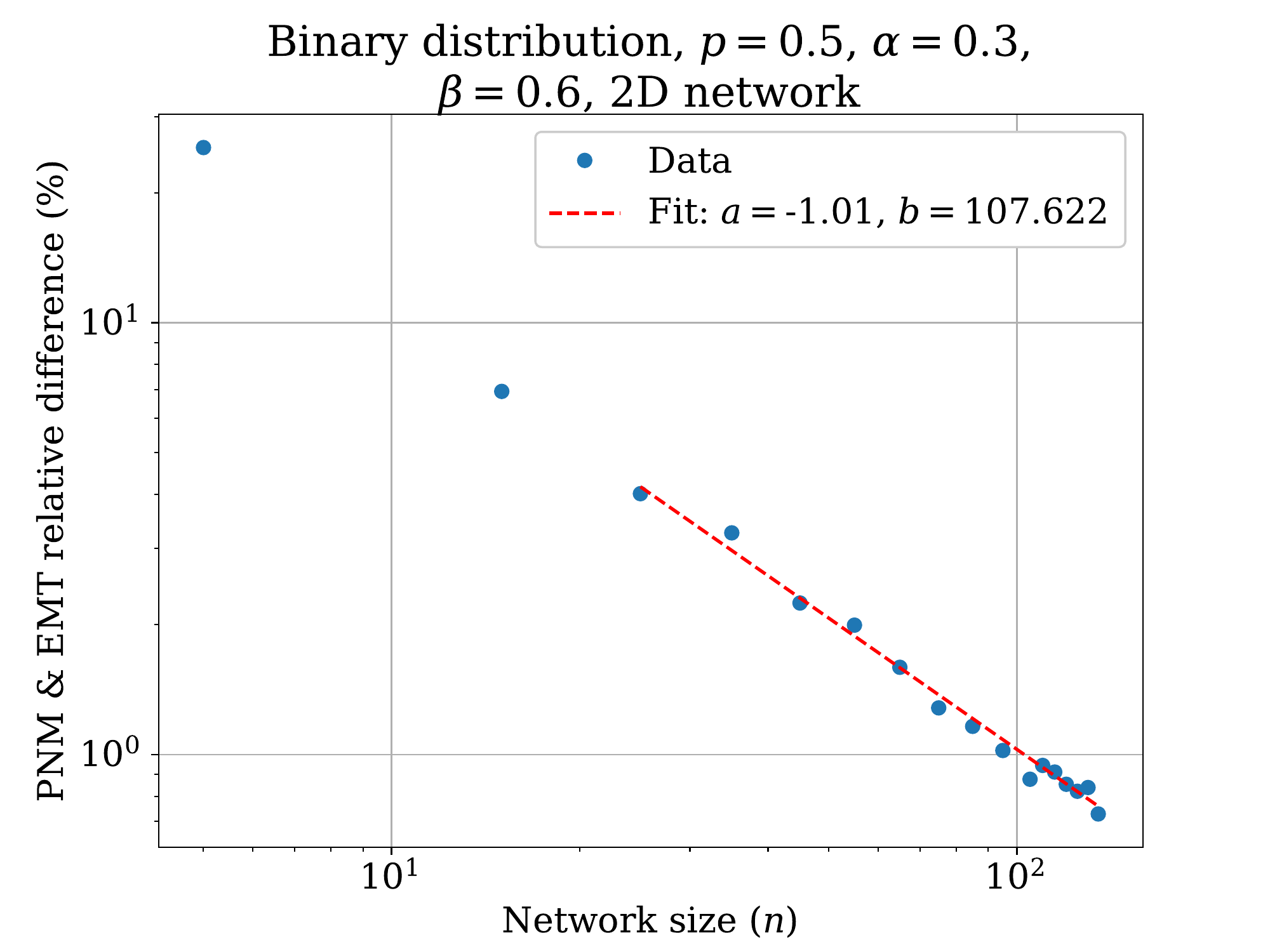}
        \caption{}
        \label{fig:app-binary-p0p5-2d-err}
    \end{subfigure}
    \begin{subfigure}[b]{\width\linewidth}
        \includegraphics[width=\linewidth]{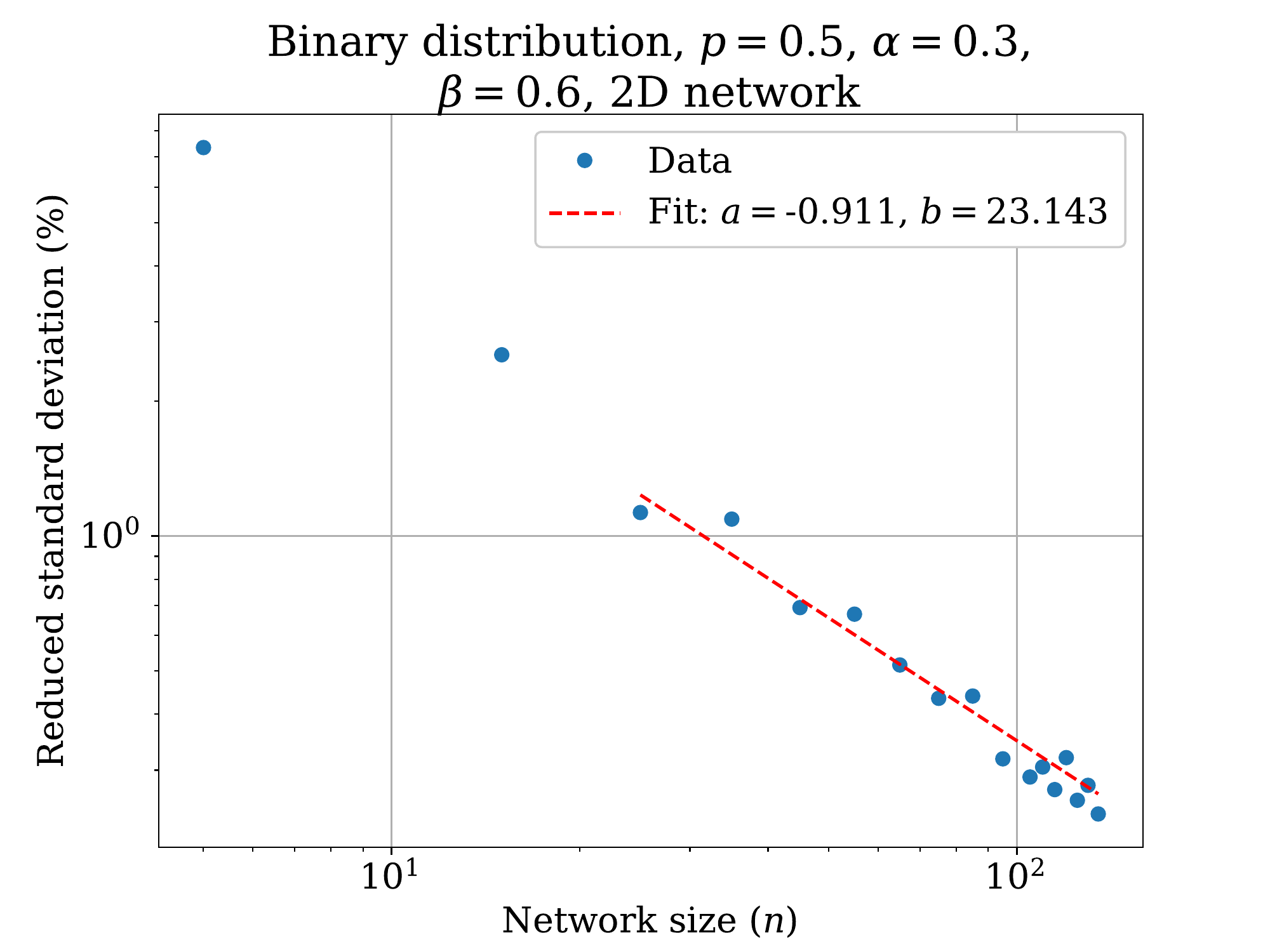}
        \caption{}
        \label{fig:app-binary-p0p5-2d-rsd}
    \end{subfigure}
    
    \begin{subfigure}[b]{\width\linewidth}
        \includegraphics[width=\linewidth]{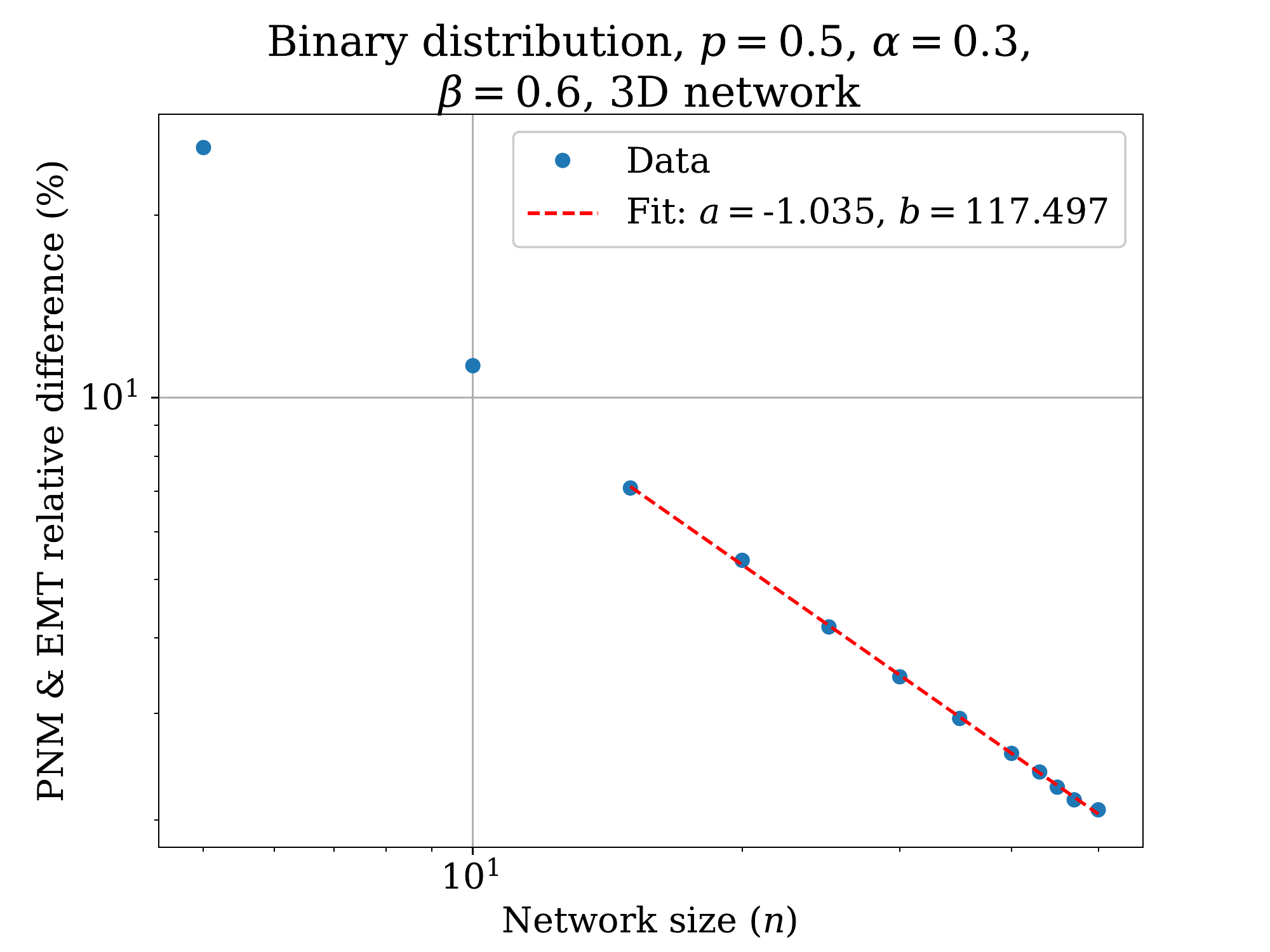}
        \caption{}
        \label{fig:app-binary-p0p5-3d-err}
    \end{subfigure}
    \begin{subfigure}[b]{\width\linewidth}
        \includegraphics[width=\linewidth]{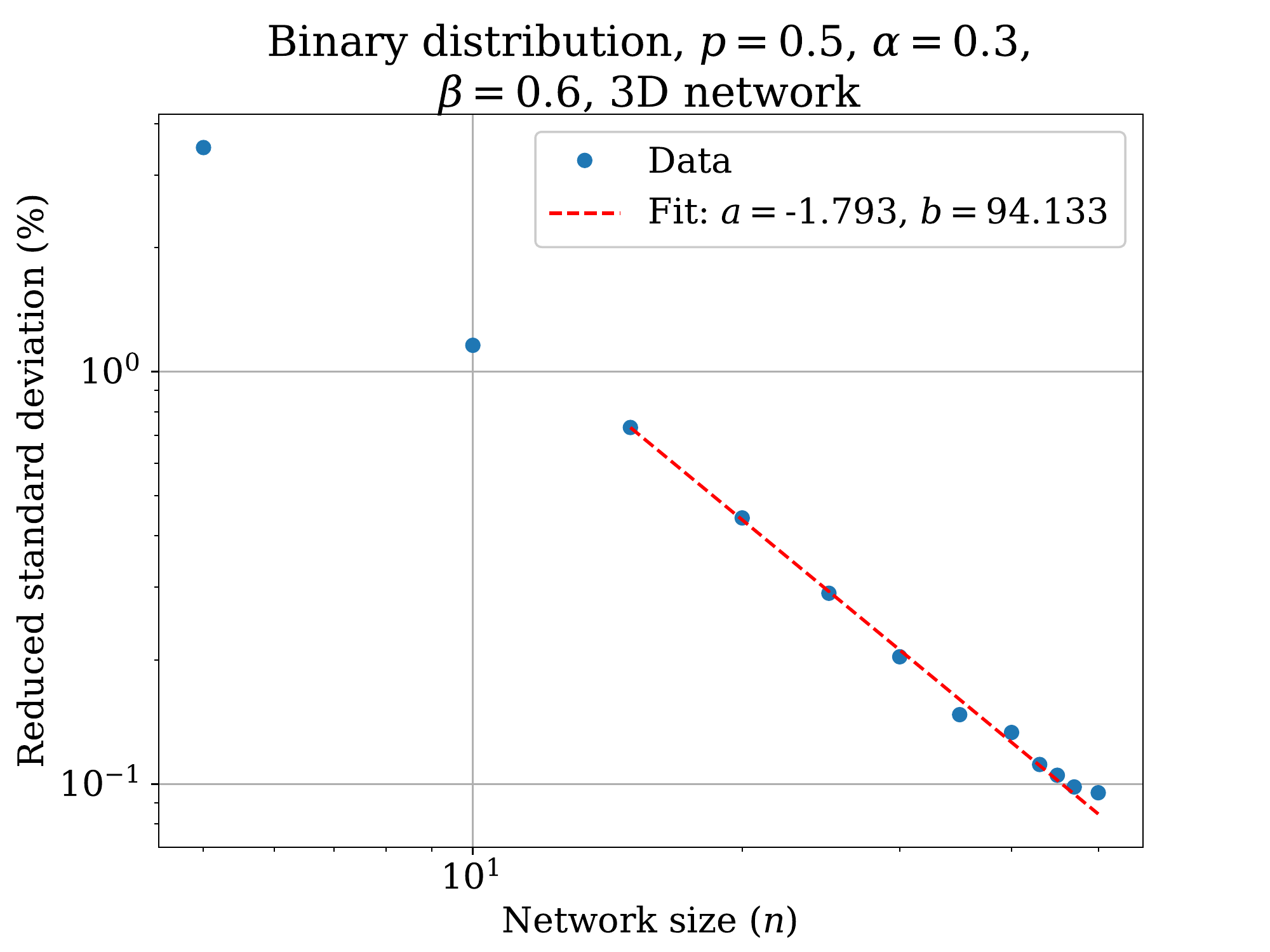}
        \caption{}
        \label{fig:app-binary-p0p5-3d-rsd}
    \end{subfigure}
    
    \begin{subfigure}[b]{\width\linewidth}
        \includegraphics[width=\linewidth]{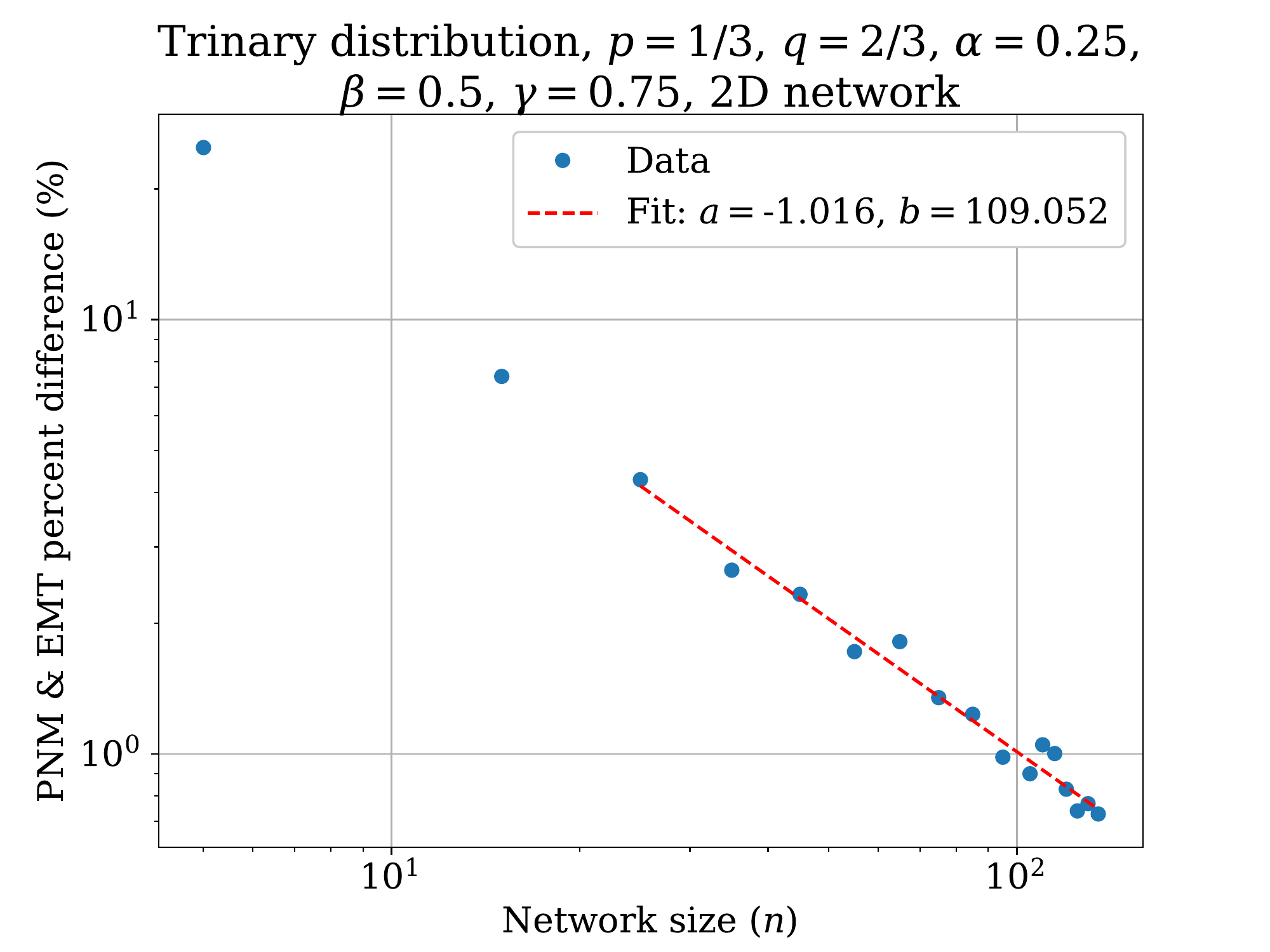}
        \caption{}
        \label{fig:app-trinary-2d-err}
    \end{subfigure}
    \begin{subfigure}[b]{\width\linewidth}
        \includegraphics[width=\linewidth]{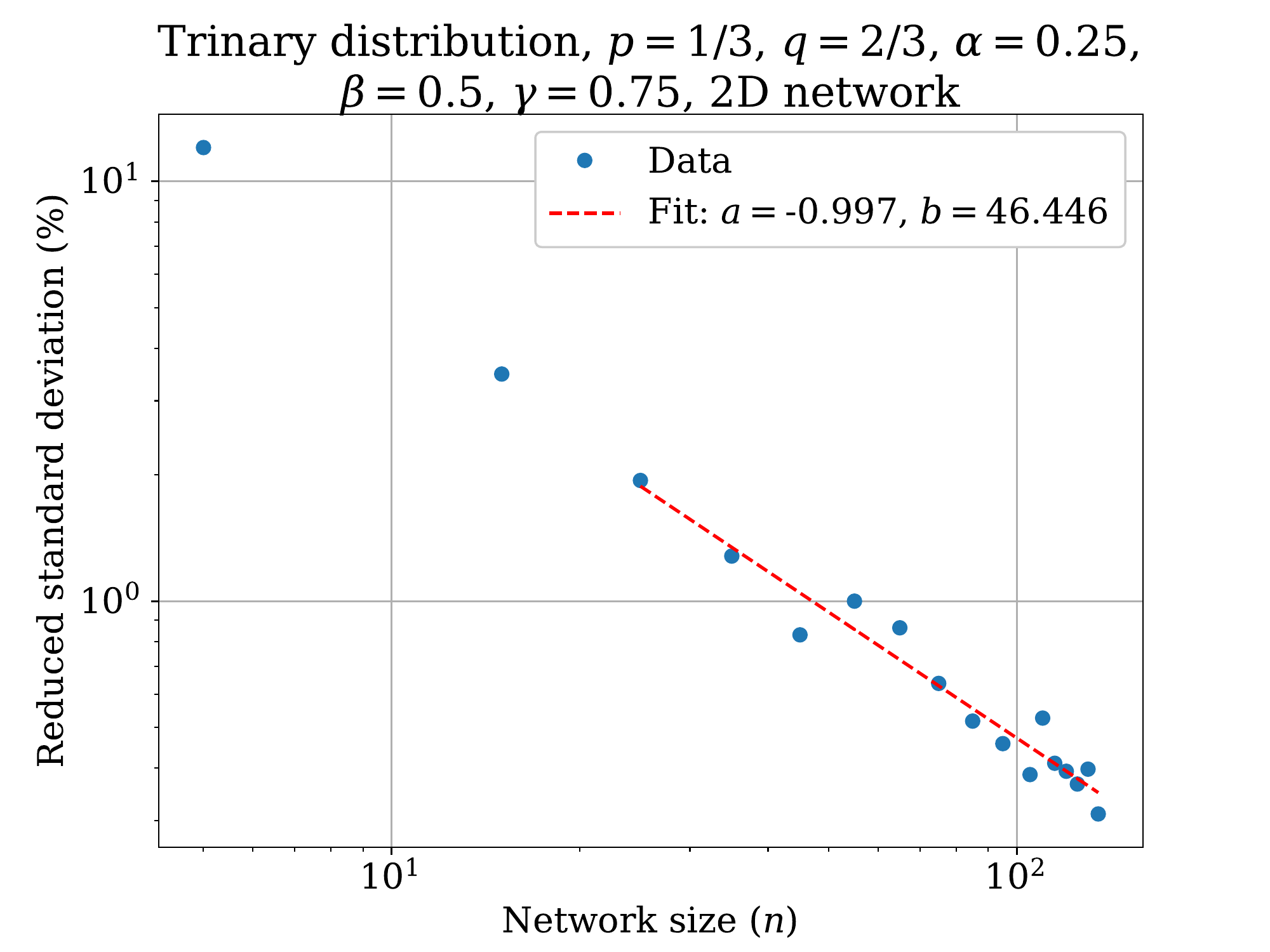}
        \caption{}
        \label{fig:app-trinary-2d-rsd}
    \end{subfigure}
    
\caption[]{Power law fits to relative difference (\acs{RD}) and reduced standard deviation (\acs{RSD}) of \acs{PNM} solution, as a function of $n$.}
\end{figure}

\begin{figure}[H]\ContinuedFloat
    \centering
    
    \begin{subfigure}[b]{\width\linewidth}
        \includegraphics[width=\linewidth]{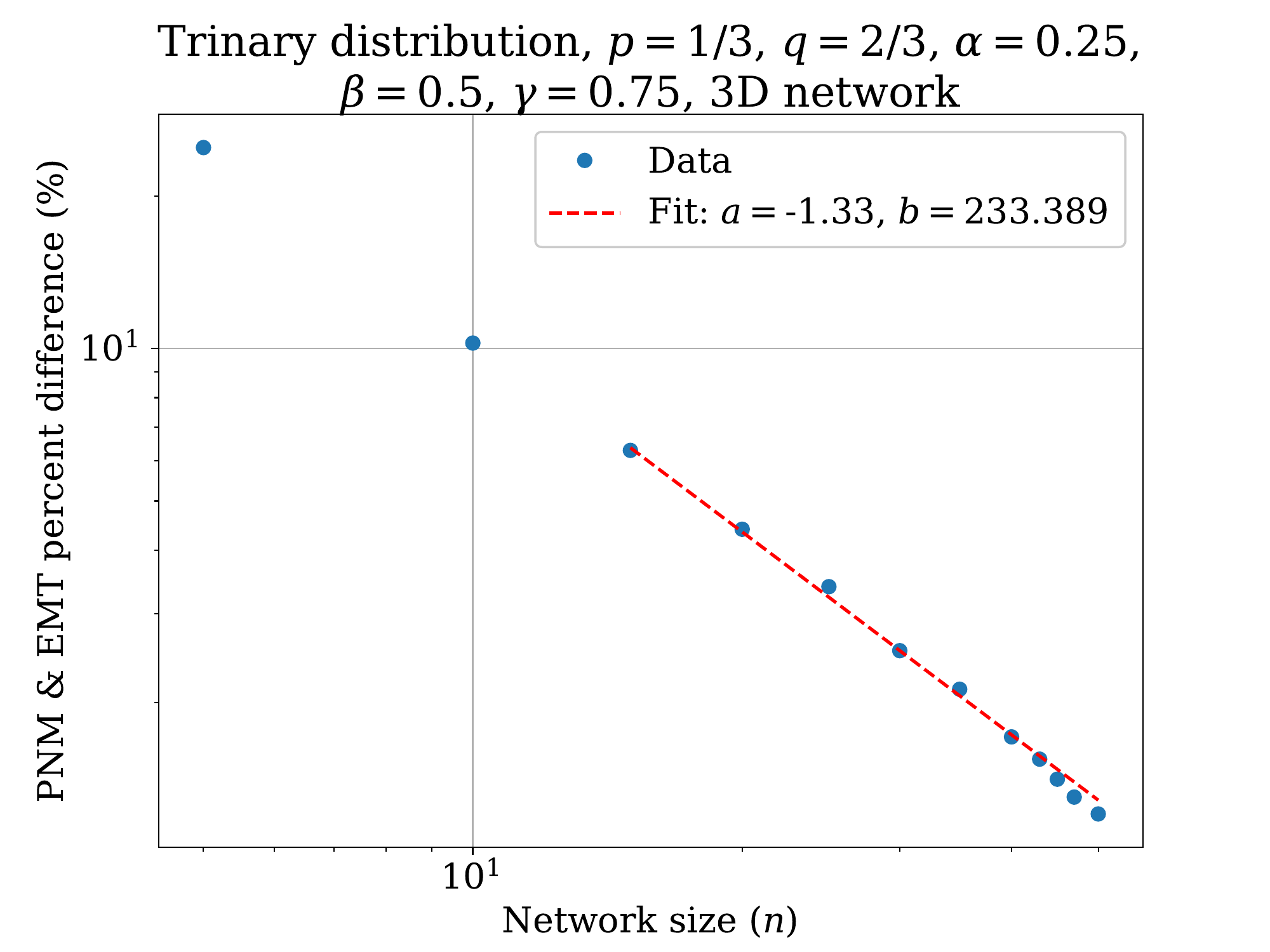}
        \caption{}
        \label{fig:app-trinary-3d-err}
    \end{subfigure}
    \begin{subfigure}[b]{\width\linewidth}
        \includegraphics[width=\linewidth]{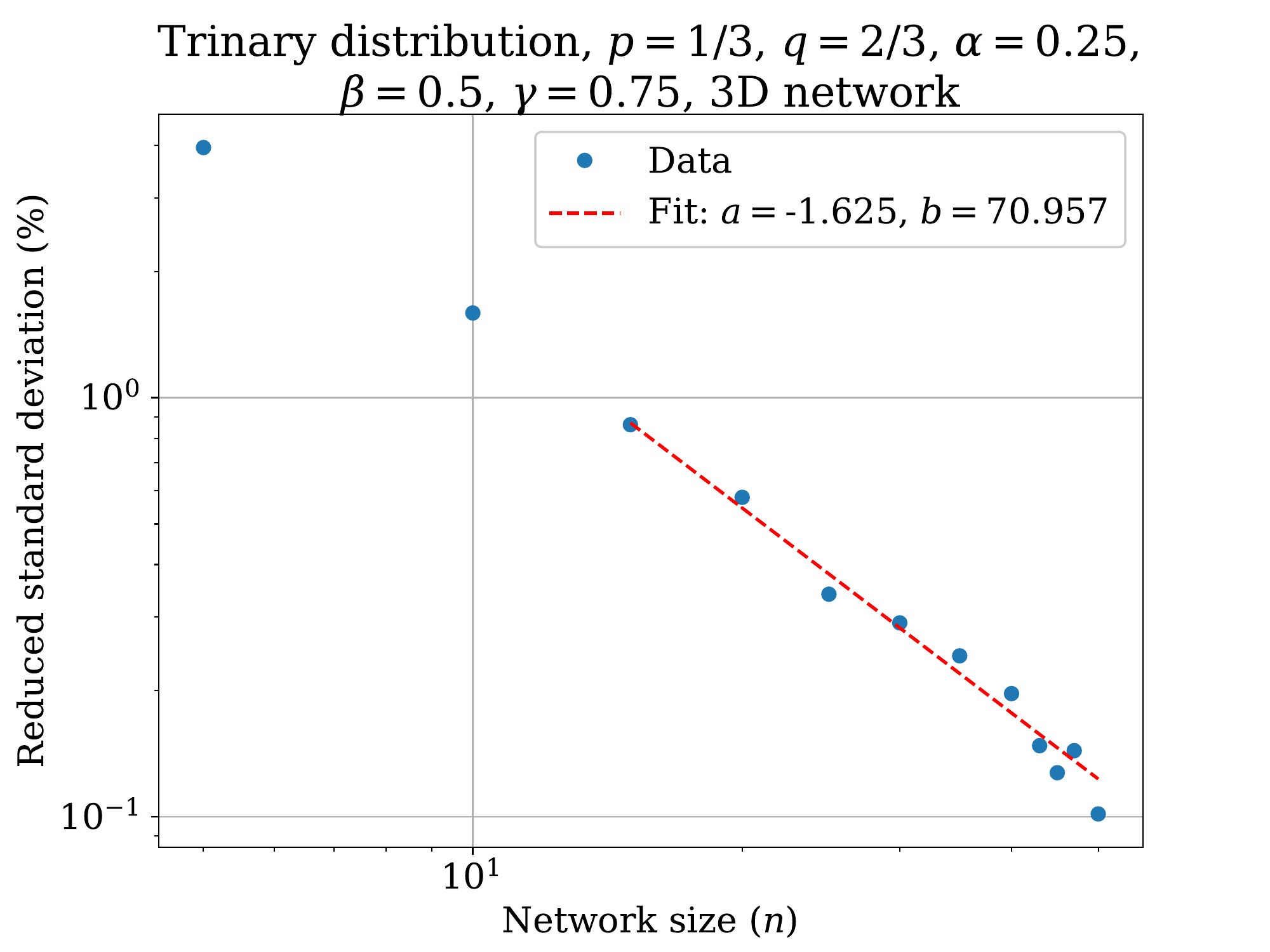}
        \caption{}
        \label{fig:app-trinary-3d-rsd}
    \end{subfigure}
    
    \begin{subfigure}[b]{\width\linewidth}
        \includegraphics[width=\linewidth]{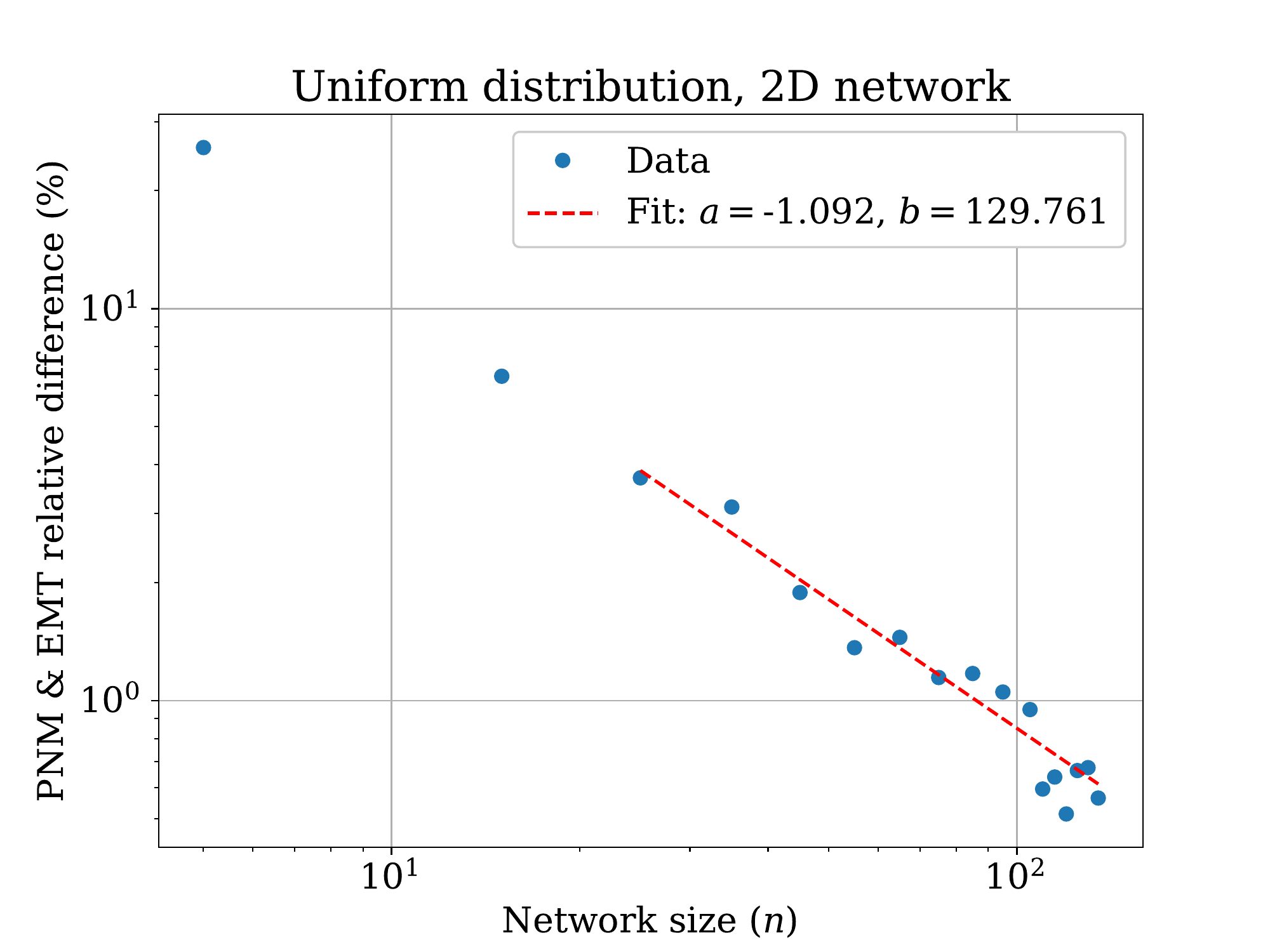}
        \caption{}
        \label{fig:app-uniform-2d-err}
    \end{subfigure}
    \begin{subfigure}[b]{\width\linewidth}
        \includegraphics[width=\linewidth]{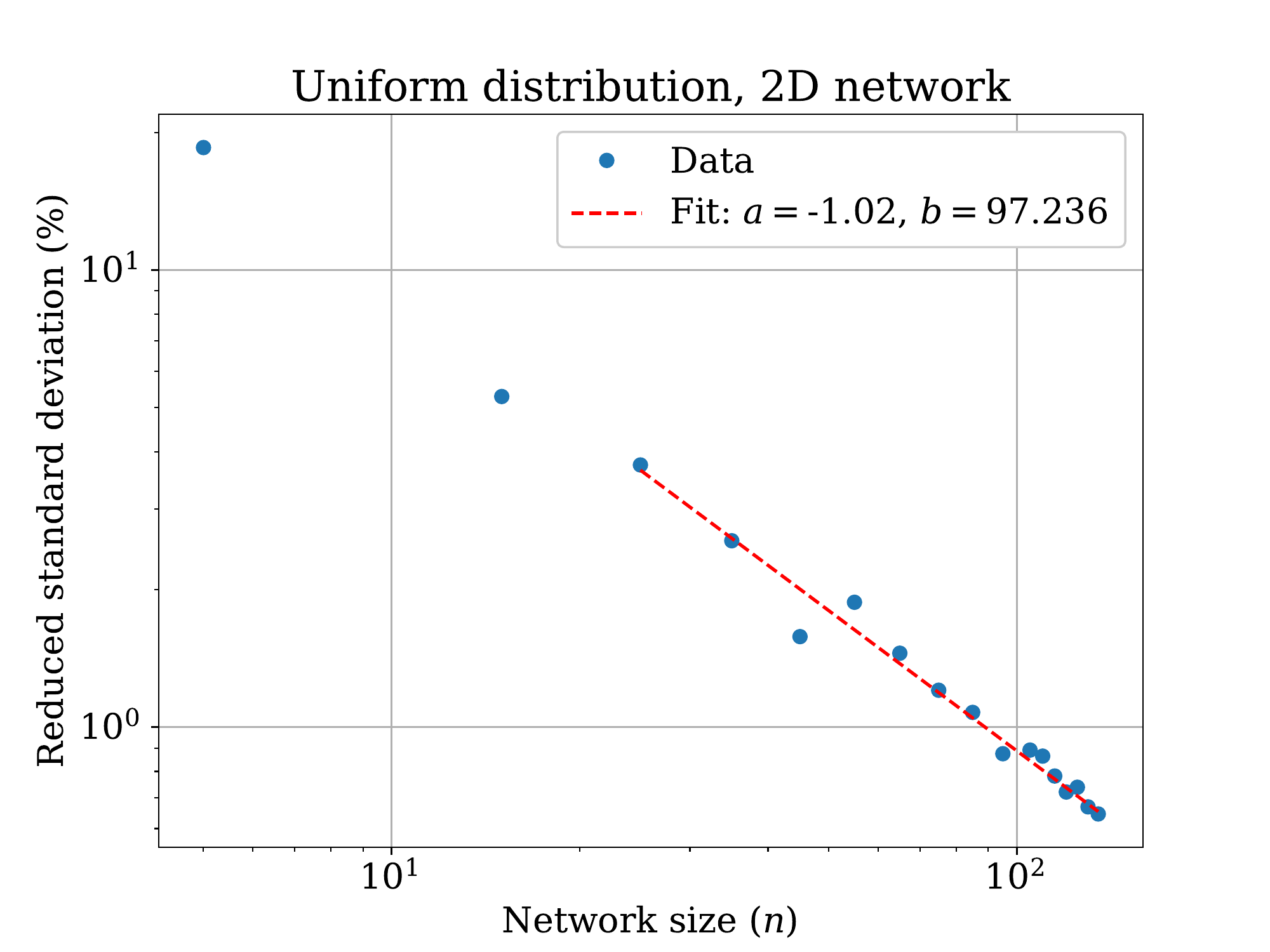}
        \caption{}
        \label{fig:app-uniform-2d-rsd}
    \end{subfigure}

    \begin{subfigure}[b]{\width\linewidth}
        \includegraphics[width=\linewidth]{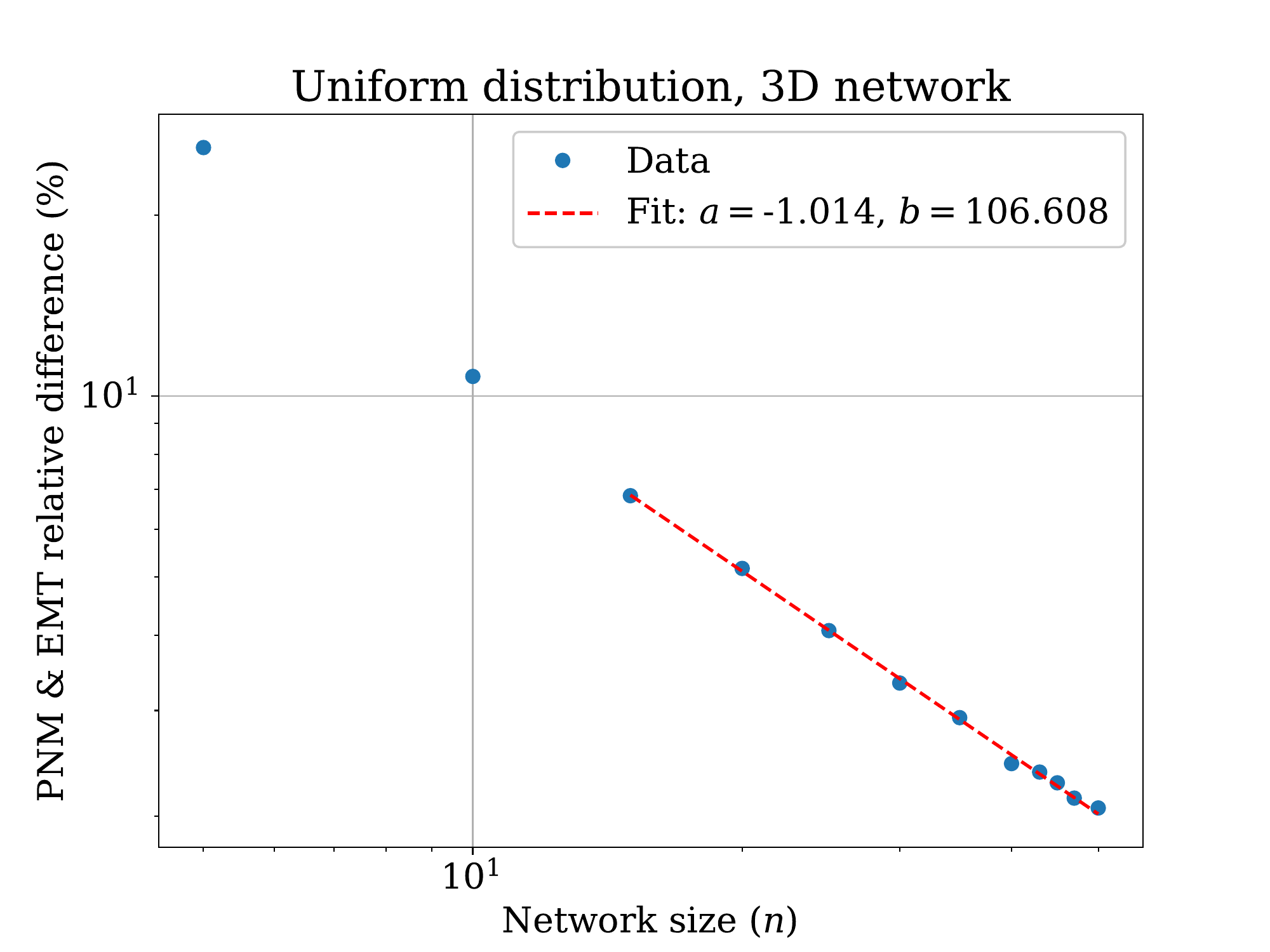}
        \caption{}
        \label{fig:app-uniform-3d-err}
    \end{subfigure}
    \begin{subfigure}[b]{\width\linewidth}
        \includegraphics[width=\linewidth]{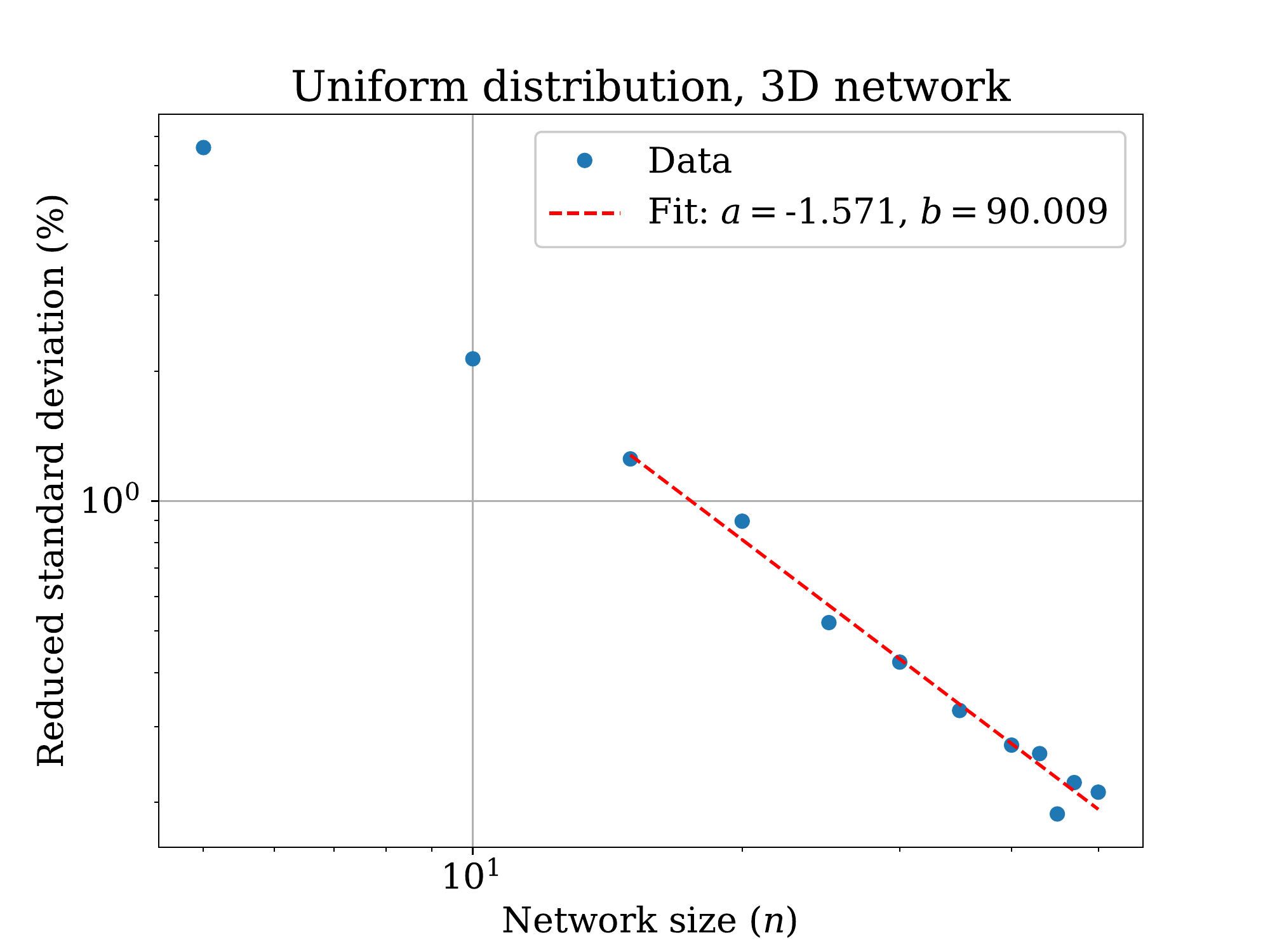}
        \caption{}
        \label{fig:app-uniform-3d-rsd}
    \end{subfigure}
    
\caption[]{Power law fits to relative difference (\acs{RD}) and reduced standard deviation (\acs{RSD}) of \acs{PNM} solution, as a function of $n$.}
\end{figure}

\begin{figure}[H]\ContinuedFloat
    \centering

    \begin{subfigure}[b]{\width\linewidth}
        \includegraphics[width=\linewidth]{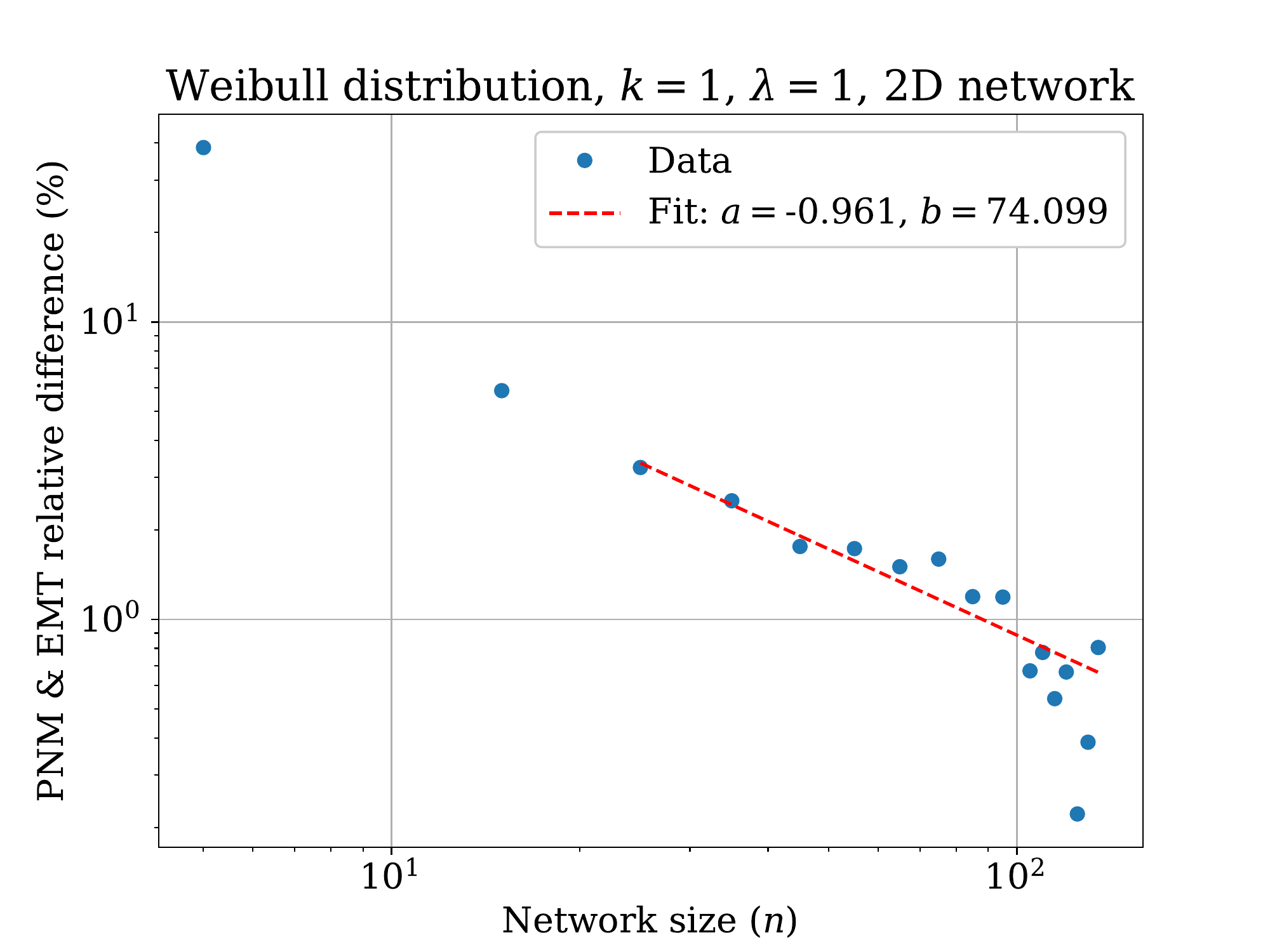}
        \caption{}
        \label{fig:app-weibull-1-2d-err}
    \end{subfigure}
    \begin{subfigure}[b]{\width\linewidth}
        \includegraphics[width=\linewidth]{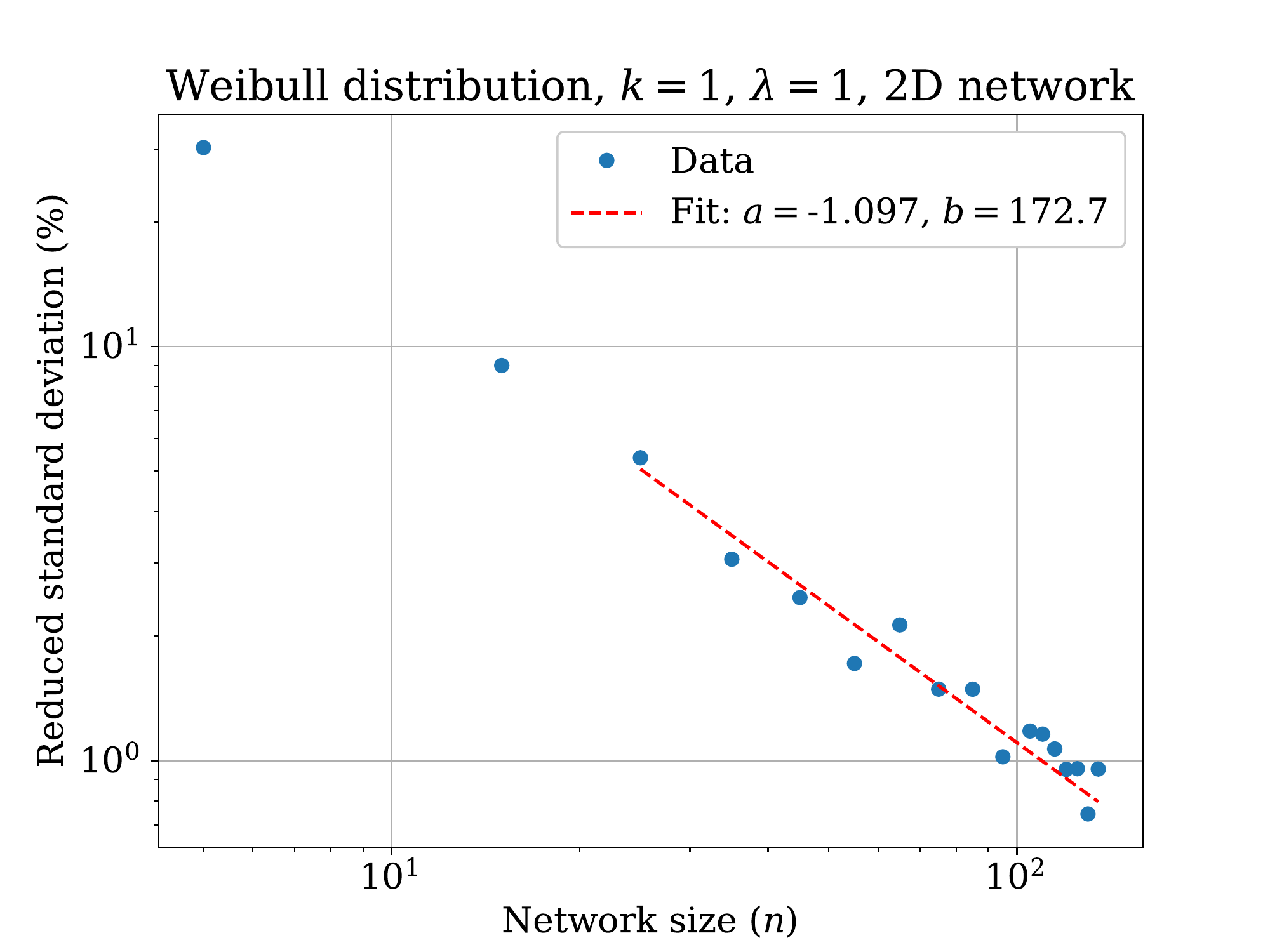}
        \caption{}
        \label{fig:app-weibull-1-2d-rsd}
    \end{subfigure}

    \begin{subfigure}[b]{\width\linewidth}
        \includegraphics[width=\linewidth]{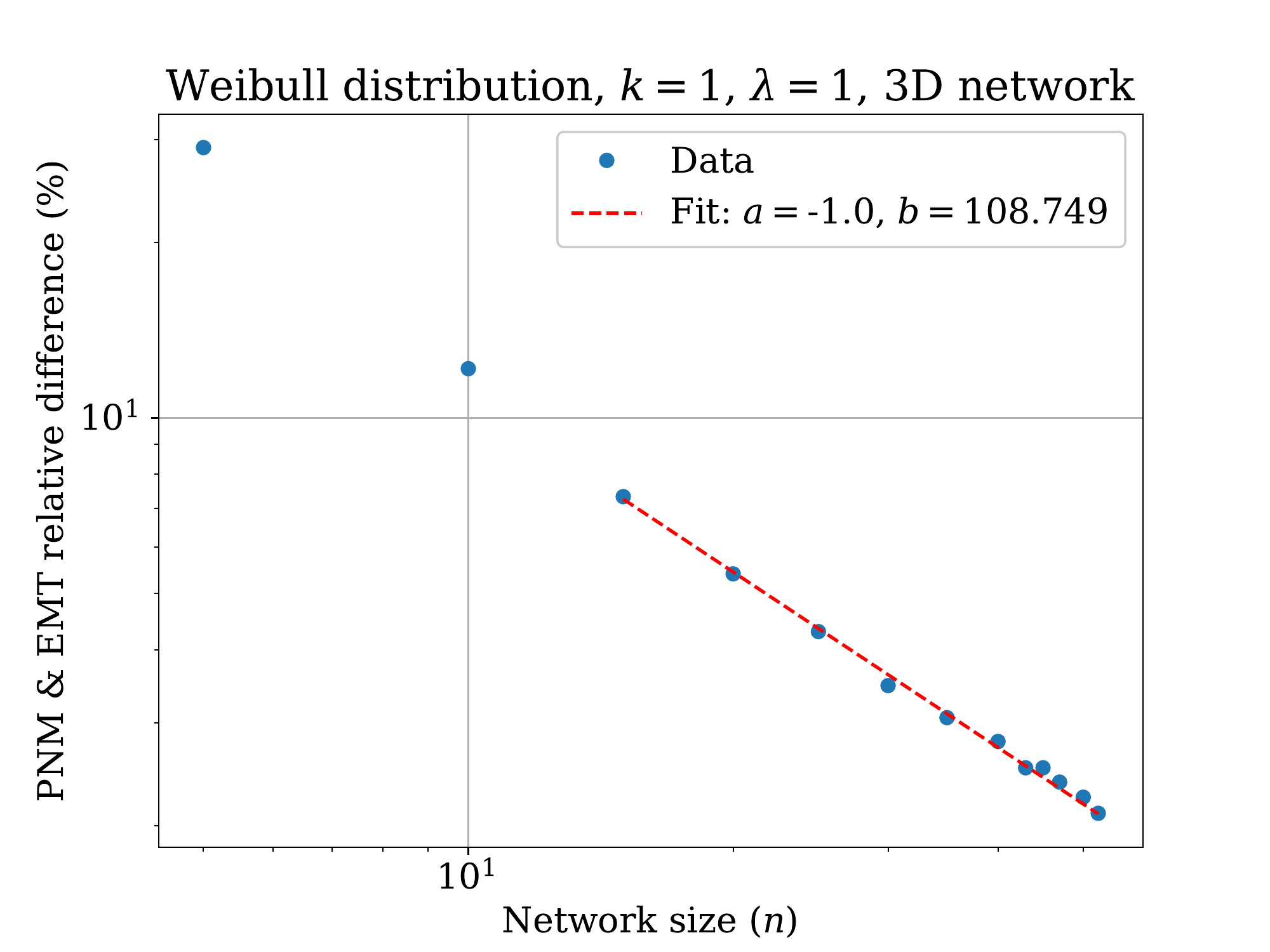}
        \caption{}
        \label{fig:app-weibull-1-3d-err}
    \end{subfigure}
    \begin{subfigure}[b]{\width\linewidth}
        \includegraphics[width=\linewidth]{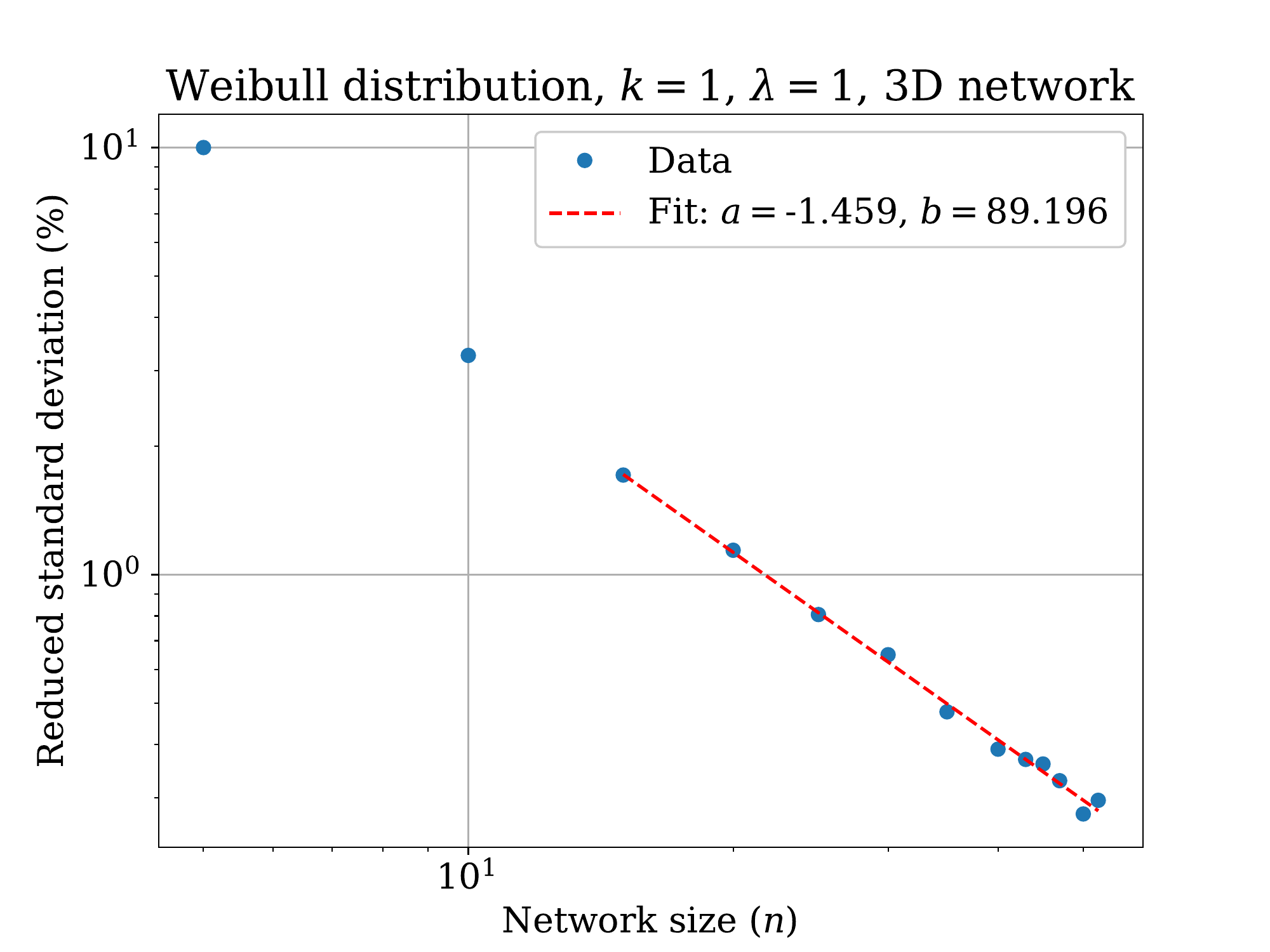}
        \caption{}
        \label{fig:app-weibull-1-3d-rsd}
    \end{subfigure}

    \begin{subfigure}[b]{\width\linewidth}
        \includegraphics[width=\linewidth]{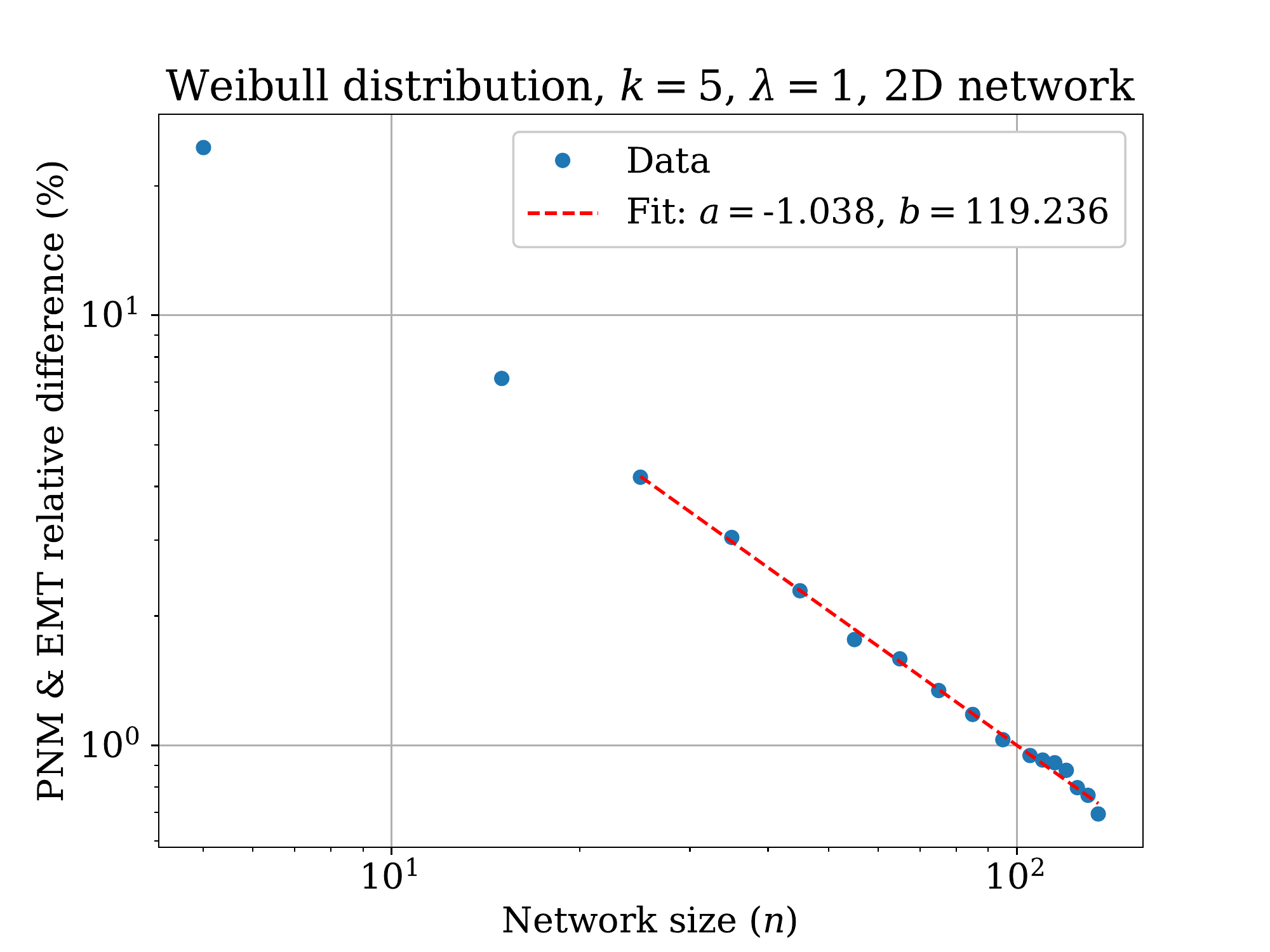}
        \caption{}
        \label{fig:app-weibull-5-2d-err}
    \end{subfigure}
    \begin{subfigure}[b]{\width\linewidth}
        \includegraphics[width=\linewidth]{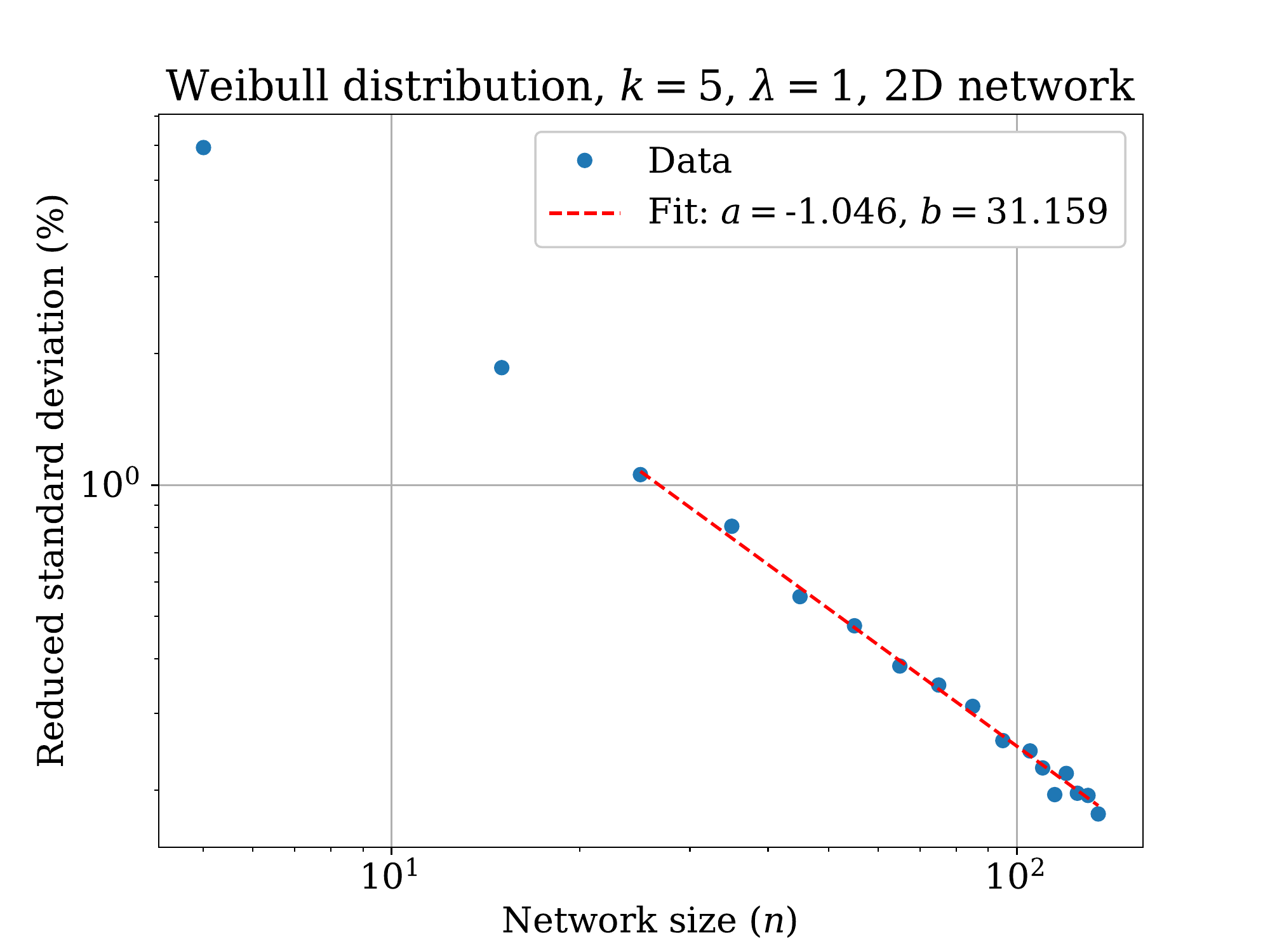}
        \caption{}
        \label{fig:app-weibull-5-2d-rsd}
    \end{subfigure}
    
\caption[]{Power law fits to relative difference (\acs{RD}) and reduced standard deviation (\acs{RSD}) of \acs{PNM} solution, as a function of $n$.}
\end{figure}

\begin{figure}[H]\ContinuedFloat
    \centering

    \begin{subfigure}[b]{\width\linewidth}
        \includegraphics[width=\linewidth]{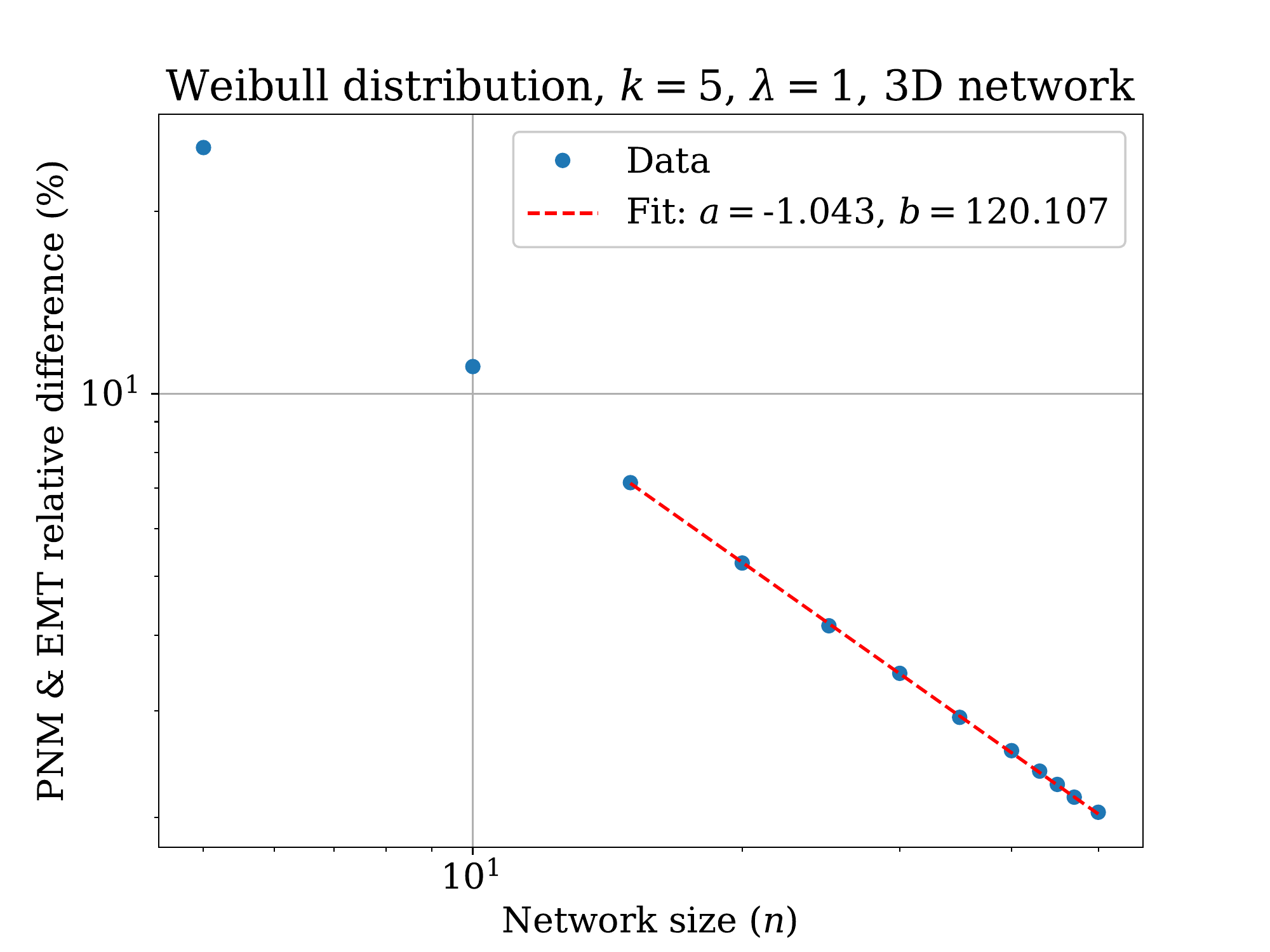}
        \caption{}
        \label{fig:app-weibull-5-3d-err}
    \end{subfigure}
    \begin{subfigure}[b]{\width\linewidth}
        \includegraphics[width=\linewidth]{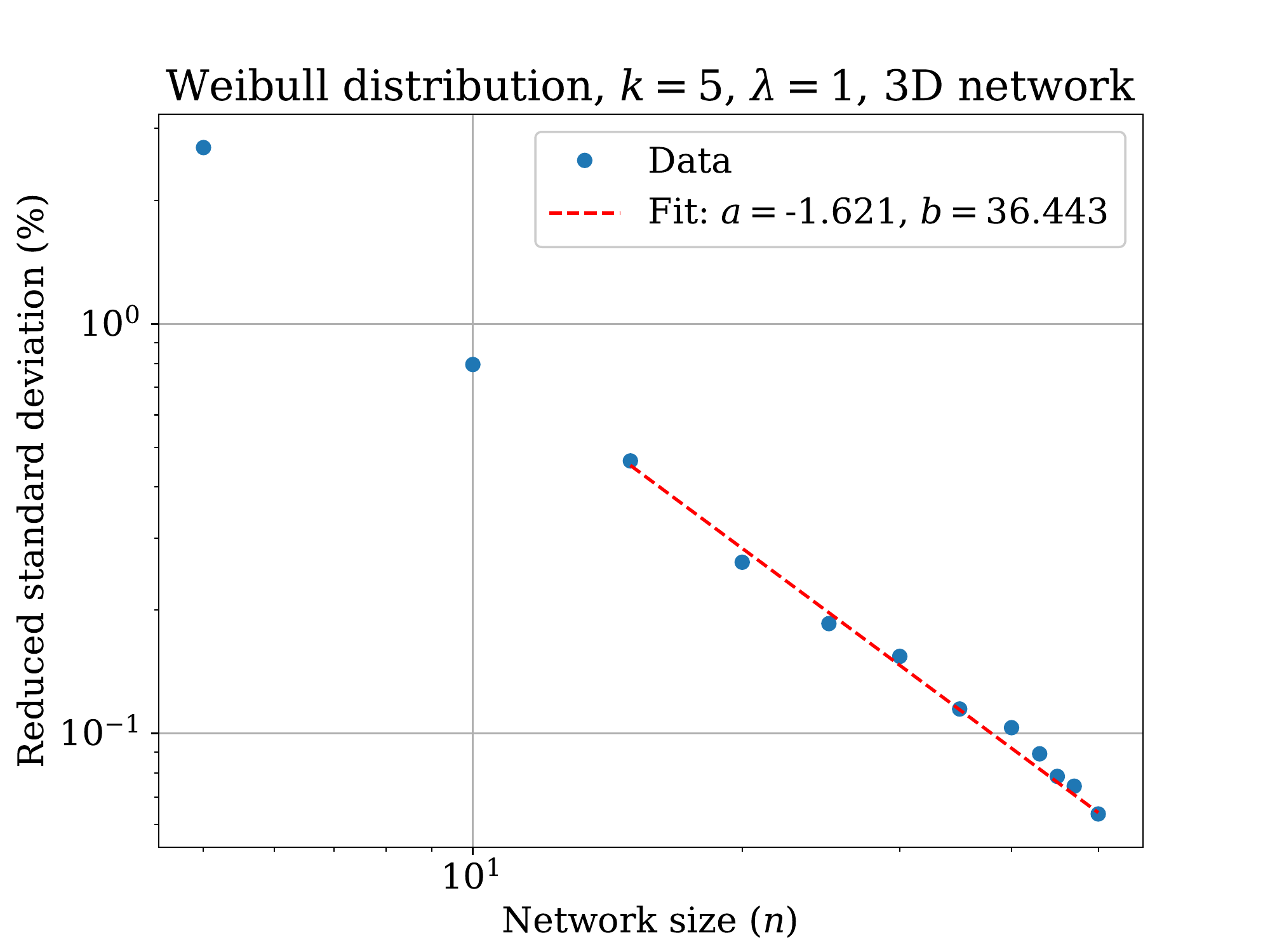}
        \caption{}
        \label{fig:app-weibull-5-3d-rsd}
    \end{subfigure}
    
\caption[]{Power law fits to relative difference (\acs{RD}) and reduced standard deviation (\acs{RSD}) of \acs{PNM} solution, as a function of $n$.}
\end{figure}

\section{Least-square fit parameter errors}
\label{app:errors}

\begin{table}[htbp]
  \footnotesize
  \centering
  \caption{Least-square fit errors (\%) for fit parameters in \autoref{tab:fits}.}
    \begin{tabular}{l|r|r|r|r|r|r|r|r}
    \textit{Distribution} & \multicolumn{1}{l|}{2D $a_{\mathrm{RD}}$} & \multicolumn{1}{l|}{2D $b_{\mathrm{RD}}$} & \multicolumn{1}{l|}{2D $a_{\mathrm{RSD}}$} & \multicolumn{1}{l|}{2D $b_{\mathrm{RSD}}$}& \multicolumn{1}{l|}{3D $a_{\mathrm{RD}}$} & \multicolumn{1}{l|}{3D $b_{\mathrm{RD}}$} & \multicolumn{1}{l|}{3D $a_{\mathrm{RSD}}$} & \multicolumn{1}{l}{3D $b_{\mathrm{RSD}}$} \\
    \hline
    Arcsine & 6.922 & 60.533 & 1.808 & 7.049 & 1.360 & 4.656 & 4.543 & 21.478 \\
    Bimodal sine & 2.745 & 11.329 & 3.126 & 11.298 & 1.080 & 3.685 & 2.245 & 10.053 \\
    Binary 1 & 1.391 & 5.232 & 4.361 & 16.710 & 0.466 & 1.522 & 5.346 & 20.638 \\
    Binary 2 & 2.931 & 11.192 & 6.241 & 21.839 & 0.798 & 2.612 & 1.361 & 7.196 \\
    Trinary & 3.665 & 14.067 & 5.583 & 21.084 & 1.757 & 7.171 & 3.640 & 17.674 \\
    Uniform & 5.495 & 22.402 & 4.421 & 17.024 & 0.961 & 3.090 & 4.090 & 19.290 \\
    Weibull 1 & 8.423 & 30.852 & 5.394 & 22.076 & 1.469 & 4.700 & 1.267 & 5.628 \\
    Weibull 2 & 5.738 & 22.832 & 2.734 & 10.710 & 2.981 & 9.765 & 7.632 & 34.744 \\
    Weibull 3 & 1.102 & 4.307 & 1.941 & 7.633 & 0.364 & 1.199 & 3.211 & 15.564 \\
    \end{tabular}%
  \label{tab:app-fits-error}%
\end{table}%

\begin{table}[htbp]
  \centering
  \caption{Least-square fit errors (\%) for fit parameters in \autoref{tab:fits-quasi-2d}.}
    \footnotesize
    \begin{tabular}{l|r|r|r|r|r|r}
    \textit{Distribution} & \multicolumn{1}{l|}{$a_{\mathrm{RD}}$, 2D \acs{EMT}} & \multicolumn{1}{l|}{$b_{\mathrm{RD}}$, 2D \acs{EMT}} & \multicolumn{1}{l|}{$a_{\mathrm{RD}}$, 3D \acs{EMT}} & \multicolumn{1}{l|}{$b_{\mathrm{RD}}$, 3D \acs{EMT}} & \multicolumn{1}{l|}{$a_{\mathrm{RSD}}$} & \multicolumn{1}{l}{$b_{\mathrm{RSD}}$} \\
    \hline
    Weibull 2 & 5.436 & 4.961 & 10.100 & 61.340 & 3.087 & 12.332 \\
    \end{tabular}%
  \label{tab:addlabel}%
\end{table}%

\end{document}